\documentclass[12pt,letterpaper]{article}
\pdfoutput=1
\usepackage{jheppub}
\usepackage{color,float,graphicx,hyperref}
\usepackage{amsmath,amssymb,subfig}
\usepackage{soul}

\newcommand{\cO}{\mathcal{O}}
\newcommand{\pts}{H_{T,J}}
\newcommand{\rjet}{R_{\rm jet}}

\begin{document}
\title{Jet Timing}
\author[a]{Wen Han Chiu,}
\author[b]{Zhen Liu,}
\author[c]{Matthew Low,}
\author[a,d,e]{and Lian-Tao Wang}
\affiliation[a]{Department of Physics, University of Chicago, Chicago, Illinois 60637, U.S.A.}
\affiliation[b]{School of Physics and Astronomy, University of Minnesota, Minneapolis, MN 55455, U.S.A.}
\affiliation[c]{Theoretical Physics Department, Fermilab, P.O. Box 500, Batavia, Illinois 60510, U.S.A.}
\affiliation[d]{Enrico Fermi Institute, University of Chicago, Chicago, Illinois 60637, U.S.A}
\affiliation[e]{Kavli Institute for Cosmological Physics, University of Chicago, Chicago, Illinois 60637, U.S.A.}
\emailAdd{wenhan@uchicago.edu}
\emailAdd{zliuphys@umn.edu}
\emailAdd{mattlow@fnal.gov}
\emailAdd{liantaow@uchicago.edu}
\preprint{FERMILAB-PUB-21-372-T}
\date{\today}

\abstract{The measurement of the arrival time of a particle, such as a lepton, a photon, or a pion, reaching the detector provides valuable information. A similar measurement for a hadronic final state, however, is much more challenging as one has to extract the relevant information from a collection of particles.  In this paper, we explore various possibilities in defining the time of a jet through the measurable arrival times of the jet constituents.  We find that a definition of jet time based on a transverse momentum weighted sum of the times of the constituents has the best performance.  For prompt jets, the performance depends on the jet trajectory.  For delayed jets, the performance depends on the trajectory of the jet, the trajectory of the mother particle, and the location of the displaced vertex.  Compared to the next-best-performing jet time definition, the transverse momentum weighted sum has roughly a factor of ten times better jet time resolution.  We give a detailed discussion of the relevant effects and characterize the full geometrical dependence of the performance.  These results highlight the critical importance of using a proper definition of jet time with its corresponding detector-dependent calibration and the exciting possibility of deepening our understanding of jets in the time domain.}

\maketitle

%%%%%%%%%%%%%%%%%%%%%%%%%%%%%%%%%%%%%%%%%%%%%%%%%%%%%%%%%%%%%%%%%%
%%%%%%%%%%%%%%%%%%%%%%%%%%%%%%%%%%%%%%%%%%%%%%%%%%%%%%%%%%%%%%%%%%
\section{Introduction}
\label{sec:intro}

The time at which a particle arrives at a particular detector layer is a piece of independently measurable and valuable information. Measuring the time of a lepton, a photon, or a hadron has been used extensively at the Large Hadron Collider (LHC) to great effect.\footnote{Timing has been used in existing searches for heavy stable charged particles~\cite{CMS:2016kce,ATLAS:2019gqq}, stopped particles~\cite{CMS:2017kku,ATLAS:2021mdj}, and non-pointing photons~\cite{ATLAS:2014kbb,CMS:2019zxa}, where spatial information is unavailable or ineffective.}  Recently, it was shown that timing information is vital in the search for long-lived particles (LLPs)~\cite{Liu:2018wte}.  The upgraded electronics at the high-luminosity LHC will significantly improve the timing resolution for various subdetectors, reaching tens of picoseconds in some cases, extending the sensitivity of LLP searches even further.  For instance, particle timing can improve prompt detection of beyond the Standard Model (BSM) physics~\cite{Klimek:2019cny}, enable LLP mass and lifetime determination~\cite{Flowers:2019gvj,Kang:2019ukr,Banerjee:2019ktv,Bae:2020dwf}, and enhance other various BSM searches~\cite{ElHedri:2018atj,Cerri:2018rkm,Abada:2018sfh,Frugiuele:2018coc,Kribs:2018ilo,Berlin:2018jbm,Xu:2018ofw,Belanger:2018sti,Evans:2018jmd,Kilic:2018sew,Delgado:2018qxq,Liu:2019ayx,Chakraborti:2019ohe,Serra:2019omd,Mason:2019okp,Du:2019mlc,Zabi:2020gjd,Shuve:2020evk,Yuan:2020eeu,Liu:2020vur,Fuchs:2020cmm,Gershtein:2020mwi,Borsato:2021aum,Cheung:2021utb,Dienes:2021cxr,Bhattacherjee:2020nno}.

Obtaining similar information for final-state quarks and gluons, however, is much more challenging.  These particles undergo showering and hadronization before arriving at the detector as a collection of particles, with a corresponding collection of arrival times.  A jet is the standard object that combines these particles into a single object that can be used in analyses and searches.  In momentum-space variables, summing the four-vectors of the constituents provides a natural definition of the four-vector of the jet.  Unfortunately, there is not an obvious choice for the definition of the arrival time of a jet.

The selection of a proper jet time definition is pivotal.  A proper definition will enable efficient separation of the Standard Model (SM) prompt background and BSM long-lived signatures.  A poor definition, on the other hand, will not allow us to take full advantage of the precision timing capabilities at the level of $30-40$ picoseconds, that will be part of upgrades to ATLAS~\cite{Allaire:2018bof}, CMS~\cite{Collaboration:2296612}, and LHCb~\cite{Aaij:2244311}.  Already, CMS has demonstrated sensitivity to delayed jets in their search for displaced gluinos~\cite{Sirunyan:2019gut}.

Beyond just performance, a proper definition of jet time may help identify exciting properties of quantum chromodynamics (QCD), enable new jet tagging possibilities, and provide additional inputs for machine learning applications.  Even pileup suppression may benefit substantially from an effective usage of jet time because generically pileup vertices have a spread both in space and in time.  At the high-luminosity LHC any improvements to pileup suppression are indispensable.

The purpose of this paper is to explore a variety of definitions of jet time and characterize their performances. As with any measurement tool, there are two aspects: accuracy and precision.  For jet time, as we will discuss in detail later, the ``correct'' time is somewhat ambiguous.  The precision, or resolution, is well-defined and will be the main criterion in comparing different approaches.

The structure of the paper is as follows.  In Sec.~\ref{sec:defns} we provide a brief overview of various possible definitions of jet time.  The general behavior, both for prompt jets and delayed jets, is discussed in Sec.~\ref{sec:analytic}.  In Sec.~\ref{sec:numerical} we perform an in-depth numerical study of the behavior of each jet time definition, paying special attention to the dependence on the event geometry.  Finally, our conclusions are in Sec.~\ref{sec:outlook} along with outlook for future studies.  Several appendices are included for cross-checks and studies of additional effects.  We discuss the behavior of jet time when endcaps are also used to measure arrival times in App.~\ref{app:finitedet}, the impact of pileup and jet grooming in App.~\ref{app:pileup}, the effects of detector resolution in App.~\ref{app:detector}, and the impact of hadronization in App.~\ref{app:hadronization}.

%%%%%%%%%%%%%%%%%%%%%%%%%%%%%%%%%%%%%%%%%%%%%%%%%%%%%%%%%%%%%%%%%%
%%%%%%%%%%%%%%%%%%%%%%%%%%%%%%%%%%%%%%%%%%%%%%%%%%%%%%%%%%%%%%%%%%
\section{Definitions for Jet Time}
\label{sec:defns}

In this section we briefly describe the definitions of jet time that we study.  More detailed descriptions will follow in Sec.~\ref{sec:analytic} and simulation results will be shown in Sec.~\ref{sec:numerical}.

We first define our notation.  A single particle $i$ has a four-momentum $(E_i, \vec{p}_i)$ and a particle time $t_i$, which is the time 
that it crosses a particular layer of the detector.
A jet $J$ is a set of particles which we write as $J = \{ i \}$.  While a particle has an unambiguous time, the jet has a set of times $\{ t_i \}$ associated to it.  In the same way that it is often useful to treat the jet as a single four-vector, {\it e.g.} in new physics searches, it is also useful to be able to assign a jet a single time $t_J$, that we call the jet time.

There are a number of possibilities that can be used.  One can choose a single constituent $i'$ in the jet and use its particle time $t_{i'}$ to represent the jet time.  Jet time definitions of this type include:
\begin{itemize}

\item {\it median time}: take $t_J$ to be the median value of the particle times $\{ t_i \}$,

\item {\it hardest time}: take $t_J$ to be the time $t_{i_h}$, that corresponds to the time of the constituent $i_h$ with the largest transverse momentum,

\item {\it random time}: take $t_J$ to be a randomly-drawn value of the particle times $\{ t_i \}$.

\end{itemize}
The median time has been used by CMS in their search for gluinos with displaced decays~\cite{Sirunyan:2019gut}.  The hardest particle in a jet is likely to be very close to the jet axis so it may be a good proxy for the time of the jet.  We do not expect choosing a random particle time as the jet time to perform well, but it is useful as a baseline comparison.

Another option is to calculate $t_J$ from a weighted sum of $\{ t_i \}$, similar to a jet shape.  Generically, this would take the form\footnote{A number of alternatives are possible, such as particle energy $E$ in place of $p_T$ or angle $\theta$ in place of $\Delta R$.  In a brief survey, we did not find these to outperform the variables used in Eq.~\eqref{eq:t-shape}.}
\begin{equation} \label{eq:t-shape}
t_J^{(\alpha,\beta,\gamma)} \propto \sum_{i \in {\rm jet}} (p_{T,i})^\alpha (\Delta R_i)^\beta t_i^\gamma,
\end{equation}
where $\Delta R_i$ is the $\eta\phi$-distance between particle $i$ and the jet axis.  The two simplest versions of a weighted sum are the average time where $(\alpha,\beta,\gamma)=(0,0,1)$ and the $p_T$-weighted time where $(\alpha,\beta,\gamma)=(1,0,1)$:\footnote{We briefly studied a few additional cases such as $p_T^2$-weighted but did not see an improvement over the $p_T$-weighted time or average time.  An optimization of $\alpha$, $\beta$, and $\gamma$ is beyond the scope of this work.}
\begin{itemize}

\item {\it average time}: take $t_J$ to be
\begin{equation} \label{eq:t-average}
t_J^{\rm average} = \frac{1}{N} \sum_{i=1}^N t_i,
\end{equation}
where there are $N$ particles in the jet,

\item {\it $p_T$-weighted time}: take $t_J$ to be 
\begin{equation} \label{eq:t-ptw}
t_J^{p_T} = \frac{1}{\pts} \sum_{i=1}^N p_{T,i} t_i,
\quad\quad\quad
\pts = \sum_{i=1}^N p_{T,i},
\end{equation}
where $\pts$ is a normalization factor.

\end{itemize}
Finally, one could simply disregard the particles times, treat the jet as a single particle, and calculate its time based on the jet kinematics.  There are two variations depending on whether the jet is treated as a massless particle or massive particle:
\begin{itemize}

\item {\it null time}: treat the jet $J$ as a massless particle and calculate its crossing time using the three-momentum $\vec{p}_J$ of the jet (assuming knowledge of the production vertex),

\item {\it kinematic time}: treat the jet $J$ as a massive particle and calculate its crossing time using the four-momentum $(E_J, \vec{p}_J)$ of the jet (assuming knowledge of the production vertex).

\end{itemize}
Since these definitions do not utilize the information available from timing measurements, they do not correspond to experimentally measurable arrival times.  They are useful, however, to determine what constitutes good performance from a jet time definition.  These times are constant for fixed parent particle trajectory and jet kinematics. In addition, the null time represents the crossing time if the parton did not undergo showering and hadronization and will serve as a useful reference time.\footnote{Note that while for prompt jets the null time is computable in data, for delayed jets the null time requires the location of the displaced vertex so it is not always computable in data.}\footnote{Of the definitions studied, the $p_T$-weighted, null, and kinematic times are IRC safe. The hardest time is only IR safe. The average, median, and random times are all IRC unsafe.}

After having chosen a definition for $t_J$ we also need to choose a metric to evaluate which definition is the most useful.  For a jet time definition, we will compare the relative time difference $\Delta t/t_{\rm ref}$ defined as
\begin{equation} \label{eq:dt1}
\frac{\Delta t}{t_\text{ref}} = \frac{t_J - t_J^{\rm ref}} {t_J^{\rm ref}},
\end{equation}
to determine a good choice.

Each jet has a different value of $\Delta t / t_{\rm ref}$ so that a sample of jets will lead to a distribution for the relative time difference.  The mean of this distribution corresponds to the accuracy of the jet time definition while the width corresponds to the precision, or resolution, of the definition.  Since the choice of $t_{\rm ref}$ is arbitrary it is not obvious that the mean of the distribution is important (not to mention that constant offsets can be corrected in practice).  The width of the distribution, on the other hand, is a robust indicator of a stable time definition.  For that reason, the width of the relative time difference distribution will be used as the figure of merit when comparing definitions.

%%%%%%%%%%%%%%%%%%%%%%%%%%%%%%%%%%%%%%%%%%%%%%%%%%%%%%%%%%%%%%%%%%
%%%%%%%%%%%%%%%%%%%%%%%%%%%%%%%%%%%%%%%%%%%%%%%%%%%%%%%%%%%%%%%%%%
\section{General Behavior}
\label{sec:analytic}

In this section, we study analytically the general behavior of the various jet time definitions.  We start with the prompt case where the majority of particles originate from the origin and then we move on to the delayed case where the majority of particles originate from a displaced decay.

%%%%%%%%%%%%%%%%%%%%%%%%%%%%%%%%%%%%%%%%%%%%%%%%%%%%%%%%%%%%%%%%%%
\subsection{Prompt Particles}
\label{sec:gen-prompt}

For prompt particles we assume that the particles originate at $t=0$ from the origin of the detector $\vec{x} = \vec{0}$.\footnote{In reality and in simulation, there are displacements from processes like $B$-hadron decays.  These have a negligible impact on our analysis.}
As a detector model, we will consider an infinite cylinder with radius $r_T$.\footnote{The differences when endcaps are included are discussed in App.~\ref{app:finitedet}.}  The time $t_i$ of a particle $i$ with four-momentum $(E_i, \vec{p}_i)$ is then given by
\begin{equation} \label{eq:ti}
t_i = \frac{r_T}{c} \frac{E_i}{p_{T,i}},
\end{equation}
where $c$ is the speed of light.

For a massless particle this simplifies to
\begin{equation} \label{eq:ti-massless}
t_i = \frac{r_T}{c} \frac{|\vec{p}_i|}{p_{T,i}} = \frac{r_T}{c} \cosh \eta_i,
\end{equation}
which is a good approximation for particles in a high-momentum jet.

The jet times coming from a single particle within the jet are similarly calculated.  For the median, hardest, and random times, the time of the jet is given by the time of the median-time particle $i_m$, the hardest particle $i_h$, or a random particle $i_r$, respectively, and is
\begin{equation} \label{eq:t-single}
t_J^{\{ {\rm median, hardest, random} \} } 
= t_{ \{ i_m, i_h, i_r \} }
= \frac{r_T}{c} \cosh \eta_{ \{ i_m, i_h, i_r \} }.
\end{equation}
With the cylindrical detector, the null and kinematic times of the jet, with four-momentum $(E_J, \vec{p}_J)$, can also be calculated.  The null time of a jet is
\begin{equation} \label{eq:t-null}
t_J^{\rm null} = \frac{r_T}{c} \frac{|\vec{p}_J|}{p_{T,J}} 
= \frac{r_T}{c} \cosh \eta_J,
\end{equation}
while the kinematic time of a jet is
\begin{equation} \label{eq:t-kin}
t_J^{\rm kinematic} = \frac{r_T}{c} \frac{E_J}{p_{T,J}}
= t_J^{\rm null} \frac {E_J} {|\vec{p}_J|}.
\end{equation}
In the limit of small jet mass these definitions differ by $\cO(m_J^2 / \vec{p}_J{}^2)$.

The average and $p_T$-weighted times follow the definitions in Eqs.~\eqref{eq:t-average}~and~\eqref{eq:t-ptw}.

%%%%%%%%%%%%%%%%%%%%%%%%%%%%%%%%%%%%%%%%%%%%%%%%%%%%%%%%%%%%%%%%%%
\subsubsection*{Prompt Relative Time Difference}
\label{sec:prompt-rel}
 
For jet times using a single particle (median, hardest, and random times), with a time $t_i$, the relative time difference using Eqs.~\eqref{eq:t-single}~and~\eqref{eq:t-null} is
\begin{equation} \label{eq:dt}
\frac{\Delta t}{t_{\rm ref}}
= \frac{t_i - t_J^{\rm null}}{t_J^{\rm null}}
= \frac{\cosh \eta_i}{\cosh \eta_J} - 1.
\end{equation}
When the particle $i$ points along the same direction as the jet axis, the relative time difference is always zero.  When there is a fixed angular distance $\Delta\eta$ between the particle $i$ and the jet axis, however, the relative time difference changes with $\eta_J$.  Due to the detector geometry, as the jet becomes more forward, the relative time difference will grow.

The furthest that a particle $i$ can be from the axis of the jet is approximately given by the jet radius $\rjet$.  Therefore, for a given $\eta_J$ there is a maximum relative time difference given by
\begin{equation} \label{eq:tmax}
\frac{\Delta t}{t_{\rm ref}} \bigg|^{\rm max} 
= \frac{\cosh (\eta_J \pm \rjet)}{\cosh \eta_J} - 1,
\end{equation}
where the $+$ applies for positive $\eta_J$ and the $-$ applies for negative $\eta_J$.

\begin{figure} [t]
\begin{center}
  \includegraphics[width=0.50\textwidth]{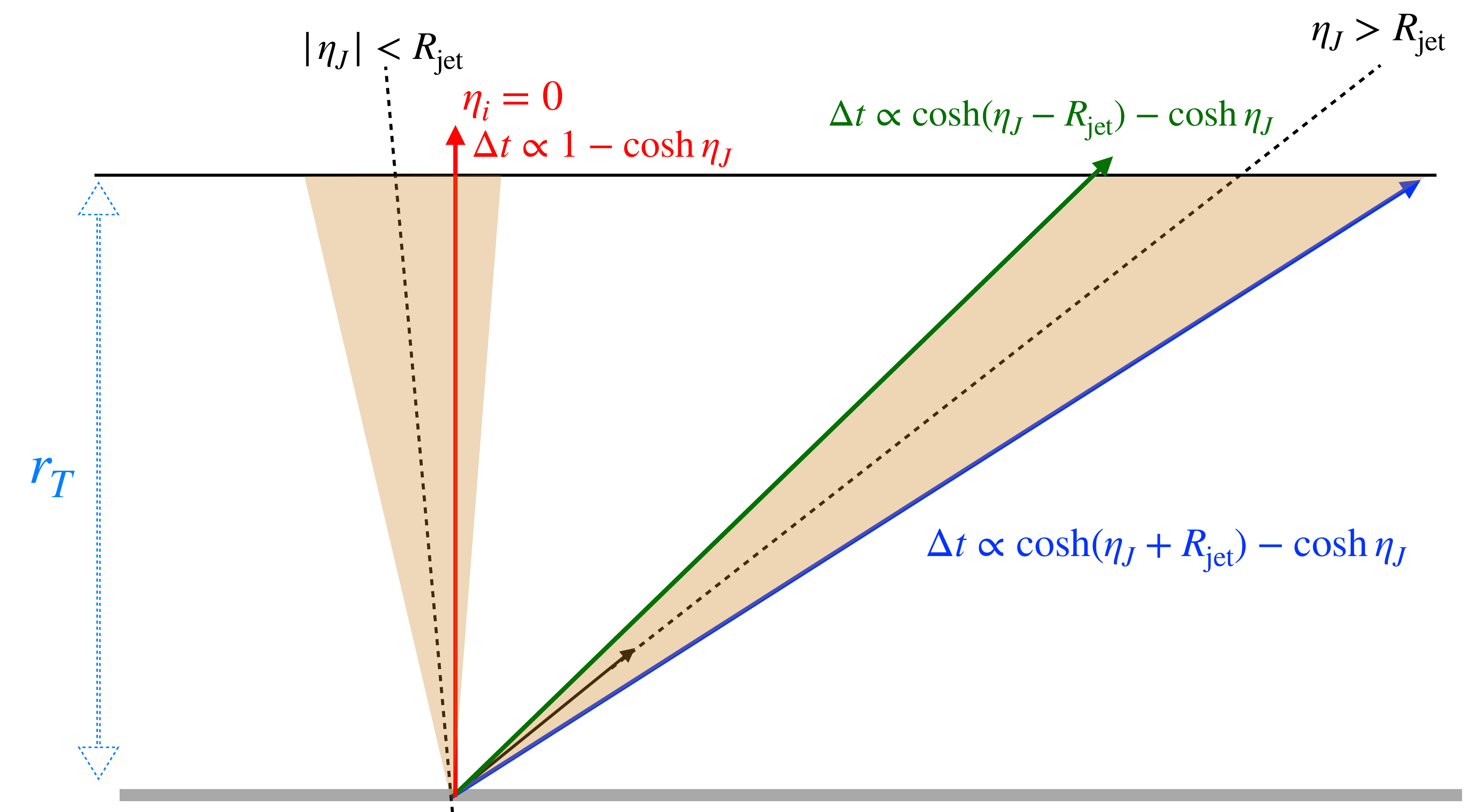} \quad\quad
  \raisebox{-1.5em}{\includegraphics[width=0.4\textwidth,trim= 0 0 0 0]{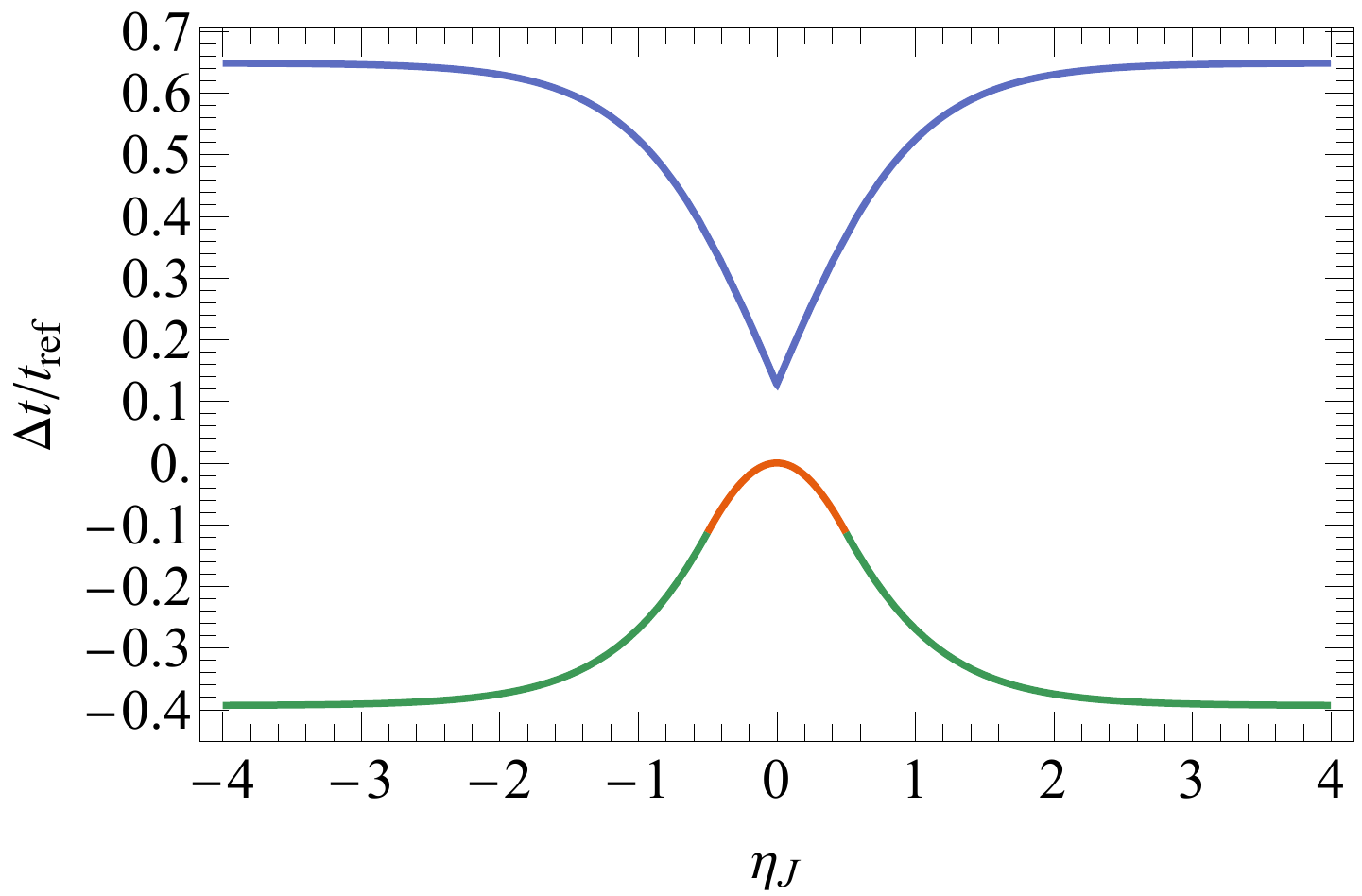}}
  \caption{Illustration of the bounds for a prompt jet (left). The slowest particle time, Eq.~\eqref{eq:tmax}, follows the high $|\eta|$ boundary of the jet (blue).  The fastest particle time, Eq.~\eqref{eq:tmin}, follows the low $|\eta|$ boundary of the jet (green) or the $\eta=0$ line (red).  Plot of the boundaries for the relative time difference for a jet with $R_\text{jet}=0.5$ (right) where the colored lines correspond to the boundaries in the left image.}
  \label{fig:diagram_prompt_jet_bounds}
\end{center}
\end{figure}

The minimum is similar except that for $|\eta_J| < \rjet$ there is a stronger bound that comes from the fact that $\eta_i=0$ for a massless particle corresponds to the fastest time possible since it is the shortest path from the origin.  The bound for $|\eta_J| < \rjet$ consequently only depends on $\eta_J$.  We find
\begin{equation} \label{eq:tmin}
\frac{\Delta t}{t_{\rm ref}} \bigg|^{\rm min} 
= 
\begin{cases}
{\rm sech} \; \eta_J - 1,                        & \quad\quad |\eta_J|<\rjet, \\
\frac{\cosh (\eta_J \mp \rjet)}{\cosh \eta_J} - 1,   & \quad\quad {\rm else}.
\end{cases}
\end{equation}
Eqs.~\eqref{eq:tmax}~and~\eqref{eq:tmin} taken together specify boundaries in the space of pseudorapidity vs. relative time difference.  The different jet time definitions will have different distributions within these boundaries.  Fig.~\ref{fig:diagram_prompt_jet_bounds} illustrates these boundaries graphically.

%%%%%%%%%%%%%%%%%%
%%%%%%%%%%%%%%%%%%

At $\eta_J=0$ the minimum of $\Delta t / t_{\rm ref}$ is 0 while the maximum is $\cosh(\rjet)-1>0$ which means that the relative time difference cannot be negative.  Therefore, for very central jets we expect the relative time difference distributions to skew towards positive values since there is more available phase space.  For less central jets we do not expect a strong preference for positive or negative values based only on phase space.

For the kinematic time rather a bound, one can relate the relative time difference to a kinematic quantity.  From Eqs.~\eqref{eq:t-null}~and~\eqref{eq:t-kin} we find that for the kinematic time
\begin{equation} \label{eq:tdiff-kin}
\frac{\Delta t}{t_{\rm ref}}
= \frac{E_J}{|\vec{p}_J|} - 1
= \frac{1}{\beta_J} - 1,
\end{equation}
where $\beta_J$ is the velocity.  Since $\beta_J \leq 1$ the relative time difference for the kinematic time is always non-negative.  This is expected since the kinematic time points in the same direction as the null time, but adjusts for the mass of the jet.

Next, we move to the relative time difference for the $p_T$-weighted time.  This has a simple form given our cylindrical detector model.  The $p_T$-weighted time is
\begin{equation}\label{eq:pt-simp}
t_J^{p_T} 
= \frac{1}{\pts} \sum_{i=1}^N p_{T,i} t_i
= \frac{1}{\pts} \sum_{i=1}^N \frac{r_T}{c} E_i
= \frac{r_T}{c} \frac{E_J}{\pts},
\end{equation}
and the corresponding relative time difference is
\begin{equation} \label{eq:tdiff-ptw}
\frac{\Delta t}{t_{\rm ref}}
= \frac{E_J}{\pts} \frac{p_{T,J}}{|\vec{p}_J|} - 1
= \frac{E_J}{|\vec{p}_J|} \frac{p_{T,J}}{\pts} - 1.
\end{equation}
Written in the form after the second equality we recognize the factor $E_J/|\vec{p}_J| \geq 1$ from Eq.~\eqref{eq:tdiff-kin}.  The other factor $p_{T,J}/\pts$ is the ratio of the jet $p_T$ to the scalar sum of the constituent $p_T$ values.   Since $p_{T,J}$ is a vector sum we have $p_{T,J}/\pts \leq 1$.  The distribution at a given $\eta_J$ is determined by the $\eta$-dependence of each of these two terms.

For small mass jets  $E_J/|\vec{p}_J| \approx 1 + m_J^2 / \vec{p}_J{}^2$ and schematically for QCD jets the mass is $\langle m_J^2 \rangle~\sim~R^2 \, p_{T,J}^2~\sim~R^2 \, \vec{p}_J{}^2 \, {\rm sech}^2 \eta_J$~\cite{Salam:2009jx}.  Consequently, $E_J/|\vec{p}_J| \sim 1 + R^2 \, {\rm sech}^2 \eta_J$ which peaks at $\eta_J=0$ and reduces as $|\eta_J|$ grows.  The other quantity $p_{T,J}/\pts$ depends on the energy distribution in the jet and is not strongly correlated with $\eta_J$.  Therefore, we expect that the relative time distribution for the $p_T$-weighted time to be positively-skewed for central jets and switch over to negatively-skewed as the jets become more forward.

%%%%%%%%%%%%%%%%%%%%%%%%%%%%%%%%%%%%%%%%%%%%%%%%%%%%%%%%%%%%%%%%%%
\subsection{Delayed Particles}
\label{sec:gen-delayed}

Next, we study the jet time behavior for delayed particles.  Our benchmark scenario involves a mother particle $M$ that travels a macroscopic distance, then decays into two daughter particles $D$ and $\widetilde{D}$.  We assume that $M$ and $\widetilde{D}$ are unobserved while $D$ is colored and results in a jet due to showering and hadronization.\footnote{
If $M$ or $\widetilde{D}$ (or both) are colored, they will propagate as color-neutral $R$-hadrons.  They will be unobserved if the resulting $R$-hadrons are electrically-neutral.}
In our numerical study we take $M$ as a gluino, $\widetilde{D}$ as a gravitino, and $D$ as a gluon.

Let the mother $M$ have four-momentum $(E_M, \vec{p}_M)$ and decay at the displaced vertex $\vec{x}_M$ at a time $t_M$.  The daughter $D$ showers and hadronizes into a delayed jet.  A particle $i$ in the delayed jet has a four-momentum $(E_i, \vec{p}_i)$ and originates from $\vec{x}_M$ at $t_M$.  Let the vector pointing from $\vec{x}_M$ to where $i$ crosses the detector be $\vec{x}_i$.\footnote{For simplicity we neglect the effects of curvature in the magnetic field of the detector.  For a magnetic field of $3.8~{\rm T}$ the effect on the measurement of time or momentum is less than $1\%$ for particles with $p_T>2.5~{\rm GeV}$.  In App.~\ref{app:pileup} where we study pileup, we do include curvature induced by the magnetic field.} Since $|\vec{x}_i|$ only makes up a fraction of the total traversed distance, the impact of the per-particle time spread is generically lessened for delayed jets. We will refer to this effect as the daughter time fraction.

%%%%%%%%%%%%%%%%%%
\begin{figure} [t]
\begin{center}
  \includegraphics[width=0.4\textwidth,trim=0 100 0 100]{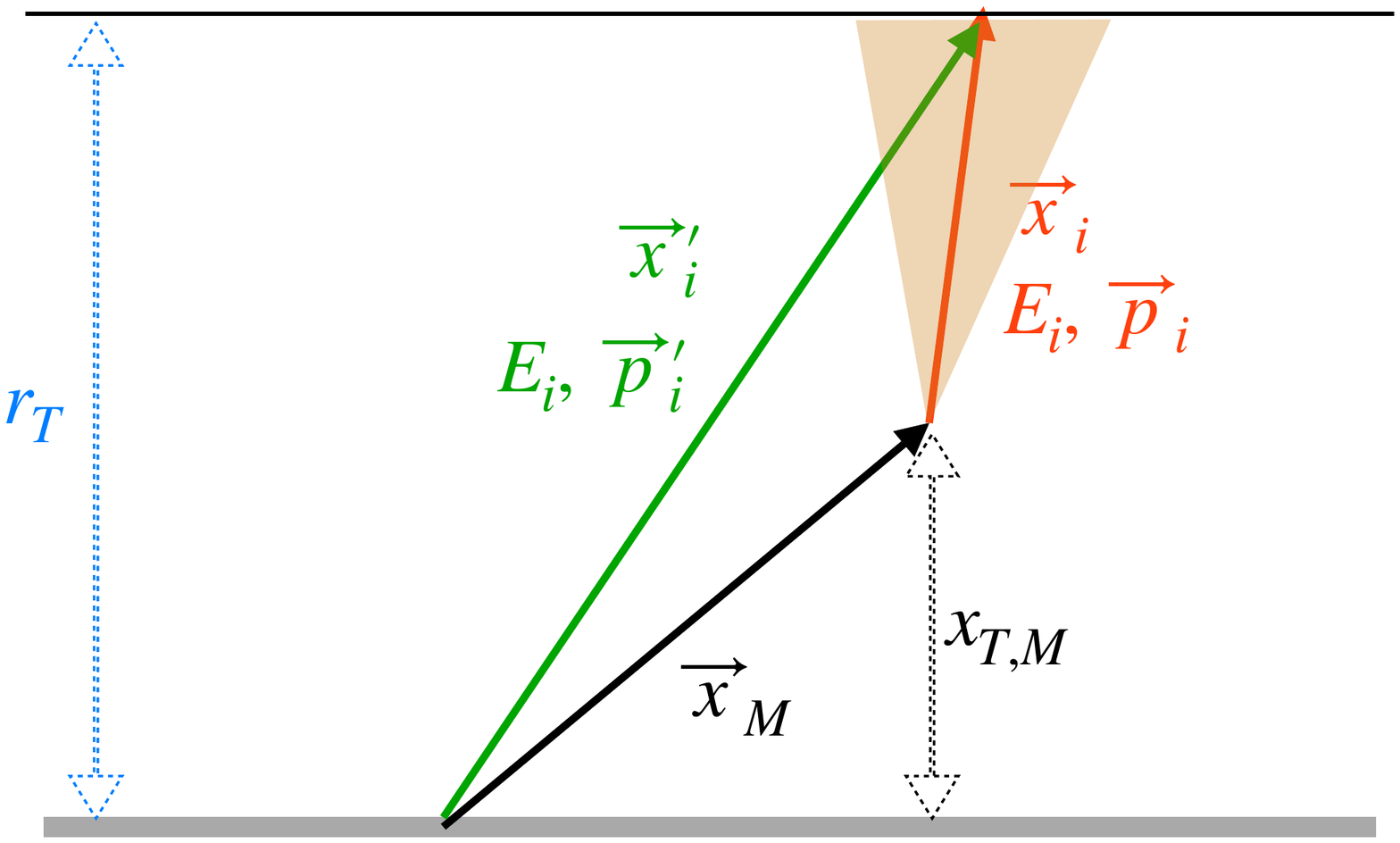} \quad\quad\quad
  \includegraphics[width=0.4\textwidth,trim=0 100 0 100]{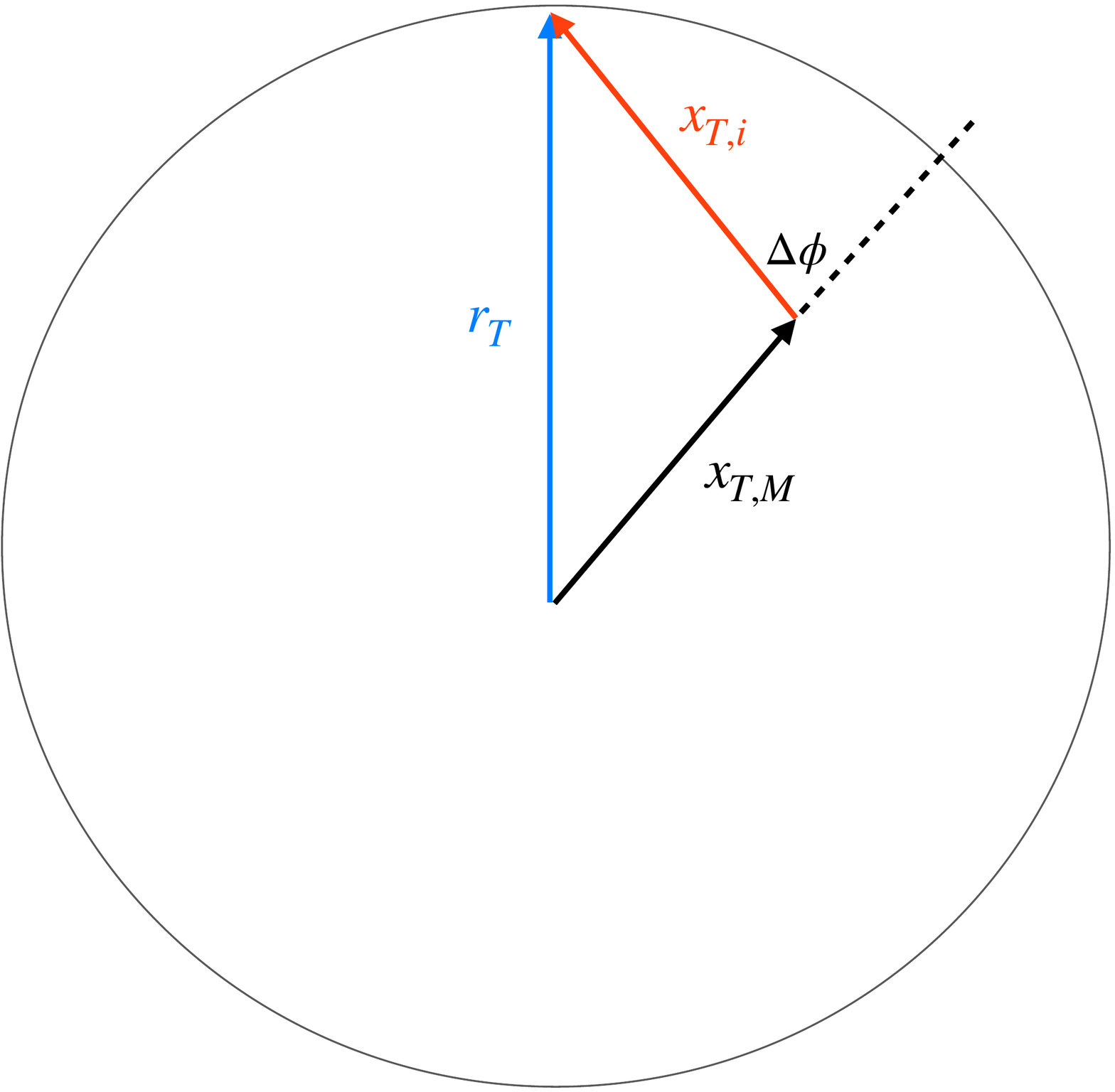}
  \caption{The mother particle $M$ travels along $\vec{x}_M$ and decays to a delayed jet (shaded brown).  The daughter particle $i$ travels along $\vec{x}_i$ until it crosses the detector.  If the displaced vertex is not identified then $i$ is assumed to have traveled along $\vec{x}_i{'}$.}
  \label{fig:diagram_detector_geometry}
\end{center}
\end{figure}
%%%%%%%%%%%%%%%%%%

If particle $i$ is measured, but the displaced vertex is not identified, then $i$ will be assumed to have come from the origin, having traveled along $\vec{x}_i{'} = \vec{x}_M + \vec{x}_i$.  This is illustrated in Fig.~\ref{fig:diagram_detector_geometry}.  We call the kinematics computed using $\vec{x}_i{'}$ the observed kinematics and those using $\vec{x}_i$ the truth kinematics.
\begin{equation} \label{eq:observed}
\begin{array}{c|c|c}
                           & \quad\text{four-vector}\quad  & \quad\text{assumed trajectory}\quad \\ \hline
\text{truth}     & (E_i, \vec{p}_i)    & \vec{x}_i                 \\ \hline
\text{observed}  & (E_i, \vec{p}_i{'}) & \vec{x}_i{'}
\end{array}
\end{equation}
For transverse displacements $\gtrsim 10~{\rm cm}$ the tracking efficiency is $\lesssim 40\%$ in CMS and drops off further above $50~{\rm cm}$~\cite{CMS:TrackEff}. Conservatively, we assume that the displaced vertex is not identified and work with the observed kinematics. The most direct impact of this is that the transverse momentum of any given constituent can be incorrectly assigned. This occasionally leads to noticeable broadening of time definitions that directly depend on the constituents' $p_T$. While those that do not, {\it e.g.} the median time, are mostly insensitive to whether the truth or observed kinematics are used. A more subtle consequence of using observed kinematics is that the observed opening angle does not correspond to the true opening angle, leading to the inclusion/exclusion of particles with extreme arrival times.  We will refer to this discrepancy in opening angles as the effective radius.

In the following, we study in detail these three primary effects which control the performance of the timing of a delayed jet: the daughter time fraction, observed kinematics, and the effective radius. 

%%%%%%%%%%%%%%%%%%%%%%%%%%%%%%%%%%%%%%%%%%%%%%%%%%%%%%%%%%%%%%%%%%
\subsection*{Observed Kinematics}

For a massless particle $i$, if it is prompt its time is fully specified by its pseudorapidity $\eta_i$ (see Eq.~\eqref{eq:ti-massless}).  When $i$ is delayed, its time depends now on its pseudorapidity $\eta_i$, the pseudorapidity of the mother $\eta_M$, the azimuthal angle difference $\Delta\phi = \phi_i - \phi_M$,  the speed of the mother $\beta_M$, and the transverse decay location of the mother $x_{T,M}$:
\begin{equation} \label{eq:delvars}
\eta_M, \; \eta_i, \; \Delta\phi, \; \beta_M, \; x_{T,M}.
\end{equation}
The transverse distance $x_{T,i}$ traveled by $i$ is calculated to be
\begin{equation} \label{eq:xti}
x_{T,i} = \sqrt{r_T^2 - x_{T,M}^2 \sin^2 (\Delta\phi)} - x_{T,M} \cos(\Delta\phi).
\end{equation}
The observed kinematics, $(p_{T,i}{'}, \eta_i{'}, \phi_i{'})$, can be computed in terms of the variables in Eq.~\eqref{eq:delvars} and $x_{T,i}$.  The observed pseudorapidity $\eta_i{'}$ is found from solving
\begin{equation} \label{eq:etaObs}
r_T \; {\rm sinh} (\eta_i{'}) = x_{T,M} \; {\rm sinh} (\eta_M) + x_{T,i} \; {\rm sinh} (\eta_i).
\end{equation}
In terms of the true transverse momentum $p_{T,i}$, the observed transverse momentum $p_{T,i}{'}$ is
\begin{equation} \label{eq:ptObs}
p_{T,i}{'} \sqrt{1 + \left(\frac{x_{T,M}}{r_T} \; {\rm sinh}(\eta_M) + \frac{x_{T,i}}{r_T} \; {\rm sinh}(\eta_i)  \right)^2} 
= p_{T,i} \; {\rm cosh}(\eta_i).
\end{equation}
Finally, the observed azimuthal angle $\phi_i{'}$ is
\begin{equation} \label{eq:phiObs}
\tan(\phi_i{'})
= \frac{x_{T,M} \sin(\phi_M) + x_{T,i} \sin(\phi_i)}{x_{T,M} \cos(\phi_M) + x_{T,i} \cos(\phi_i)}.
\end{equation}
Jets are clustered using the observed kinematics.  The time $t_i$ of a particle $i$ is not impacted by using observed kinematics because the arrival time of a particle is an independent measurement.  Since $\vec{x}_i$ and $\vec{x}_i{'}$ cross the detector at the same location, the effect on clustering using observed kinematics is nearly negligible (comparable to the difference between different jet algorithms).

The primary impact of using observed kinematics is on jet time definitions that utilize $p_T$ information.  We expect the $p_T$-weighted time to be impacted at a noticeable level (the size of this effect will be studied in Sec.~\ref{sec:numerical}).  The hardest time could be affected if which jet constituent is the hardest changes under the observed kinematics.  In practice, this is rare due to the hierarchical nature of the parton shower.  The median time, likewise, is minimally affected.

%%%%%%%%%%%%%%%%%%%%%%%%%%%%%%%%%%%%%%%%%%%%%%%%%%%%%%%%%%%%%%%%%%
\subsection*{Effective Radius}

The radius of a jet, $\rjet$, is a parameter in the jet finding algorithm that determines which particles are included in the same jet.  It determines the catchment area of a jet in $\eta\phi$-space which is approximately a circle with radius $\rjet$ for cone-like algorithms used on isolated jets~\cite{Cacciari:2008gn}.

When choosing a jet radius there are trade-offs.  If the radius is too small, then particles coming from the showering of a hard particle could fall outside a particular jet and the jet will not be a useful proxy for the underlying hard quark or gluon.  If the radius is too large, the jet is more susceptible to contamination like underlying event and pileup~\cite{Salam:2009jx}.

One consequence for prompt jets of using a fixed $\rjet$ for jet finding is that an optimal jet radius may be different for central jets as compared to forward jets.  This is because for a fixed $\rjet$ in $\eta\phi$-space, the corresponding angular distance, $\Delta\theta$, between a pair of particles is smaller for forward jets than for central jets.  Physically, if a set of central particles within a jet with radius $\rjet$ were shifted to larger $|\eta|$ values, then they may not all fit within a radius $\rjet$ anymore.  

%%%%%%%%%%%%%%%%%%
\begin{figure} [t]
\begin{center}
  \includegraphics[width=0.55\textwidth,trim=0 100 0 100]{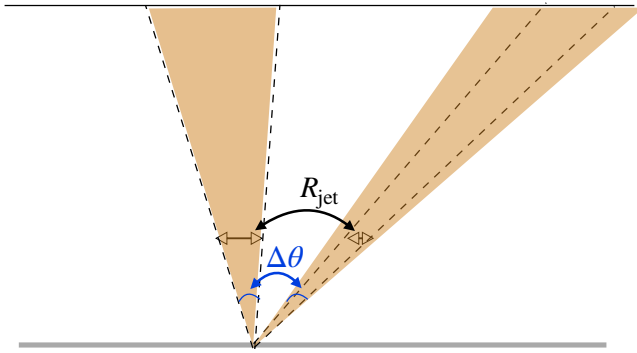} 
  \caption{Illustration of $\Delta\theta$ effect for prompt jets.  The dashed lines depict jets with the same $R_\text{jet}$ while the shaded regions depict jets with the same $\Delta\theta$.  }
  \label{fig:diagram_prompt_effradius}
\end{center}
\end{figure}
%%%%%%%%%%%%%%%%%%

In the prompt case, we consider the effective radius of the jet to be the angular distance $\Delta\theta$ that is required to keep $\Delta R$ fixed.  This means that forward jets have a smaller effective radius than central jets because for fixed $\Delta R$ the required $\Delta\theta$ distance shrinks.  See Fig.~\ref{fig:diagram_prompt_effradius} for an illustration.  Variable $R$ jets were proposed to account for this by letting the jet radius grow at larger $|\eta|$ by scaling the radius inversely with transverse momentum~\cite{Krohn:2009zg}.

\begin{figure} [t]
\begin{center}
  \includegraphics[width=0.55\textwidth]{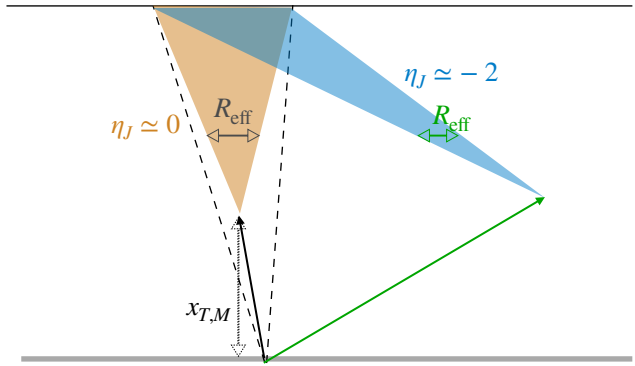}
  \caption{Illustration of $R_{\rm eff}$ effect for delayed jets.  The brown shaded region has a larger $R_{\rm eff}$ compared to the dashed lines because $x_{T,M}$ is larger.  The blue shaded region shows both the effect of shifting and tilting.}
  \label{fig:diagram_delayed_effradius}
\end{center}
\end{figure}

In contrast, for delayed jets we consider the effective radius of a jet to be the true $\Delta R$ distance that is required to keep observed $\Delta R$ fixed.  There are two factors that alter the effective radius for delayed jets.  The first is that a non-zero value of $x_{T,M}$ means the jet originates closer to the detector radius $r_T$.  The same way that the image from a projector is smaller as you move the projector closer to the screen, a fixed observed $\Delta R$ value corresponds to a larger true $\Delta R$ value as $x_{T,M}$ grows.  See Fig.~\ref{fig:diagram_delayed_effradius} for an illustration.

%%%%%%%%%%%%%%%%%%
%%%%%%%%%%%%%%%%%%

The second effect is that both $\eta_J$ and $\eta_M$ can vary.  Changing $\eta_J$ tilts the direction of the jet and generally causes the effective radius to shrink with $\eta_J$ similar to the prompt case.  Changing $\eta_M$ is not a tilt, but rather a shift of the origin point of the particles.  Due to the geometry of $\eta\phi$-space the effective radius generally increases as $|\eta_M|$ grows.

\begin{figure} [t]
\begin{center}
  \includegraphics[width=0.45\textwidth]{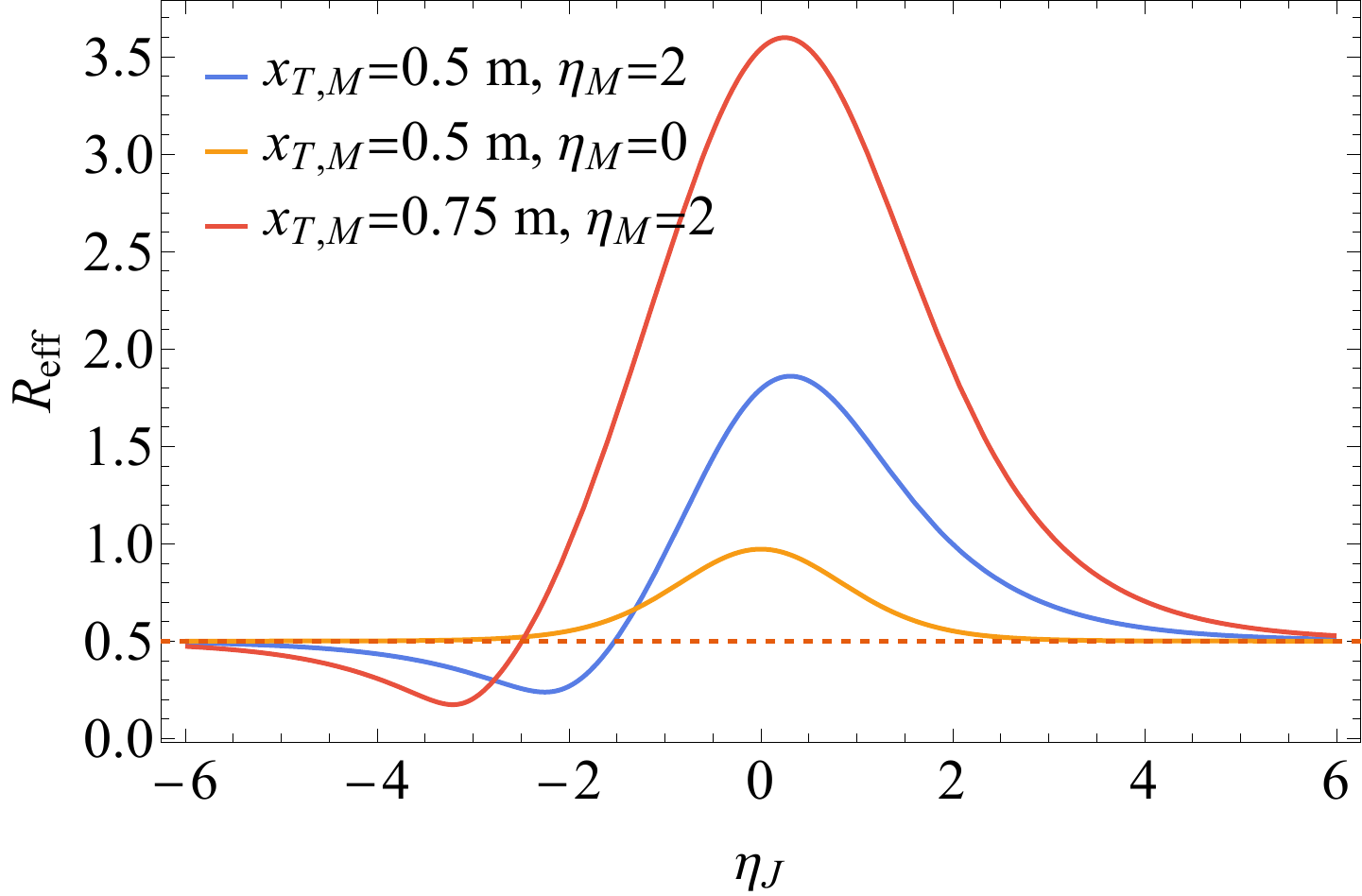}
  \caption{The effective radius, $R_\text{eff}$, as a function of $\eta_J$ with $x_{T,M} =0.5~{\rm m}$ and $\eta_M=2$ (blue solid line), $x_{T,M} =0.5~{\rm m}$ and $\eta_M=0$ (yellow solid line), and $x_{T,M} =0.75~{\rm m}$ and $\eta_M=2$ (red solid line).  A fixed value of 0.5 is also shown (red dashed line).}
  \label{fig:reff_vs_etaJ}
\end{center}
\end{figure}

The effective radius can be estimated numerically.  As shorthand we write the observed pseudorapidity of a jet as $\eta_J{'} = f(\eta_M, x_{T,M}, \eta_J)$ where the function is found in Eq.~\eqref{eq:etaObs}.  We define the effective jet radius as $R_{\rm eff}$ and find it by solving
\begin{equation} \label{eq:reff}
\rjet = 
f\left(\eta_M, x_{T,M}, \eta_J + \frac{1}{2} R_{\rm eff} \right)
- f\left(\eta_M, x_{T,M}, \eta_J - \frac{1}{2} R_{\rm eff} \right).
\end{equation}
The definition is not rigorous but rather is meant to provide intuition for the general behavior. We also set $\Delta\phi$ zero in the above for simplicity.  In Fig.~\ref{fig:reff_vs_etaJ} we plot $R_{\rm eff}$ as a function of $\eta_J$ for several sample points of $x_{T,M} $ and $\eta_M$ with fixed jet radius $\rjet = 0.5$.

%%%%%%%%%%%%%%%%%%
%%%%%%%%%%%%%%%%%%

For isolated QCD jets, we expect the jet properties to change slowly with respect to increasing $R_{\rm eff}$.  In a typical parton shower there are both more and higher momentum particles near the center of the jet.  Including additional soft particles further from the jet axis will not perturb the jet four-vector by much.  When $R_{\rm eff}$ decreases the jet properties should change faster as more and higher momentum particles are excluded.

%%%%%%%%%%%%%%%%%%%%%%%%%%%%%%%%%%%%%%%%%%%%%%%%%%%%%%%%%%%%%%%%%%
\subsection*{Daughter Time Fraction}

\begin{figure} [H]
\begin{center}
  \includegraphics[width=0.55\textwidth,trim=0 150 0 200]{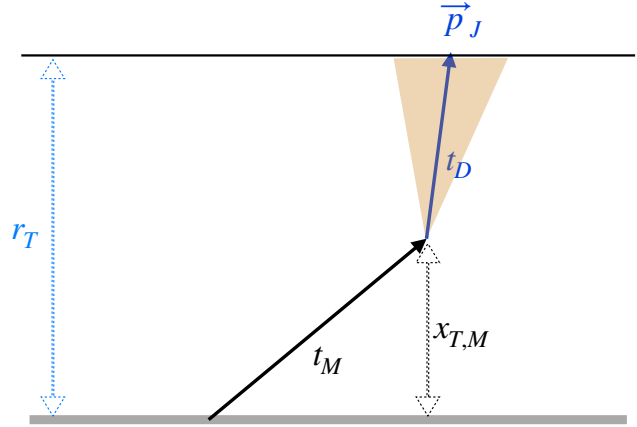}
  \caption{The time of flight of the mother particle is $t_M$ and the time of flight of a daughter particle is $t_D$.}
  \label{fig:diagram_tM_tD}
\end{center}
\end{figure}

The third effect is the fraction of time that comes from time of flight of the daughter $i$ as compared to the time of flight of the mother, as shown in Fig.~\ref{fig:diagram_tM_tD}.  Intuitively, when the mother travels more of the distance from the origin to the detector, there is less variation among the times of the particles in a jet.  Consequently, the distribution of jet times becomes narrower.

%%%%%%%%%%%%%%%%%%
%%%%%%%%%%%%%%%%%%

\begin{figure}
  \centering \includegraphics[width=0.55\textwidth]{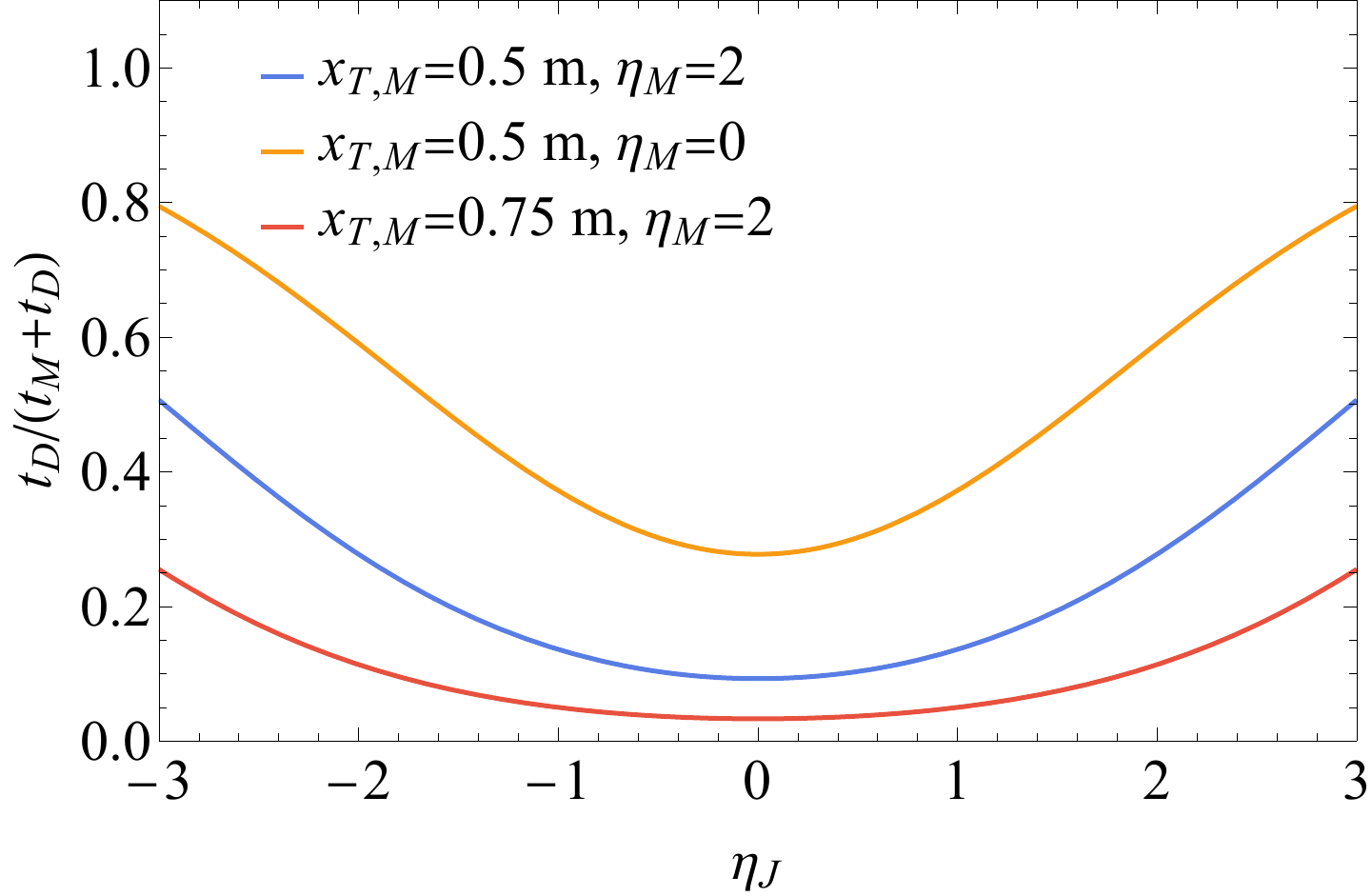}
  \caption{The overall suppression in the width due to the daughter time fraction as a function of $\eta_J$ using $\beta_M=0.4$ with $x_{T,M}=0.5~{\rm m}$ and $\eta_M=2$ (blue), $x_{T,M}=0.5~{\rm m}$ and $\eta_M=0$ (yellow), and $x_{T,M}=0.75~{\rm m}$ and $\eta_M=2$ (red).}
  \label{fig:daughter_time_frac_vs_etaJ}
\end{figure}

The time of a delayed particle $i$ is
\begin{equation} \label{eq:ti-delay}
t_i 
= t_M + \frac{|\vec{x}_{i}|}{c} \frac{E_i}{|\vec{p}_i|}
= t_M + \frac{x_{T,i}}{c} \frac{E_i}{p_{T,i}}.
\end{equation}
The null time of a jet, which we continue to use as the reference time, is
\begin{equation} \label{eq:tnull-delay}
t_J^{\rm null} 
= t_M + \frac{x_{T,J}}{c} \frac{|\vec{p}_J|}{p_{T,J}}
= t_M + \frac{x_{T,J}}{c} \cosh \eta_J.
\end{equation}
Let us first consider the relative time difference for the median time so that $i = i_m$.  We have
\begin{equation}
\frac{\Delta t}{t_{\rm ref}}
= \frac{t_i - t_J^{\rm null}}{t_J^{\rm null}}
= \frac
{\left(t_M + \frac{x_{T,i}}{c} \frac{E_i}{p_{T,i}}\right) - \left(t_M + \frac{x_{T,J}}{c} \frac{|\vec{p}_{J}|}{p_{T,J}}\right)  }
{t_M + \frac{x_{T,J}}{c} \frac{|\vec{p}_{J}|}{p_{T,J} }}.
\end{equation}
%. 
We set $t_D \equiv (x_{T,J}/c)(|\vec{p}_J|/p_{T,J})$, representing the ``null" time from the daughter segment.
If we approximate $x_{T,i} = x_{T,J}$ ({\it i.e.} the jet is narrow) and particle $i$ as massless, then we find
\begin{equation} \label{eq:tdiff-delayed}
\frac{\Delta t}{t_{\rm ref}}
= \frac{t_D}{t_M + t_D} \left( \frac{\cosh \eta_i}{\cosh \eta_J} - 1 \right).
\end{equation}
The first factor is the fraction of the particle's time that is traveled by the daughter and the second factor we recognize from Eq.~\eqref{eq:dt} as the prompt distribution evaluated at particle $i$'s true pseudorapidity.  As the distance the daughter travels, $x_{T,J}$, shrinks, so does the spread in $\Delta t/t_{\rm ref}$. In Fig.~\ref{fig:daughter_time_frac_vs_etaJ}, we plot the suppression factor, $t_D/(t_M+t_D)$, as a function of $\eta_J$ for several sample points of $x_{T,M}$ and $\eta_M$ using $\beta_M=0.4$

%%%%%%%%%%%%%
%%%%%%%%%%%%%

%%%%%%%%%%%%%%%%%%%%%%%%%%%%%%%%%%%%%%%%%%%%%%%%%%%%%%%%%%%%%%%%%%
\subsubsection*{Delayed Relative Time Difference}

Here we briefly review our expectations for the relative time difference in delayed jets.  Recall that for single particle measures, like the median or the hardest, the prompt relative time is given by
\begin{equation} \label{eq:dtt-prompt}
\frac{\Delta t}{t_{\rm ref}}
= \frac{\cosh \eta_{i}}{\cosh \eta_J} - 1.
\end{equation}
The three effects that cause the delayed distribution to differ from Eq.~\eqref{eq:dtt-prompt} are:
\begin{itemize}
\item the daughter time fraction,
\item the effective radius of the jet,
\item the difference between observed and truth kinematics.
\end{itemize}

\paragraph{}
Let us now contrast a few jet time definitions to assess the impact of each delayed effect on the relative time distribution.  We start with the hardest time.  The difference in observed kinematics should have a negligible effect except in rare instances when the hardest particle in a jet changes between observed and truth kinematics.  The effective radius should also have a minimal effect because the hardest particle in a jet tends to be near the jet axis.  The daughter time fraction, however, is an irreducible effect.

From Eq.~\eqref{eq:tdiff-delayed} we see that the delayed distribution inherits the prompt dependence on the daughter's true pseudorapidity, but with an additional suppression from the fact that spread between particles occurs over a smaller distance.  The suppression comes from the prefactor
\begin{equation} \label{eq:dtf-prefactor}
\frac{t_D ( x_{T,M}, \eta_i)}
{t_D ( x_{T,M}, \eta_i) + t_M ( x_{T,M}, \eta_M, \beta_M ) }.
\end{equation}
The times scale with their respective pseudorapidities, $t_M\propto{\rm cosh} (\eta_M)$ and $t_D~\propto{\rm cosh} (\eta_i)$, so that the prefactor is closest to 1 when $\eta_D \approx 0$ and $|\eta_i|$ is large, and closest to 0 when $\eta_i \approx 0$ and $|\eta_D|$ is large.  The prefactor can range from 0 to 1 and it plays a large role in the relative time difference distribution.

\paragraph{}
Next, we consider the median time.  Again, we expect the observed kinematics to have a negligible effect on the relative time difference.  The effective radius, however, can now have an impact because each particle has an equal effect on the median time of a jet.  As $R_{\rm eff}$ grows, particles further from the jet axis are included in the jet and in the calculation of the jet time.  Being far from the jet axis, these particles act like noise for the particle time distribution leading to more variation in the relative time difference distribution.  Conversely, a shrinking $R_{\rm eff}$ will tend to narrow the distribution somewhat. 
The daughter time fraction is irreducible and has an $\cO(1)$ effect on the median time.

\paragraph{}
Finally, we consider the $p_T$-weighted time.  The daughter time fraction is again a driving effect.  
The impact of the effective radius should be smaller than in the median case, because particles far from the jet axis are typically soft so their contribution to the $p_T$-weighted time is suppressed by their $p_T$.
The observed kinematics, however, can now have a large effect.  The $p_T$-weighted time for delayed jets is
\begin{equation}
t_J^{p_T} = \frac{1}{\pts{}'} \sum_{i=1}^N p_{T,i}{}' t_i,
\quad\quad\quad
\pts{}' = \sum_{i=1}^N p_{T,i}{}'.
\end{equation}
From Eq.~\eqref{eq:ptObs} we see that the ratio $p_{T,i}{}' / p_{T,i}$ is independent of momentum.  This means in the infinitely-narrow jet limit the $p_T$-weighted time is not affected by the observed kinematics.  Beyond this limit, the effect of using the observed $p_T$ can be large if the variation of $p_{T,i}{}' / p_{T,i}$ is large over the area of the jet. 

The ability to accurately identify displaced vertices can eliminate the impact of the observed kinematics.  Such an upgrade would be expected to improve the performance of the $p_T$-weighted time, but have a small effect on the hardest time and the median time.

%%%%%%%%%%%%%%%%%%%%%%%%%%%%%%%%%%%%%%%%%%%%%%%%%%%%%%%%%%%%%%%%%%
%%%%%%%%%%%%%%%%%%%%%%%%%%%%%%%%%%%%%%%%%%%%%%%%%%%%%%%%%%%%%%%%%%
\section{Numerical Results}
\label{sec:numerical}

In this section, we compute the relative time differences for several jet time definitions in simulated data.  Results will be compared with the derived behaviors from Sec.~\ref{sec:analytic} and are found to follow the predicted trends.

%%%%%%%%%%%%%%%%%%%%%%%%%%%%%%%%%%%%%%%%%%%%%%%%%%%%%%%%%%%%%%%%%%
\subsection{Simulation Details}
\label{sec:sim}

For prompt jets, we generate $pp \to Z' \to q\bar{q}$ events, where $q=u,d$, using Pythia v8.240~\cite{Sjostrand:2014zea} at a center of mass energy of $\sqrt{s} = 14~{\rm TeV}$ and with a $Z'$ mass of $m_{Z'} = 1~{\rm TeV}$. 
Initial state radiation (ISR) and multiparton interactions (MPI) are turned off.  Particles with $p_T < 0.5~{\rm GeV}$ or with $|\eta|>4$ are discarded.  Particles are clustered into anti-$k_T$ jets~\cite{Cacciari:2008gp} with $\rjet = 0.5$ using FastJet v3.3.2~\cite{Cacciari:2011ma}.   The results are presented at particle-level without any detector resolution or time resolution included.  The impact of these effects is shown in App.~\ref{app:detector} to be small.  

For delayed jets, we generate $pp \to \tilde{g} \tilde{g} \to (g\tilde{G}) (g\tilde{G})$ at parton-level using MadGraph5 v2.7.3~\cite{Alwall:2014hca} at a center of mass energy of $\sqrt{s} = 14~{\rm TeV}$ and with particle masses of $m_{\tilde{g}} = 2~{\rm TeV}$ and $m_{\tilde{G}} = 10^{-16}~{\rm TeV}$.\footnote{
  Whether using gluon-initiated or light quark-initiated jets does not give rise to qualitative differences.}  Events are showered with Pythia with ISR and MPI turned off.
Particles with an observed transverse momentum below $p_T{}' < 0.5~{\rm GeV}$ or with an observed pseudorapidity $|\eta'|>4$ are discarded, where a cylindrical detector with a radius of $r_T = 1~{\rm m}$ is used. 

In both the prompt and delayed samples, hadronization is turned off.  In Pythia, when there are both prompt and displaced particles, due to the hadronization procedure, some particles that descend from the displaced gluon can be assigned to the prompt vertex.  Using unhadronized events avoids the issue of determining to which vertex a hadron should be assigned.  In App.~\ref{app:hadronization} we compare the relative time distributions, in a prompt sample, with and without hadronization and find that the impact is at most a few percent.

%%%%%%%%%%%%%%%%%%%%%%%%%%%%%%%%%%%%%%%%%%%%%%%%%%%%%%%%%%%%%%%%%%
\subsection{Prompt Jets}
\label{sec:sim-prompt}

We first look at prompt jets because the prompt distributions are inputs to understanding the delayed distributions.  In each event we only consider the hardest jet and require that it has $p_T > 250~{\rm GeV}$. 

\begin{figure} [t]
\begin{center}
  \includegraphics[width=0.45\textwidth]{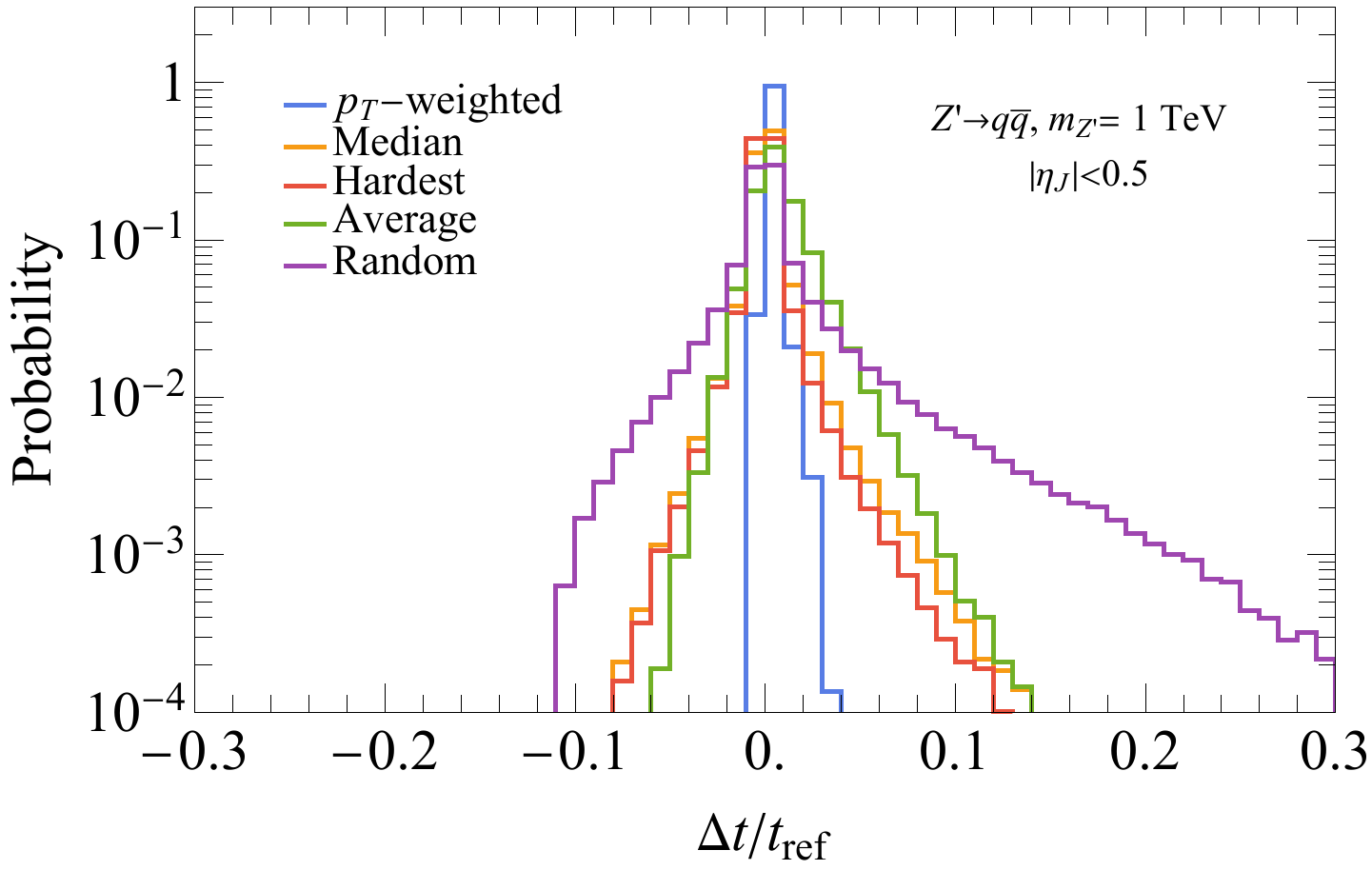} \quad\quad\quad
  \includegraphics[width=0.45\textwidth]{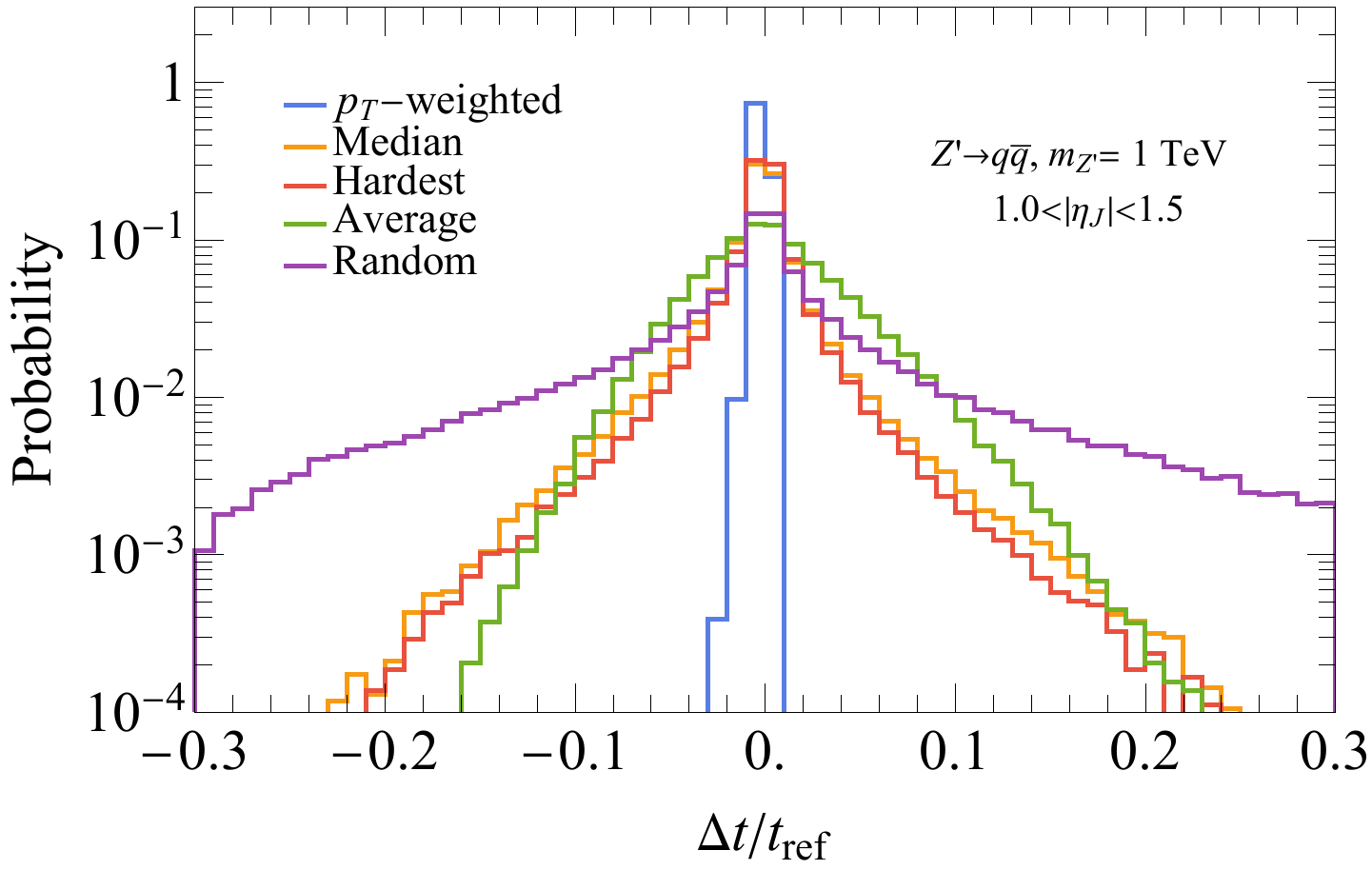}
  \caption{The relative time difference distribution in prompt jets for $|\eta_J|<0.5$ (left) and for $1.0 < |\eta_J| < 1.5$ (right).}
  \label{fig:relTD_prompt}
\end{center}
\end{figure}

In Fig.~\ref{fig:relTD_prompt} (left) we show the distribution of $\Delta t / t_{\rm ref}$ for jets with $|\eta_J| < 0.5$ for the jet time definitions of $p_T$-weighted, median, hardest, average, and random.  As expected, selecting a random particle in the jet to represent the jet time yields the widest distribution.  Its distribution skews towards positive relative times because $\eta_J~=~0$ corresponds to the fastest possible time, meaning there is more phase space for positive values.  The other time definitions have narrower distributions but still skew towards positive values.  The median, hardest, and average times have comparable performance, while the $p_T$-weighted time has the narrowest distribution. 

From Fig.~\ref{fig:relTD_prompt} we see that each $\Delta t / t_{\rm ref}$ distribution peaks near zero, but that the mean depends on the range of $\eta_J$ used.  The width of the distributions is an indicator of the resolution of a method and a useful figure of merit.  Since these distributions are non-Gaussian, the $1\sigma$ standard deviation does not fully characterize the shapes, and in particular does not provide useful information about the tails.  For that reason, we use the $3\sigma$ width ({\it i.e.} the bounds of the integral containing $99.7\%$ of events) for comparison.\footnote{In fact, we use the minimum width that contains $99.7\%$ of the events rather than width centered at the mean because of the asymmetric nature of the distributions.}  With this as the resolution, the $p_T$-weighted time performs $5$ times better than the hardest time and $6$ times better than the median time.

%%%%%%%%%%%%%%%%%%
%%%%%%%%%%%%%%%%%%
 
In Fig.~\ref{fig:relTD_prompt} (right) we look at the relative time distribution for jets with $1.0 < |\eta_J| < 1.5$.  The same pattern is present here where the random time is the widest distribution, followed by the median, hardest, and average times with similar widths, and the $p_T$-weighted time with the narrowest distribution.   Again, comparing the resolutions, we find that the $p_T$-weighted time is 16 times better than the hardest time and 17 times better than the median time.  The distributions of the median and hardest times widen noticeably in this $1.0 < |\eta_J| < 1.5$ range, as compared to $|\eta_J|<0.5$, as predicted by Eqs.~\eqref{eq:tmax}~and~\eqref{eq:tmin}.  The $p_T$-weighted time instead depends on the interplay between $E_J / |\vec{p}_J|$ and the ratio of transverse momentum to the scalar sum of the constituents' transverse momenta.

%%%%%%%%%%%%%%%%%%
\begin{figure} [t]
\begin{center}
  \includegraphics[width=0.45\textwidth]{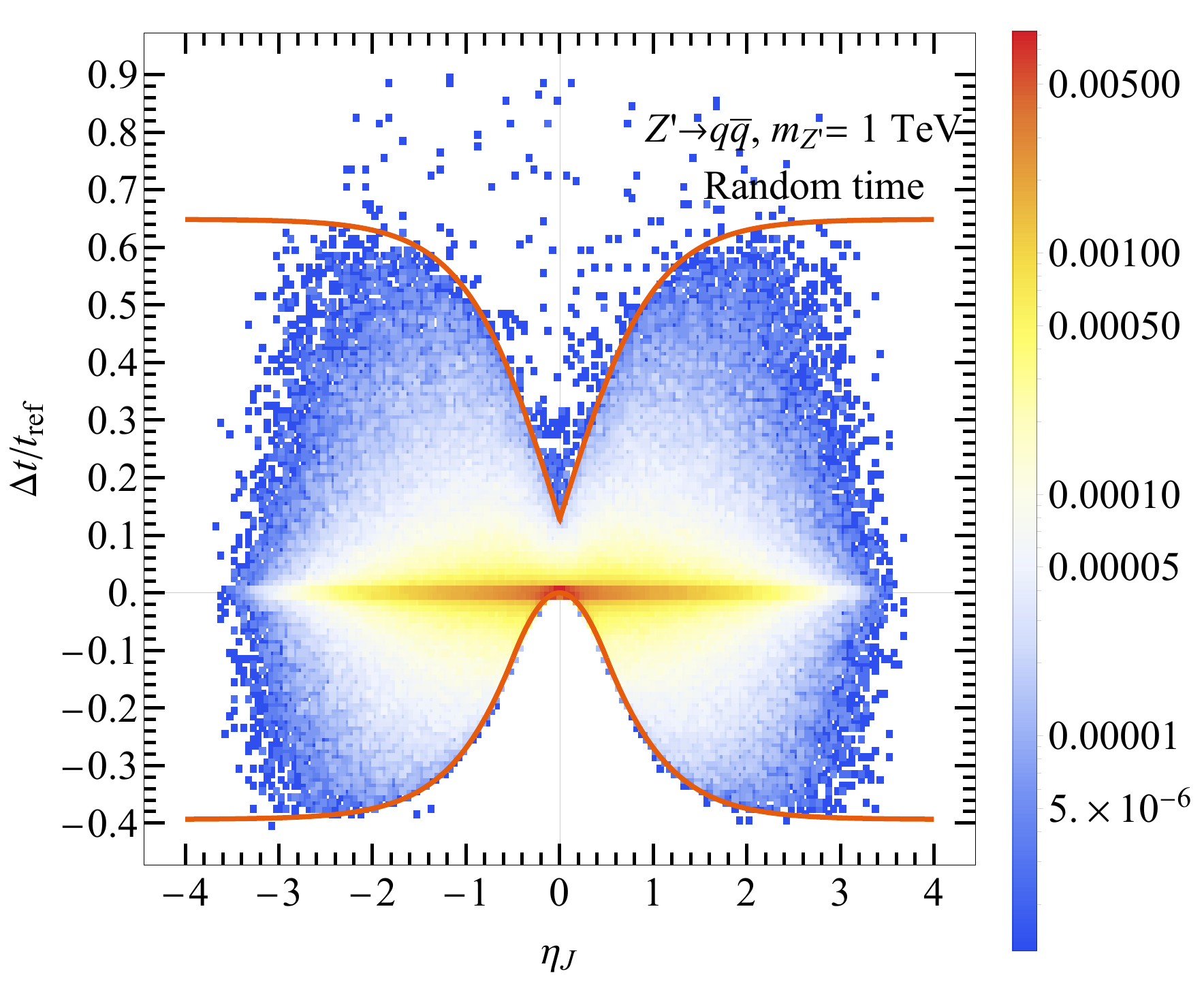} \quad\quad\quad
  \includegraphics[width=0.45\textwidth]{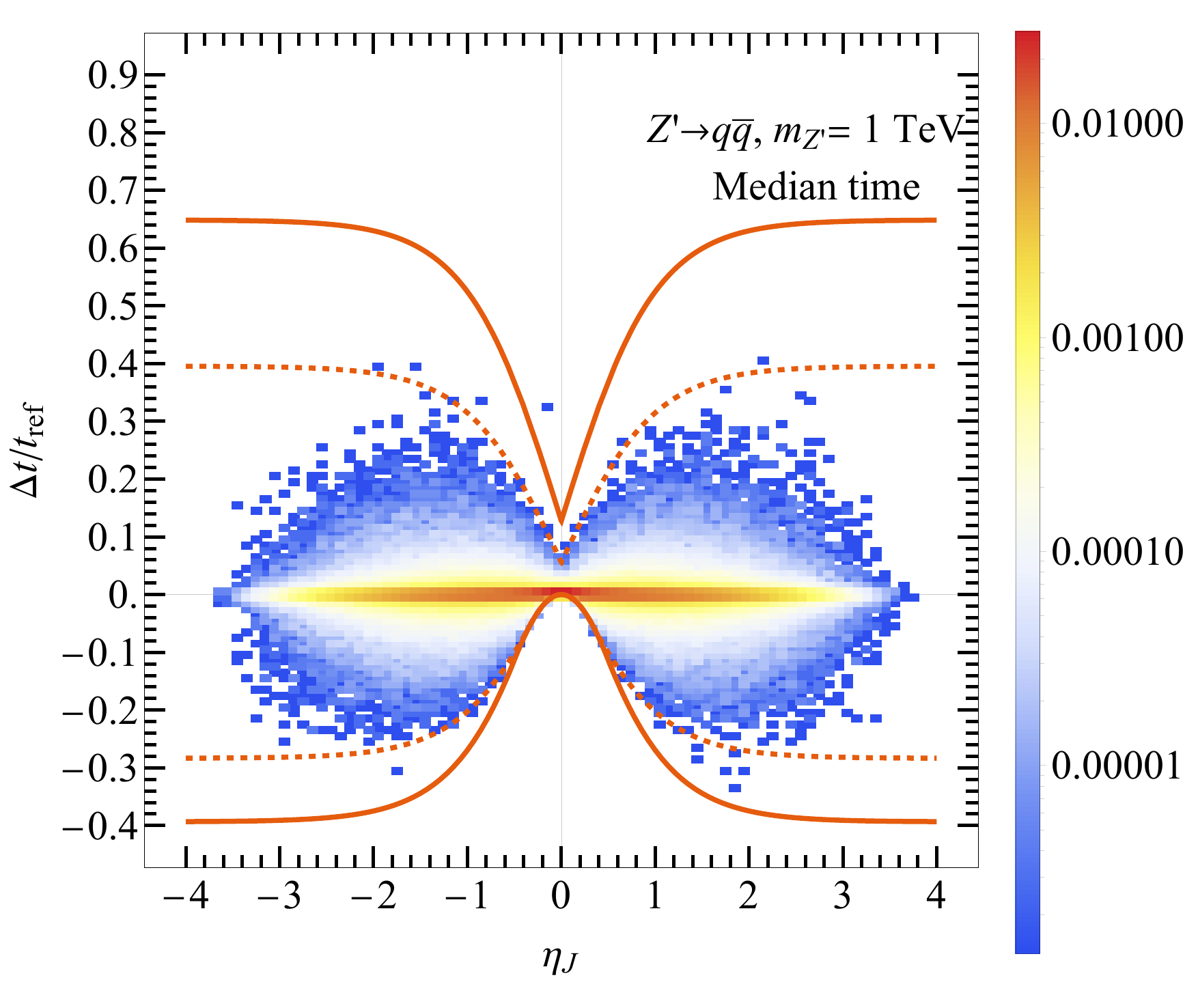} 
  \caption{The relative time difference vs. $\eta_J$ distribution for random time (left) and median time (right). The solid red curves depict the bounds from Eqs.~\eqref{eq:tmax}~and~\eqref{eq:tmin}. The dashed red curve depicts the same bounds for a jet of radius $(2/3) \rjet$. }
  \label{fig:relTD_prompt_2d_random_median}
\end{center}
\end{figure}
%%%%%%%%%%%%%%%%%%

%%%%%%%%%%%%%%%%%%
\begin{figure} [t]
\begin{center}
  \includegraphics[width=0.45\textwidth]{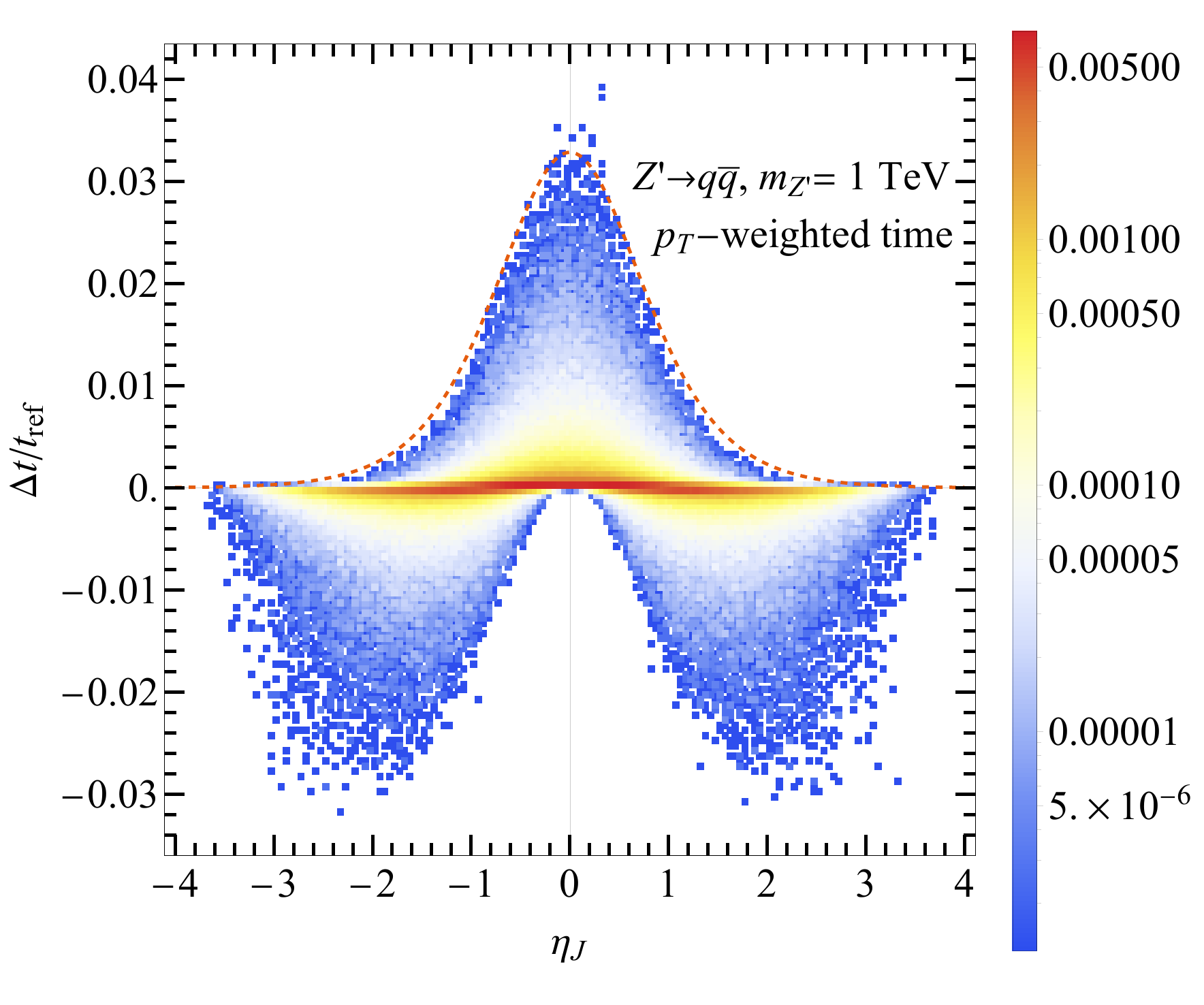}
  \caption{The relative time difference vs. $\eta_J$ distribution for $p_T$-weighted time. The dashed red curve depicts the relation in Eq.~\eqref{eq:tdiff-ptw}.}
  \label{fig:relTD_prompt_2d_ptw}
\end{center}
\end{figure}
%%%%%%%%%%%%%%%%%%

To better understand the difference between $\eta_J$ regions, we look at the two-dimensional distribution of $\Delta t / t_{\rm ref}$ vs. $\eta_J$.  Fig.~\ref{fig:relTD_prompt_2d_random_median} shows this distribution for random time with the bounds from Eqs.~\eqref{eq:tmax}~and~\eqref{eq:tmin} overlaid.  The majority of events fill out the region between the bounds with a few events above the maximum, due to mass effects, and a few events below the minimum due to the fact that particles can be slightly farther than $\rjet$ from the jet axis.

Fig.~\ref{fig:relTD_prompt_2d_random_median} (right) shows $\Delta t / t_{\rm ref}$ vs. $\eta_J$ for the median time.  The red solid lines are the boundaries from Eqs.~\eqref{eq:tmax}~and~\eqref{eq:tmin}.  This distribution clusters closer around $\Delta t / t_{\rm ref}$ values near zero.  In fact, the dashed lines are boundaries for a jet with radius $(2/3) \rjet$ which corresponds to the empirical observation that the behavior of the median time is similar to choosing a random particle from a narrower jet.

In Fig.~\ref{fig:relTD_prompt_2d_ptw} we show the same distribution for the $p_T$-weighted time.  Here, we see that the behavior predicted by Eq.~\eqref{eq:tdiff-ptw} does appear in the simulation.  The positive relative time differences near $\eta_J = 0$ result from the $E_J / |\vec{p}_J|$ factor.  The shape in that region even follows ${\rm sech}^2(\eta_J)$ as discussed in Sec.~\ref{sec:gen-prompt}.  As $|\eta_J|$ grows past $\approx 2$ the $E_J / |\vec{p}_J|$ factor approaches unity and the $p_{T,J} / \pts$ factor determines the shape.  Both of these factors have a narrow distribution leading to an overall narrow distribution for the relative time difference for the $p_T$-weighted time.

%%%%%%%%%%%%%%%%%%%%%%%%%%%%%%%%%%%%%%%%%%%%%%%%%%%%%%%%%%%%%%%%%%
\subsection{Delayed Jets}
\label{sec:sim-delayed}

For delayed jets the parameter space expands from $\eta_J$ to $\eta_M$, $\eta_J$, $\Delta\phi$, $\beta_M$, and $x_{T,M}$.  To study a delayed sample, we vary the values for $\eta_M$, $\eta_J$, and $x_{T,M}$ and fix $\Delta\phi=0$ and $\beta_M=0.4$. The effect of non-zero $\Delta \phi$ has been discussed in our analytic estimates in Section~\ref{sec:gen-delayed}.

In every event, there are two gluinos, each of which decay to a gluon leading to a hard jet.  The time of flights in the LHE file was modified such that one of these gluinos is forced to decay outside of the detector while the other gluino is set to have velocity $\beta_M$ and decays at a transverse distance $x_{T,M}$ to a gluon that points along $\eta_J$ at parton-level.  This same event is then re-showered 100,000 times to produce a sample of jets with fixed parton-level kinematics.

We consider only the hardest jet (that originates from inside the detector) in the event and require it to have an observed $p_T{}'~>~50~{\rm GeV}$.  In order to identify effects that are dependent on the event topology, we discard events that differ by more than $0.25$ in pseudorapidity or $0.25$ in azimuthal angle before and after showering.  

%%%%%%%%%%%%%%%%%%
\begin{figure} [t]
\begin{center}
  \includegraphics[width=0.45\textwidth]{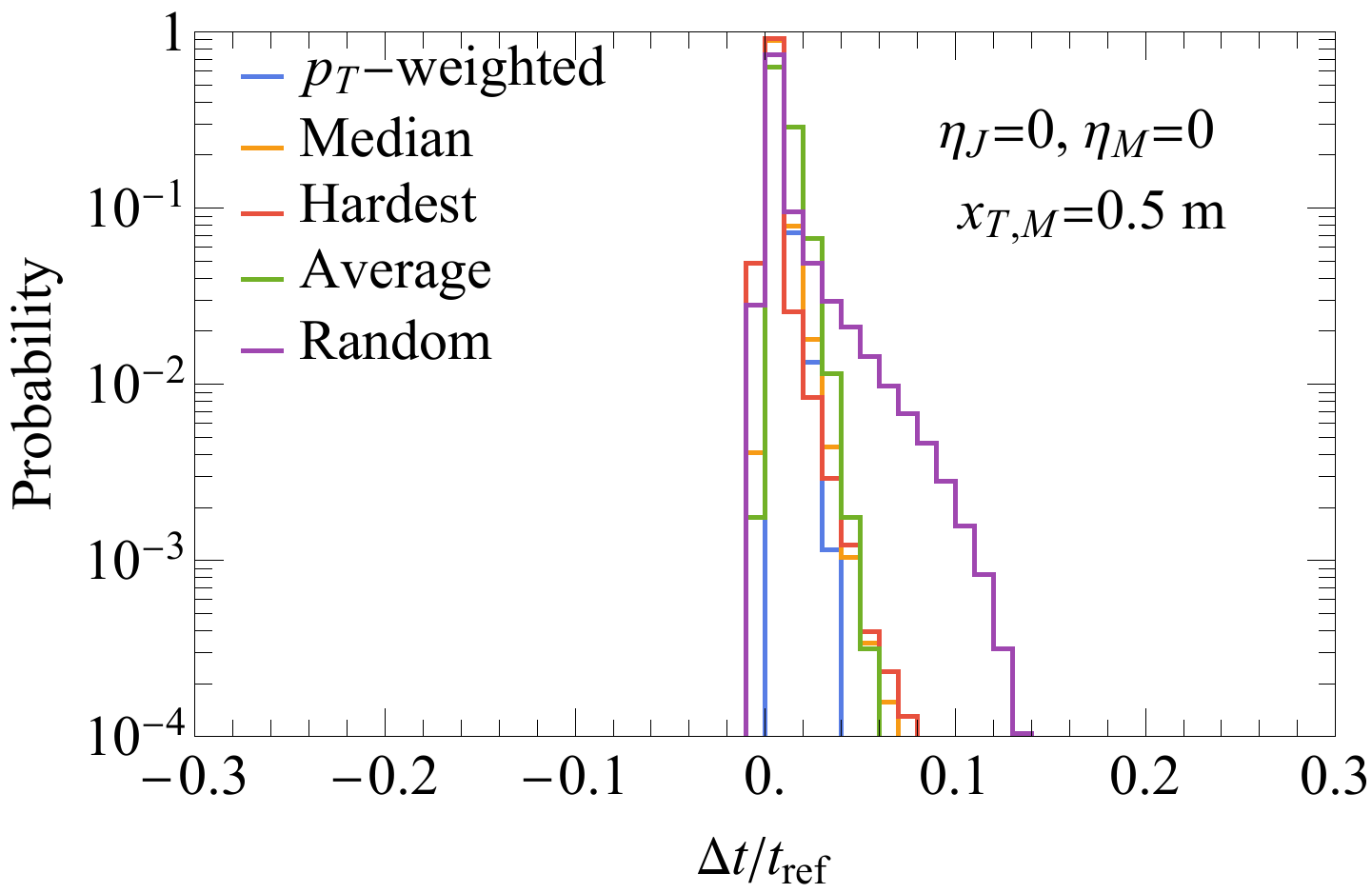} \quad\quad\quad
  \includegraphics[width=0.45\textwidth]{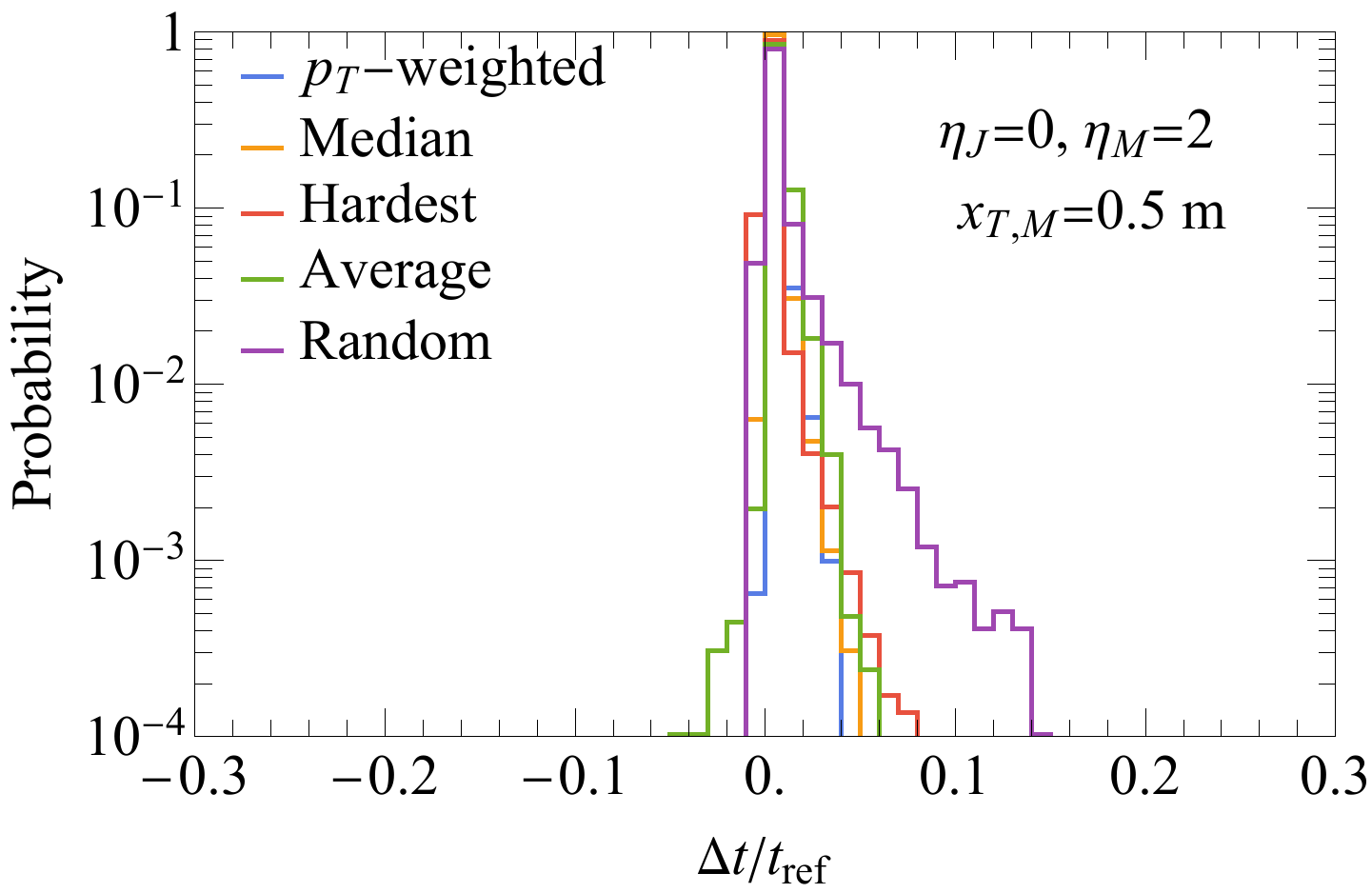} \\
  \includegraphics[width=0.45\textwidth]{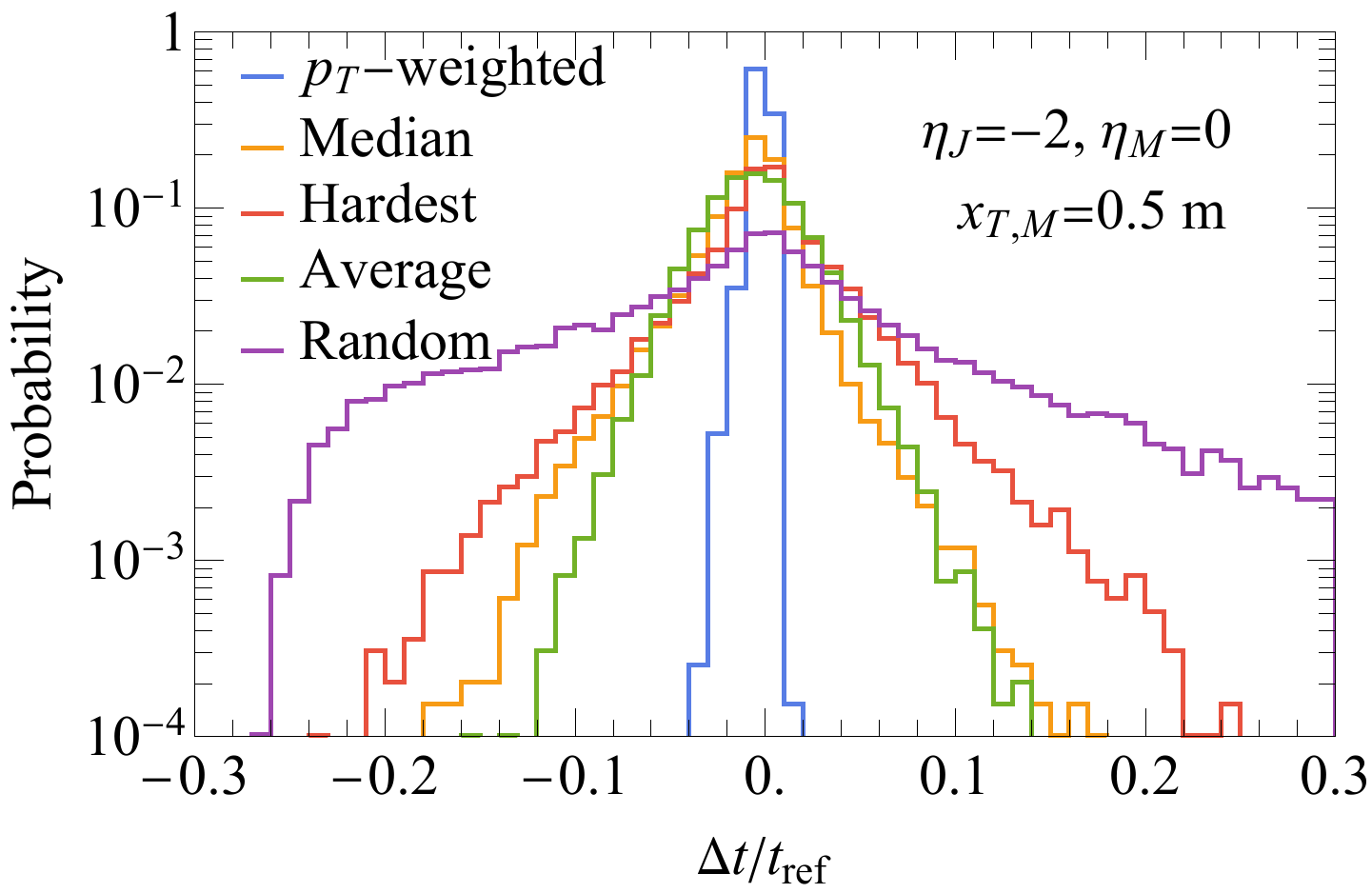} \quad\quad\quad
  \includegraphics[width=0.45\textwidth]{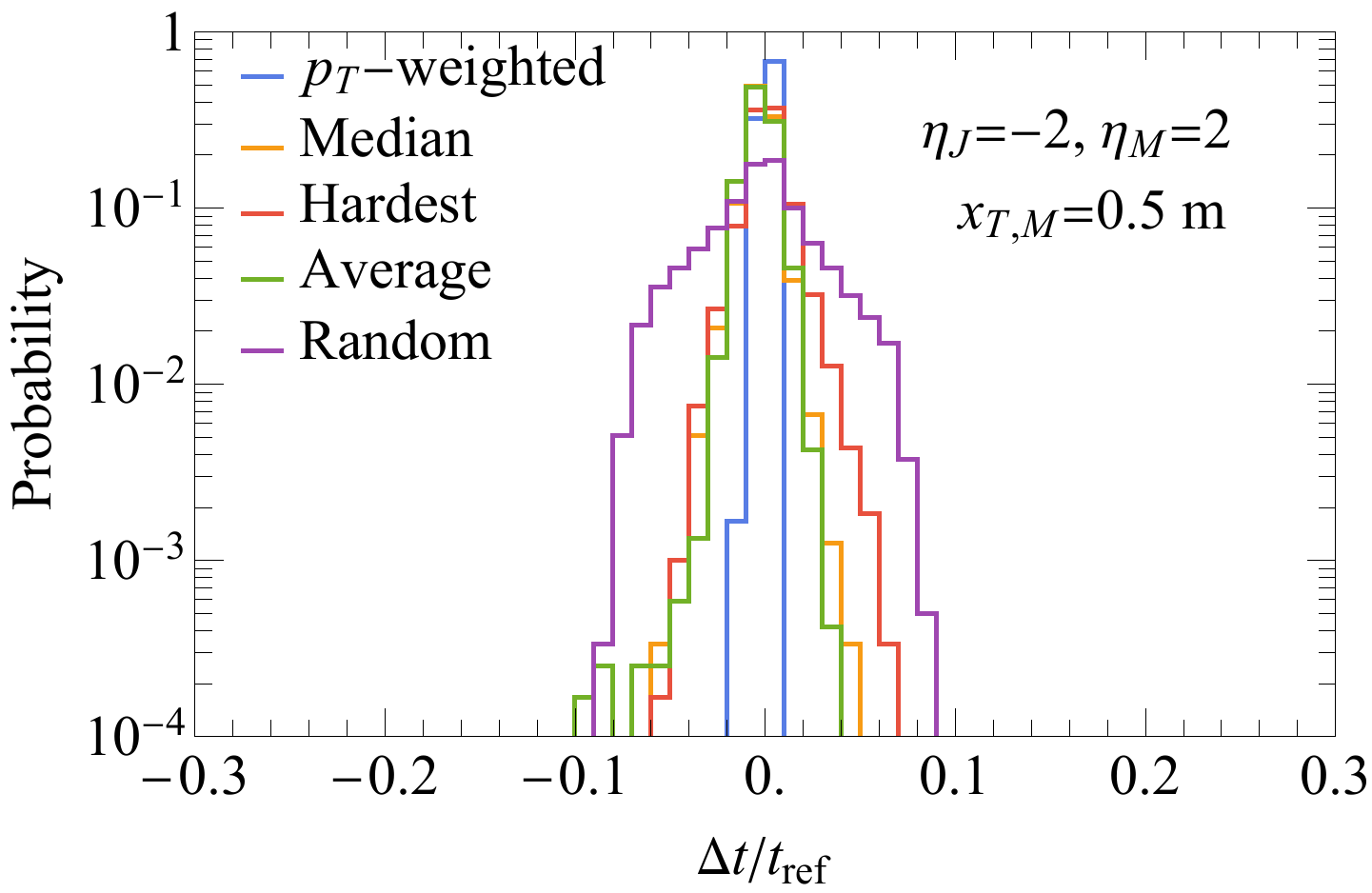}
  \caption{The relative time difference distribution for $(\eta_{J}, \eta_M) = (0, 0)$ (top left), $(\eta_{J}, \eta_M) = (0, 2)$ (top right), $(\eta_{J}, \eta_M) = (-2, 0)$ (bottom left), and $(\eta_{J}, \eta_M) = (-2, 2)$ (bottom right) for $x_{T,M}=0.5~\text{m}$.}
  \label{fig:relTD_delayed}
\end{center}
\end{figure}
%%%%%%%%%%%%%%%%%%

We first look at distributions of $\Delta t/t_{\rm ref}$ in Fig.~\ref{fig:relTD_delayed} for $x_{T,M}=0.5~{\rm m}$.  The top left plot shows the $p_T$-weighted, median, hardest, average, and random times with $(\eta_J, \eta_M) = (0, 0)$.  Much like the prompt case, every definition skews positive since $\eta_J = 0$ corresponds to fastest possible arrival time.  The distributions are narrower than the prompt case due to the decrease in variation in arrival time as captured by the daughter time fraction in Eq.~\eqref{eq:dtf-prefactor}.

Fig.~\ref{fig:relTD_delayed} (top right) shows the $\Delta t/t_{\rm ref}$ distributions for $(\eta_J,\eta_M)=(0,2)$ which corresponds to a forward gluino that decays to gluon that travels perpendicular to the beamline, directly to the detector.  From Eq.~\eqref{eq:tdiff-delayed} we expect this distribution to be similar to the prompt distribution for central jets.  Compared to the $(\eta_J,\eta_M)=(0,0)$, this point has a smaller daughter time fraction and is narrower as expected.

The bottom left of Fig.~\ref{fig:relTD_delayed} shows $\Delta t/t_{\rm ref}$ distributions for $(\eta_J,\eta_M)=(-2,0)$.  Here the gluino travels perpendicular to the beamline then decays to a backward pointing gluon.  The observed pseudorapidity for the gluon is $\eta_J{}' = -1.3$.  Focusing first on the $p_T$-weighted time, we see that the distribution is slightly wider than the prompt case, Fig.~\ref{fig:relTD_prompt_2d_ptw}, despite a slight suppression of $\approx 0.6$ from the daughter time fraction.  This is due to a sizable variation between the observed and truth kinematics.  The median and hardest distributions do not differ much from their prompt counterparts.

Fig.~\ref{fig:relTD_delayed} (bottom right) shows the distributions for $(\eta_J,\eta_M)=(-2,2)$.  In this case, each distribution is very narrow.  This is primarily a consequence of the jet having $R_\text{eff}=0.27$.\footnote{Note that because the particles in a jet are not uniformly distributed, excluding particles that are farther from the jet axis does result in a narrower relative time difference distribution.}  While the $p_T$-weighted distribution is still narrow, the difference is not as large as for other jet times because of the discrepancy between observed and truth kinematics for this configuration.

%%%%%%%%%%%%%%%%%%
\begin{figure}[h]
  \begin{center}
    \includegraphics[width=0.45\textwidth]{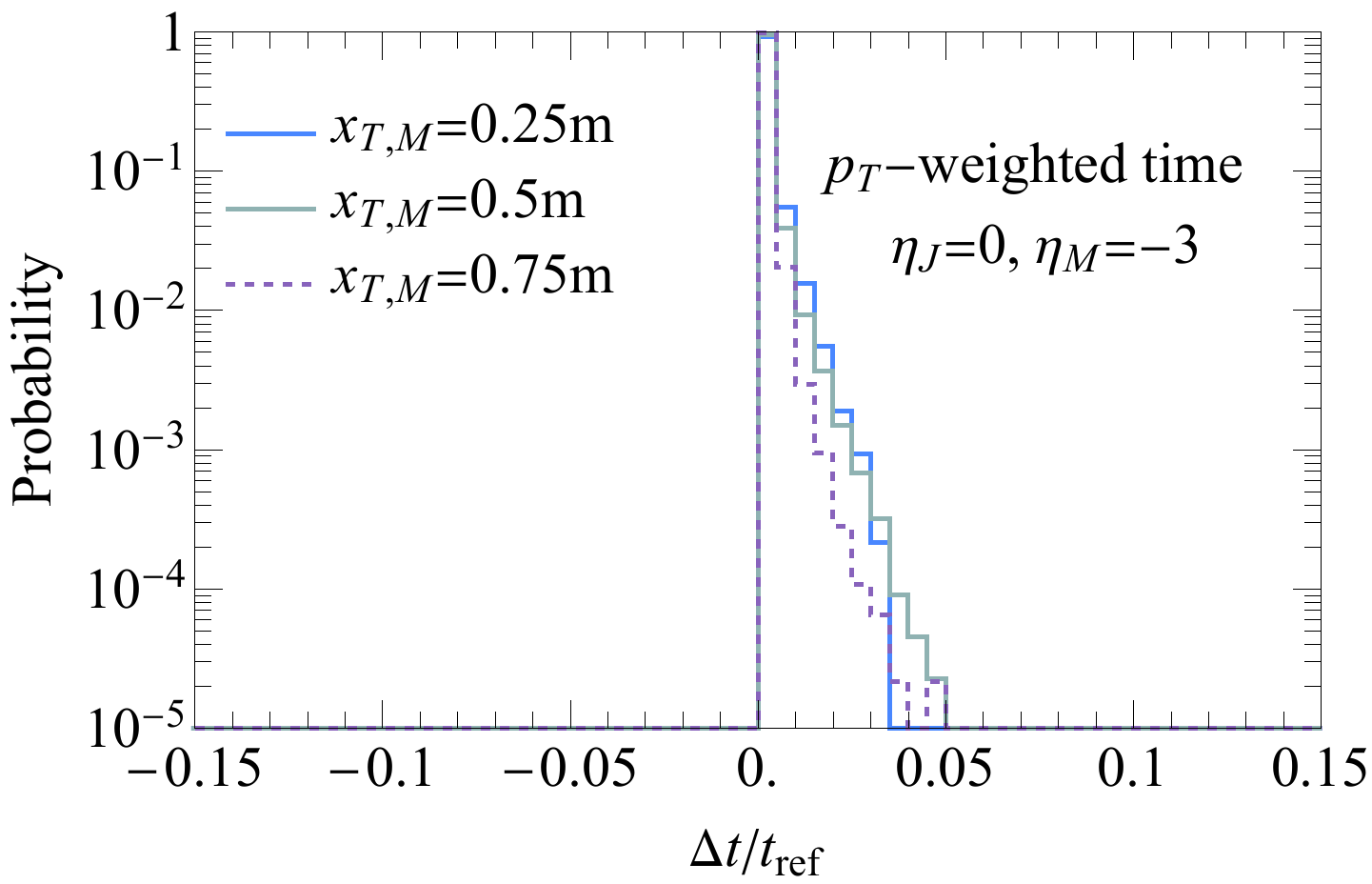}
    \includegraphics[width=0.45\textwidth]{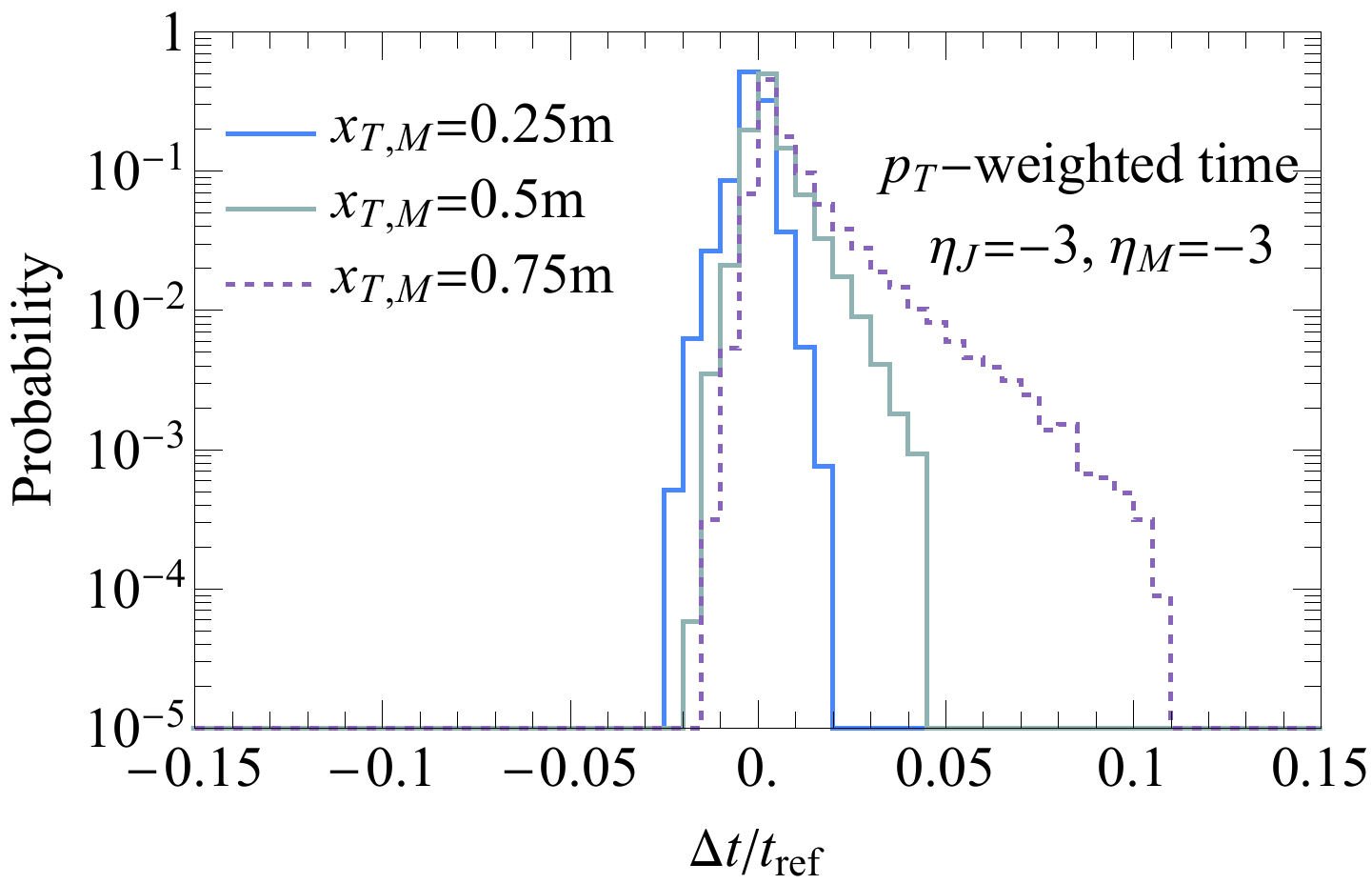}
    \caption{The relative time difference distribution for the $p_T$-weighted time at $x_{T,M}$= 0.25 m (blue solid line), 0.5 m (gray solid line), and 0.75 m (purple dashed line) for $(\eta_J,\eta_M)=(0,-3)$ (left) and $(\eta_J,\eta_M)=(-3,-3)$ (right).}
    \label{fig:diff_xt}
  \end{center}
\end{figure}
%%%%%%%%%%%%%%%%%%

In Fig.~\ref{fig:diff_xt}, we plot the distributions of the relative time difference for the $p_T$-weighted time for $x_{T,M}=0.25,0.5,0.75~\rm{m}$. The left plot shows the distributions for $(\eta_J,\eta_M)=(0,-3)$ while the right plot shows the distributions for $(\eta_J,\eta_M)=(-3,-3)$. In these two plots, we see very different behavior as $x_{T,M}$ increases. The left plot shows the distribution narrowing with $x_{T,M}$, the expected behavior from the decrease in the daughter time fraction. By contrast, the right plot shows the distribution broadening and skewing positive with $x_{T,M}$. This broadening is the expected behavior from both the larger effective cone size, and larger difference between the true and observed $p_T$. Furthermore, the relative increase in observed $p_T$ is larger forward particles, resulting in the positive skew.

The parameter scans in Figs.~\ref{fig:relTD_delayed} and~\ref{fig:diff_xt} are useful for emphasizing a few physics points.  Firstly, the $p_T$-weighted time is consistently better than other jet time definitions across the variation of key kinematics, namely $\eta_M$ and $\eta_J$.  Secondly, the $p_T$-weighted time has a different spread at different kinematical points which means one should compute calibrations and efficiencies at each point rather than using a single value over all of parameter space.\footnote{A parameter scan of $(\eta_J,\eta_M)$ with finer steps can be found in App.~\ref{app:para_scan}.  The trends shown there agree with the analysis provided here.}

%%%%%%%%%%%%%%%%%%%%%%%%%%%%%%%%%%%%%%%%%%%%%%%%%%%%%%%%%%%%%%%%%%
%%%%%%%%%%%%%%%%%%%%%%%%%%%%%%%%%%%%%%%%%%%%%%%%%%%%%%%%%%%%%%%%%%
\section{Conclusions}
\label{sec:outlook}

The time of a jet is a theoretically ambiguous and yet experimentally highly relevant quantity. The time profile of a jet provides a new independent probe of jet properties, potentially deepening our understanding of QCD.  Experimentally, the jet time is an observable with strong discrimination power in searches for long-lived particles.  Like how the jet clustering algorithm itself defines a jet using a collection of particles, the choice of jet time definition determines its properties and performance. A useful definition should have predictable behavior, give the closest representation to the parton-level information, and, more importantly, minimize the spread in arrival time.

In this work, we primarily studied five definitions of jet time. The first was the $p_T$-weighted time where the jet time is a $p_T$-weighted sum of the jet constituent arrival times. The second was the median time which uses the median constituent time as the jet time. The third was the hardest time where the time of the highest $p_T$ constituent is used as the jet time. The fourth was the average time where the jet time is taken as the average of the constituent times. The fifth was the random time where the time of a constituent was randomly chosen to be used as the jet time.

To evaluate the various definitions, we both predicted and computed in simulation the relative time difference of a definition compared to the time it would take a massless parton to travel along the jet's trajectory. The width of the relative time difference distribution tells us how precisely the jet time can be measured. For prompt jets, we showed that the performance depends on the pseudorapidity of the jet. Due to the geometry of the detector barrel, all jet time definitions have wider distributions as the jets become more forward.  We found that the $p_T$-weighted jet time consistently has the best performance.  For instance, for central jets with $1.0<|\eta|<1.5$, the $p_T$-weighted time has a 16-fold improvement over the (next-best-performing) hardest time.  For central jets with $|\eta|<0.5$, the $p_T$-weighted time has a 5-fold improvement over hardest time.

For delayed jets, the full kinematics of the event affects the performance. Specifically, the direction of the mother particle, the direction of the jet, and the transverse decay location of the mother particle determine the behavior of the jet times.  We show that delayed jet timing behavior can be understood through three effects. The first is the daughter time fraction which is the fact that as the displaced vertex gets closer to the detector there is less distance for the constituents to travel and consequently less spread in their times.  The second is the effective radius of the jet that is an effect of the displaced vertex.  The third is the that observed $p_T$ differs from the true $p_T$, which occurs when the displaced vertex is not identified.  Just as for prompt jets, the $p_T$-weighted time has the best performance over the full parameter space.  Furthermore, the strong dependence on the event kinematics emphasizes the importance of having an efficiency map that depends on the long-lived particle's direction, its decay location, and the direction of the daughter jet.

This work is the first study that looks at the impact of different definitions of jet time.  There are many related directions that can be explored.  For instance, finding the jet time definition that is most amenable to direct calculation may help reduce theory uncertainties.  More practically, given the trigger computation complexity budget, it would be useful to understand the best alternative jet time definition for a low-level delayed jet trigger.  On the analysis side, studies could be done on the interplay between jet time and pileup and grooming.  Other new physics models with different event topologies would be interesting to study.  More detailed signal-specific studies are needed to evaluate the direct impact of using the jet time in new physics searches.  We are optimistic that the jet time has the potential to be a standard tool in long-lived particle searches in the near future.

%%%%%%%%%%%%%%%%%%%%%%%%%%%%%%%%%
\acknowledgments

The authors would like to thank Matthew Citron and Nhan Tran for useful discussions.  WHC and LTW are supported by the DOE grant DE-SC0009924. The work of Z.L. is supported in part by the U.S. Department of Energy under grant No. DE-SC0022345. ML acknowledges support from the Fermi Research Alliance, LLC, under Contract No. DEAC02-07CH11359 with the U.S. Department of Energy, Office of Science, Office of High Energy Physics. ZL and LTW acknowledge Aspen Center of Physics for hospitality during the final phase of this study, which is supported by National Science Foundation grant PHY-1607611.

%%%%%%%%%%%%%%%%%%%%%%%%%%%%%%%%%%%%%%%%%%%%%%%%%%%%%%%%%%%%%%%%%%
%%%%%%%%%%%%%%%%%%%%%%%%%%%%%%%%%%%%%%%%%%%%%%%%%%%%%%%%%%%%%%%%%%
\appendix
\section{Finite Length Detectors}
\label{app:finitedet}

In the main text, we consider a detector with only a barrel capable of timing measurements. If one includes endcaps, then the timing distributions are different for jets with times that the endcaps would measure.

If the pseudorapidity at which the barrel connects to the endcap is $\eta_{\rm EC}$, then the arrival time of a particle $i$ at the endcap is
\begin{equation}
t_i^{\rm endcap} = \frac{z_{\rm EC}}{c} \frac{1}{\tanh{\eta_i}},
\end{equation}
where $z_{\rm EC} = r_T \cosh \eta_{\rm EC}$. If all of the jet's constituents lie solely in the endcap, the trajectory that yields the shortest (largest) arrival time is now the most forward (central) constituent.

For jets with constituents in the intermediate region, the trajectory that yields the largest arrival time is always the trajectory intersecting the barrel-endcap corner. Depending on the jet axis, the shortest arrival time can be a constituent that intersects the barrel or the endcap.

\begin{figure} [H]
  \begin{center}
  \includegraphics[width=0.45\textwidth]{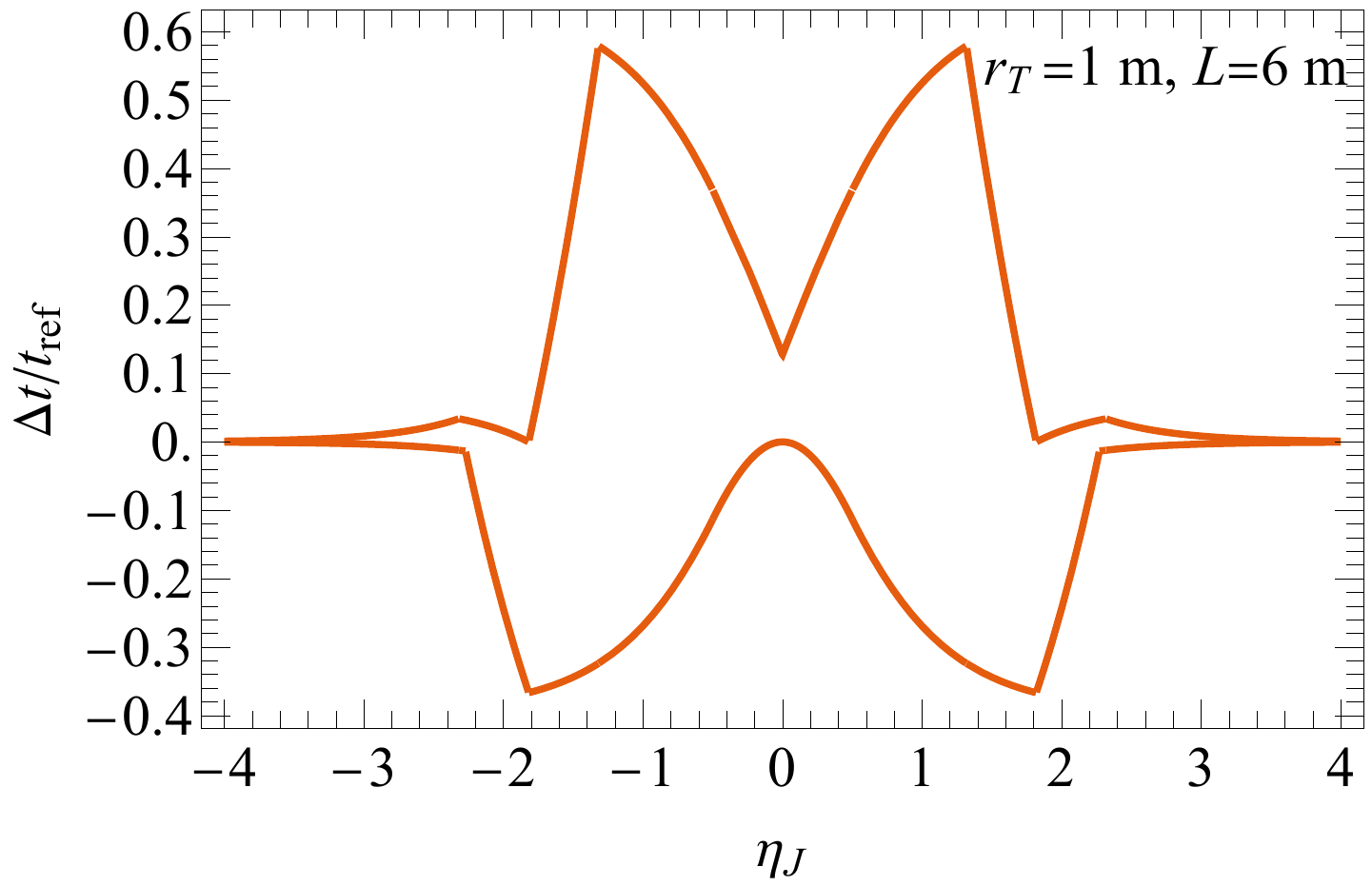}
  \caption{The boundaries for the relative time difference as a function of $\eta_J$. This can be compared with the boundaries in Fig.~\ref{fig:relTD_prompt_2d_random_median}.}
  \label{fig:relTD_prompt_with_endcaps}
  \end{center}
\end{figure}

In Fig.~\ref{fig:relTD_prompt_with_endcaps} we show the maximum and minimum relative time differences (for a single-particle measure) for $r_T=1~{\rm m}$ and a total barrel length of $L=6~{\rm m}$ which corresponds to $\eta_{\rm EC} = 1.76$.\footnote{This yields the approximate inner geometry of the CMS and ATLAS electromagnetic calorimeters~\cite{CMS:1994prop, ATLAS:1994prop}.} Once all jet constituents lie within the endcap, the allowed spread in relative time difference sharply drops for prompt jets.  

%%%%%%%%%%%%%%%%%%
%%%%%%%%%%%%%%%%%%

%%%%%%%%%%%%%%%%%%%%%%%%%%%%%%%%%%%%%%%%%%%%%%%%%%%%%%%%%%%%%%%%%%
%%%%%%%%%%%%%%%%%%%%%%%%%%%%%%%%%%%%%%%%%%%%%%%%%%%%%%%%%%%%%%%%%%
\section{Pileup and Grooming}
\label{app:pileup}

We simulate pileup by overlaying $n_\text{PU}$ soft QCD vertices onto our hard event. The number of pileup events is Poisson distributed, with $\langle n_\text{PU}\rangle=140$ and a cutoff at 200 vertices. The pileup vertices follows a Gaussian spread in both $z$ and $t$, with $\sigma_z=c \sigma_t=60~\rm{mm}$~\cite{Butler:2019rpu}. The events with pileup include all of the detector effects discussed in App.~\ref{app:detector}.

\begin{figure} [H]
\begin{center}
  \includegraphics[width=0.45\textwidth]{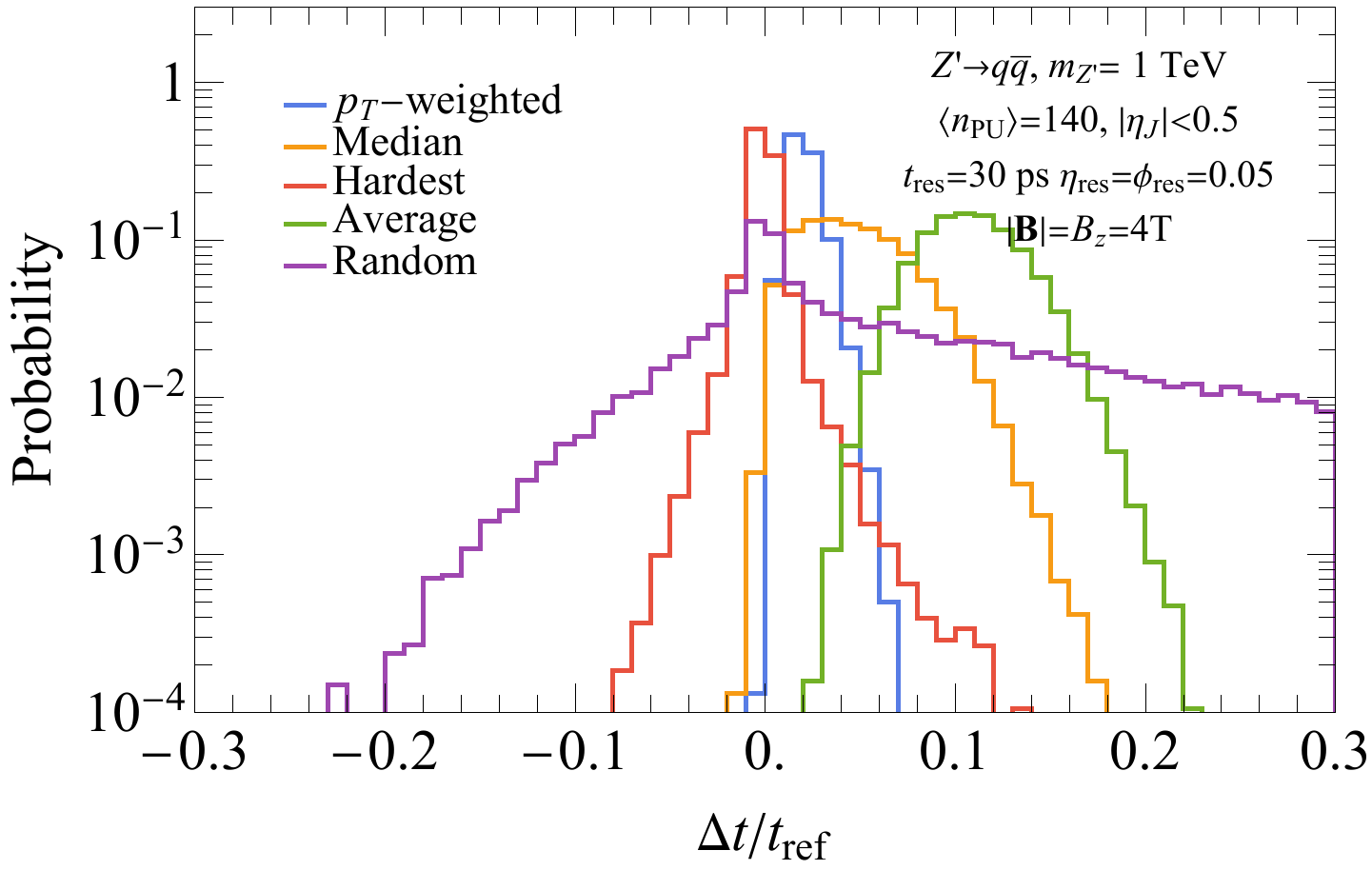} \quad\quad\quad
  \includegraphics[width=0.45\textwidth]{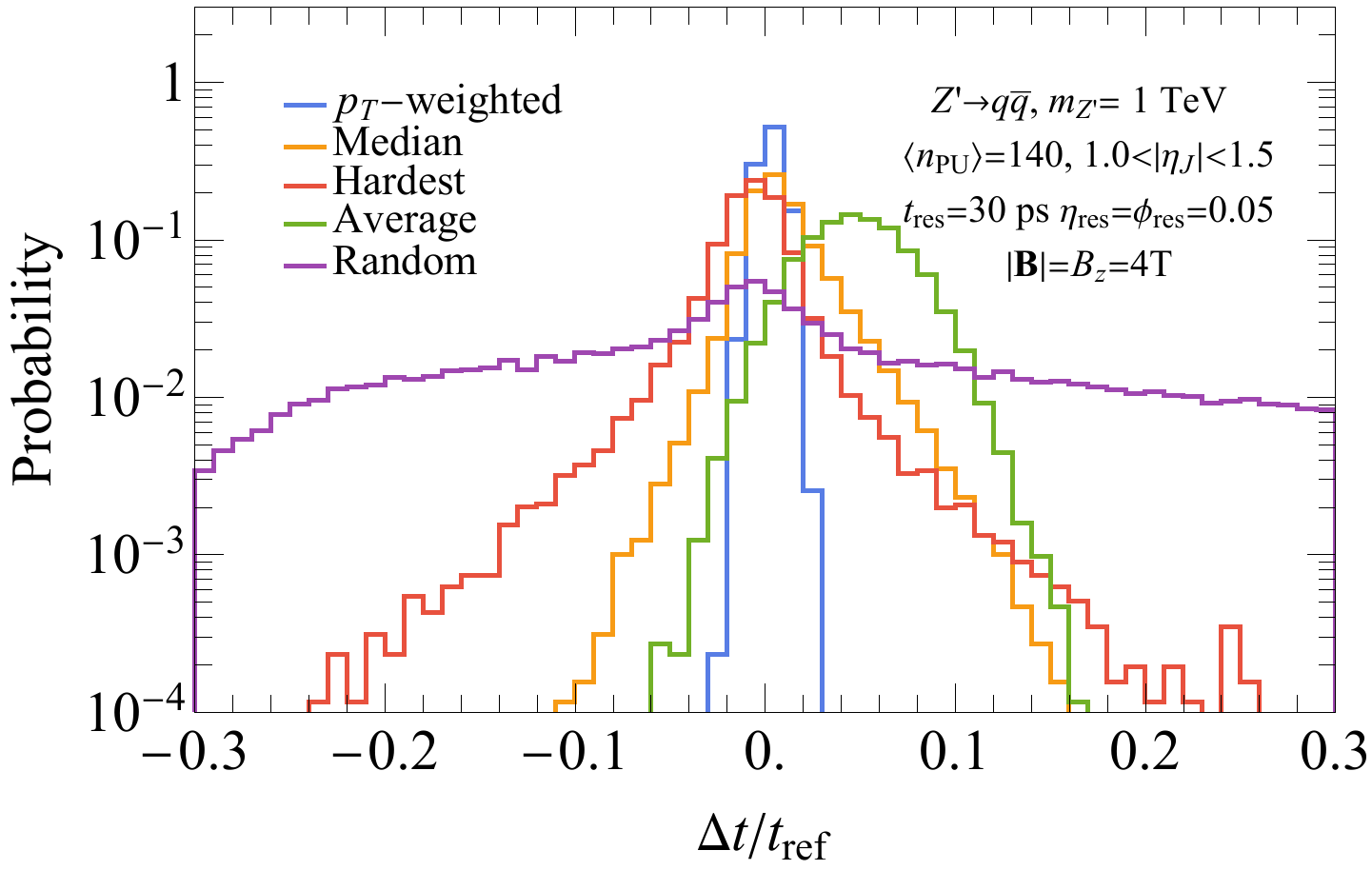}
  \caption{The relative time difference distribution for $|\eta_J|<0.5$ (left) and $1.0<|\eta_J|<1.5$ (right) with pileup and no subtraction. The $|\eta_J|<0.5$ plot can be compared with Fig.~\ref{fig:relTD_prompt_with_bfield} (left) and the $1.0<|\eta_J|<1.5$ plot can be compared with Fig.~\ref{fig:relTD_prompt_with_bfield} (right).}
  \label{fig:relTD_prompt_with_pu}
\end{center}
\end{figure}

The relative time differences without any form of pileup mitigation are shown in Fig.~\ref{fig:relTD_prompt_with_pu} for $|\eta_J|<0.5$ (left) and $1.0<|\eta_J|<1.5$ (right). The average time distribution gets distorted for the $|\eta_J|<0.5$ bin, and the peak shifts away from zero considerably. Like those from pileup, low-energy particles have a smaller curvature radius from the magnetic field and are therefore delayed more than higher-energy particles. Since a sizable fraction of a jet's constituents can come from pileup, this causes the average time to shift considerably. The same reasoning is responsible for the broadening of the distributions of the median and $p_T$-weighted times. The hardest time is affected very little.

The $1.0<|\eta_J|<1.5$ bin shows less impact from pileup as can be seen, for example, by the peak of the average time distribution remaining close to zero. Similarly, the distributions of the other times broaden slightly, but their peaks do not shift. This is the result of the $p_T$ cut restricting to more energetic particles at larger pseudorapidities.

%%%%%%%%%%%%%%%%%%
%%%%%%%%%%%%%%%%%%

For pileup mitigation, we use an idealized version of charged hadron subtraction~\cite{Sirunyan:2017ulk} where we assume all charged pileup can be removed. The remaining particles were then clustered into $R_\text{jet}=0.5$ anti-$k_T$ jets and trimmed \cite{Krohn:2009th} with $R_\text{sub}=0.2$ and $f_\text{cut}=0.03$. The choice of keeping $R_\text{jet}$ the same is to ensure that the jet times with and without pileup mitigation are directly comparable.

\begin{figure} [H]
\begin{center}
  \includegraphics[width=0.5\textwidth]{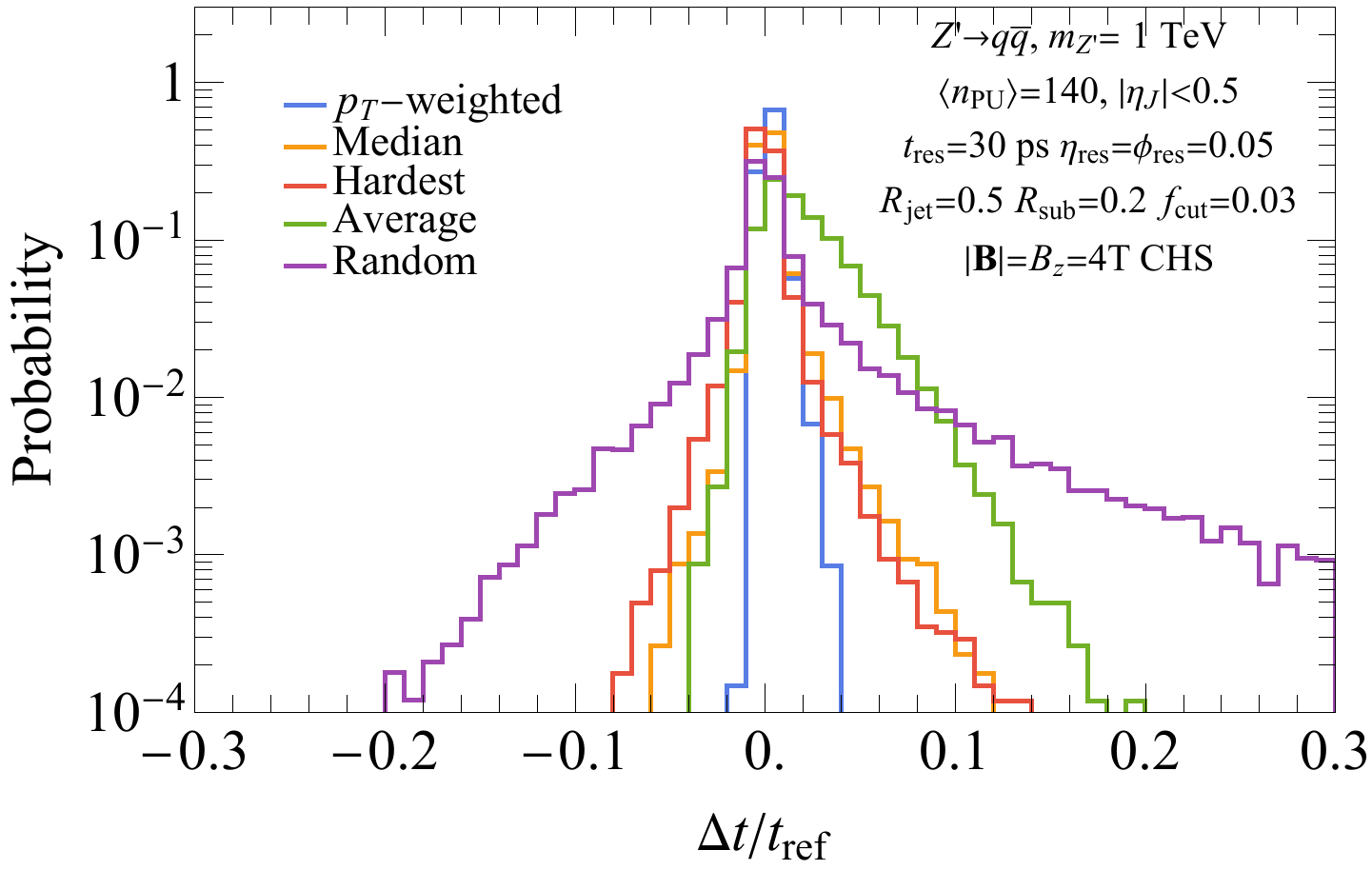}
  \caption{The relative time difference distribution for $|\eta_J|<0.5$ with pileup, charged hadron subtraction, and trimming. This can be compared with Fig.~\ref{fig:relTD_prompt_with_pu} (left).}
  \label{fig:relTD_prompt_with_pu_with_trimming}
\end{center}
\end{figure}

The distributions for the $|\eta_J|<0.5$ bin are shown in Fig.~\ref{fig:relTD_prompt_with_pu_with_trimming}. The improvement is predominantly due to the removal of soft charged pileup particles by charged hadron subtraction. These constituents are the ones that are mainly delayed by mass effects and the magnetic field. Trimming plays a minor role because $R_{\rm sub}$ is not significantly smaller than $\rjet$ and the number of pileup vertices is large. More aggressive trimming may improve results slightly.\footnote{One could also study the performance using pileup mitigation techniques that are better suited to large values of $\langle n_\text{PU} \rangle$ such as jet cleansing~\cite{Krohn:2013lba}, constituent subtraction~\cite{Berta:2014eza}, PUPPI~\cite{Bertolini:2014bba}, soft killer~\cite{Cacciari:2014gra}, or PUMML~\cite{Komiske:2017ubm}. This is beyond the scope of this work.}

%%%%%%%%%%%%%%%%%%
%%%%%%%%%%%%%%%%%%

%%%%%%%%%%%%%%%%%%%%%%%%%%%%%%%%%%%%%%%%%%%%%%%%%%%%%%%%%%%%%%%%%%
%%%%%%%%%%%%%%%%%%%%%%%%%%%%%%%%%%%%%%%%%%%%%%%%%%%%%%%%%%%%%%%%%%
\section{Detector Effects}
\label{app:detector}

\begin{figure} [H]
\begin{center}
  \includegraphics[width=0.3\textwidth]{Figs/dtt_prompt_eta05}\quad
  \includegraphics[width=0.3\textwidth]{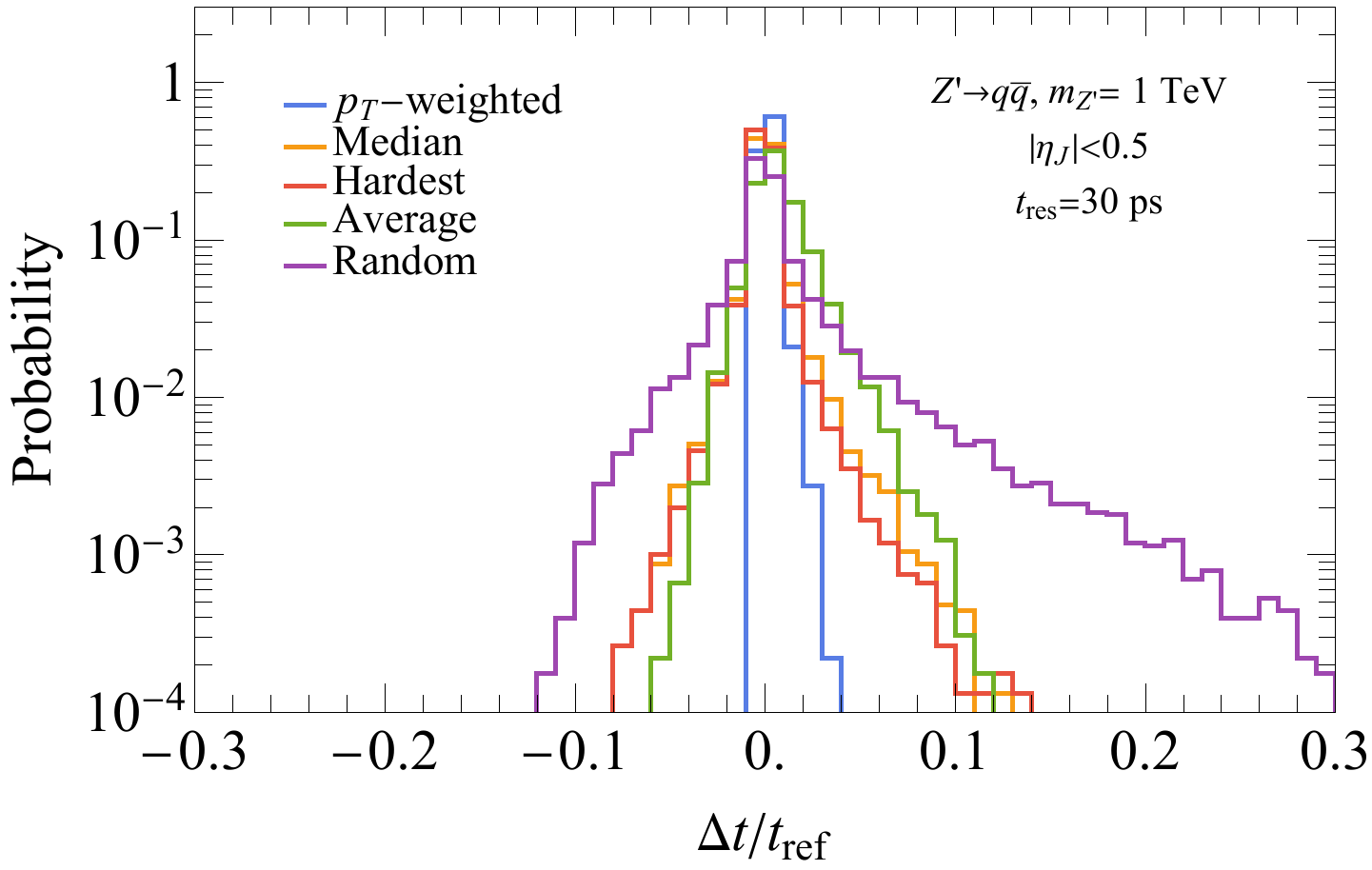} \quad
  \includegraphics[width=0.3\textwidth]{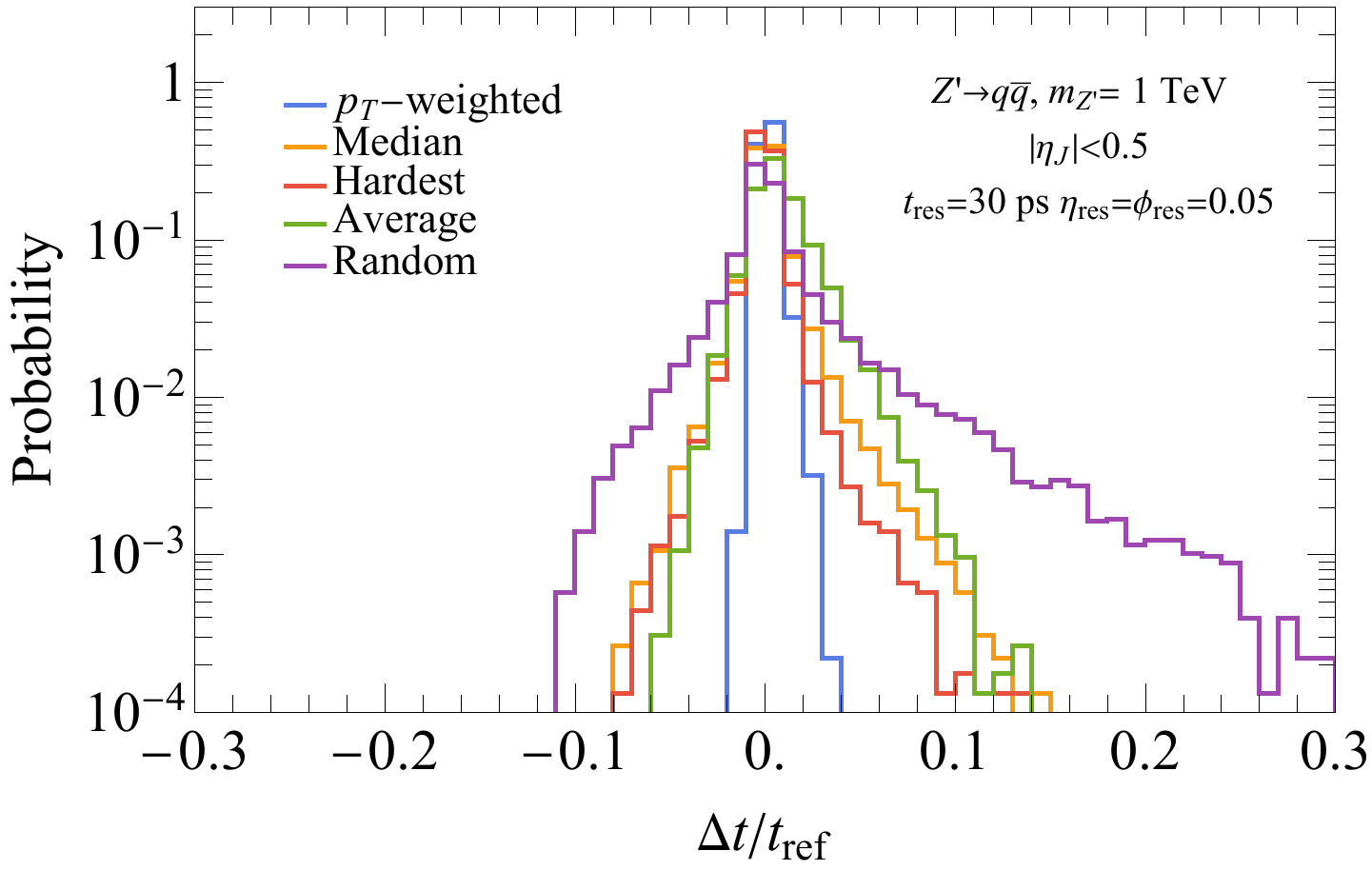}
  \caption{The relative time difference distribution for $|\eta_J|<0.5$ with no detector effects (left), with time resolution added (center), and with time and spatial resolution added (right).}
  \label{fig:relTD_prompt_05_with_resolution}
\end{center}
\end{figure}

\begin{figure} [H]
\begin{center}
  \includegraphics[width=0.3\textwidth]{Figs/dtt_prompt_eta15}\quad
  \includegraphics[width=0.3\textwidth]{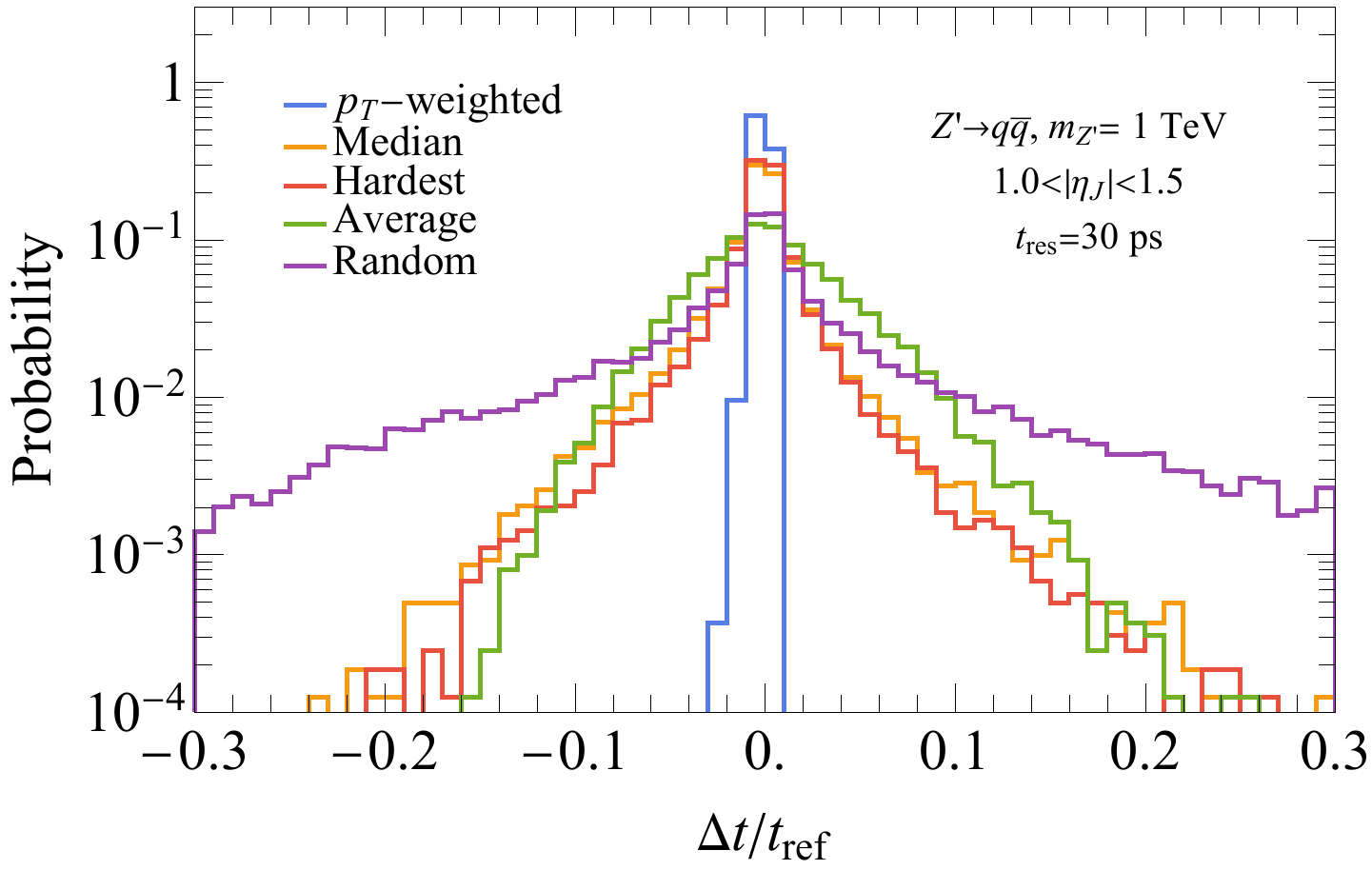} \quad
  \includegraphics[width=0.3\textwidth]{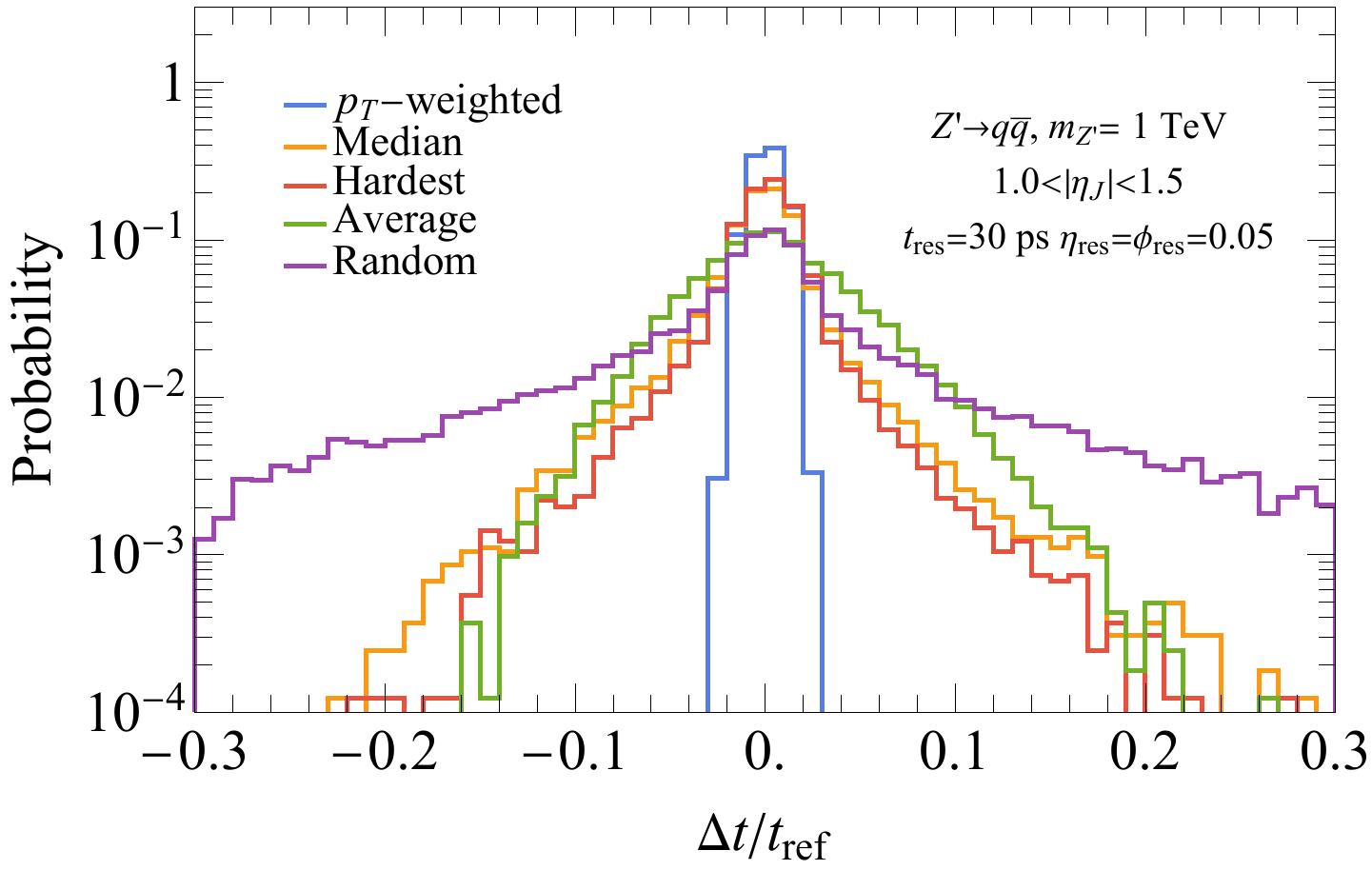}
  \caption{The relative time difference distribution for $1.0<|\eta_J|<1.5$ with no detector effects (left), with time resolution added (center), and with time and spatial resolution added (right).}
  \label{fig:relTD_prompt_20_with_resolution}
\end{center}
\end{figure}

In this section we show the effects of implementing a simple detector model. We first implement time resolution, followed by time and spatial resolution. The effects are shown in a prompt sample for $|\eta_J|<0.5$ in Fig.~\ref{fig:relTD_prompt_05_with_resolution} and for $1.5 < |\eta_J| < 2.0$ in Fig.~\ref{fig:relTD_prompt_20_with_resolution}.

For time resolution, we round each particle's time to the nearest multiple of $30~{\rm ps}$.  This is the expected resolution of LHC upgrades~\cite{Allaire:2018bof,Collaboration:2296612,Aaij:2244311}.\footnote{This time resolution does not include non-Gaussian tail effects (e.g. hot cells) which may affect per-particle arrival time measurements. These effects are very detector-dependent and is beyond the scope of this work.}  The effect on the $\Delta t / t_{\rm ref}$ distribution can be seen by comparing the left plots (no timing resolution) to the center plots ($30~{\rm ps}$ timing resolution) in Figs.~\ref{fig:relTD_prompt_05_with_resolution}~and~\ref{fig:relTD_prompt_20_with_resolution}. In both cases, the time resolution has a negligible effect on the shape of the distribution.

For spatial resolution, we consider an $\eta \times \phi$ grid of $0.05 \times 0.05$ cells. The four-momenta are replaced with a massless four-vector with the same energy as the particle, and the direction shifted pointing to the center of the corresponding $\eta\phi$-cell. If multiple particles fall into the same cell and the same time window, their energies are added, and they are combined into a single cell. The effect on the $\Delta t / t_{\rm ref}$ distribution can be seen in Fig.~\ref{fig:relTD_prompt_05_with_resolution} (right) and Fig.~\ref{fig:relTD_prompt_20_with_resolution} (right).

In the $|\eta_J|<0.5$ bin, we see that spatial resolution does not significantly affect the timing distributions. By contrast, the $1.0<|\eta_J|<1.5$ bin has a noticeable broadening in the $p_T$-weighted distribution and moderate broadening in the median and hardest distributions. This is because the fractional momentum resolution induced by the spatial resolution increases with $|\eta|$. This impacts both the momentum of the jets and their constituents.

%%%%%%%%%%%%%%%%%%
%%%%%%%%%%%%%%%%%%

%%%%%%%%%%%%%%%%%%
%%%%%%%%%%%%%%%%%%

\begin{figure} [H]
\begin{center}
  \includegraphics[width=0.45\textwidth]{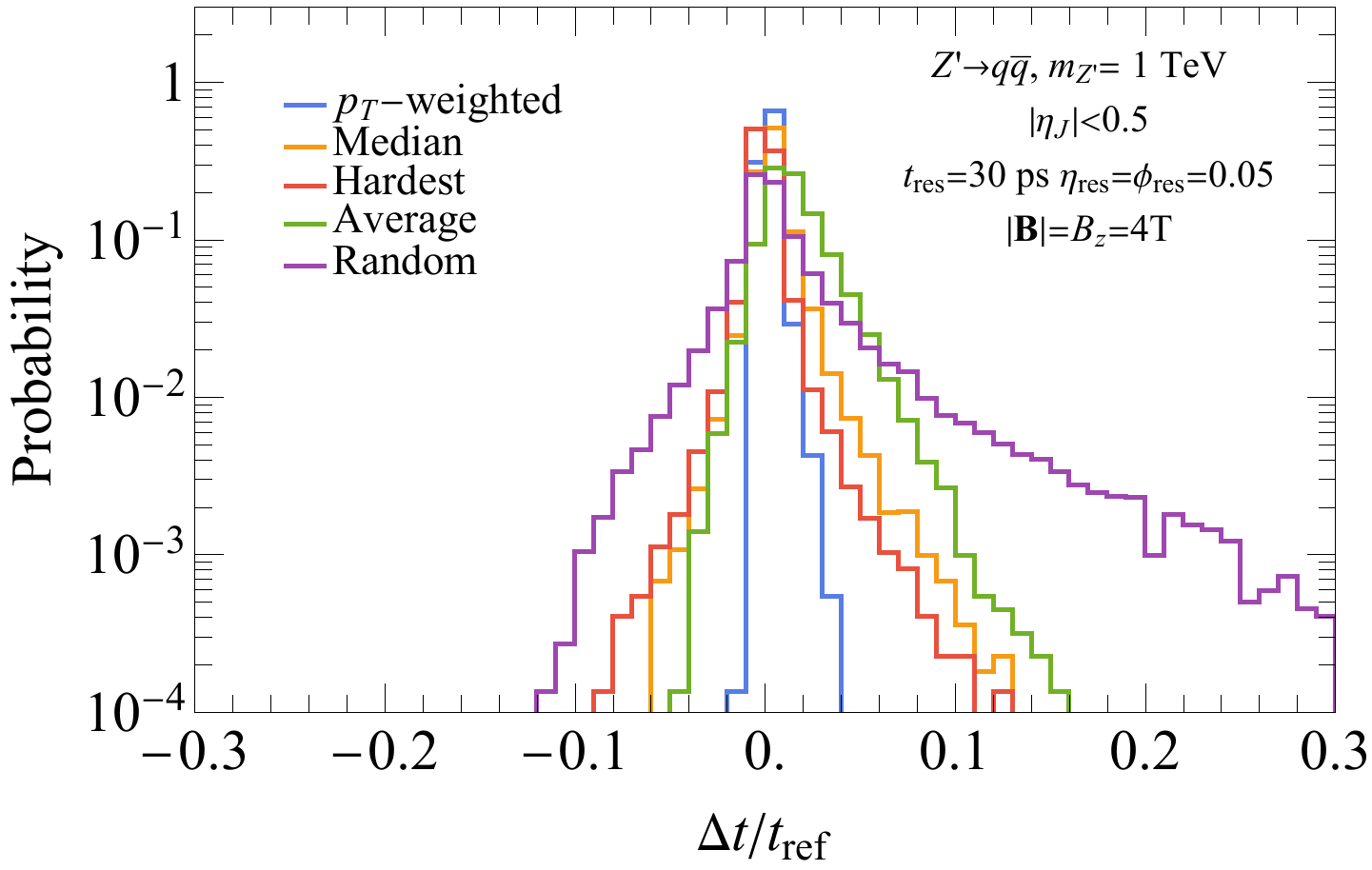} \quad\quad\quad
  \includegraphics[width=0.45\textwidth]{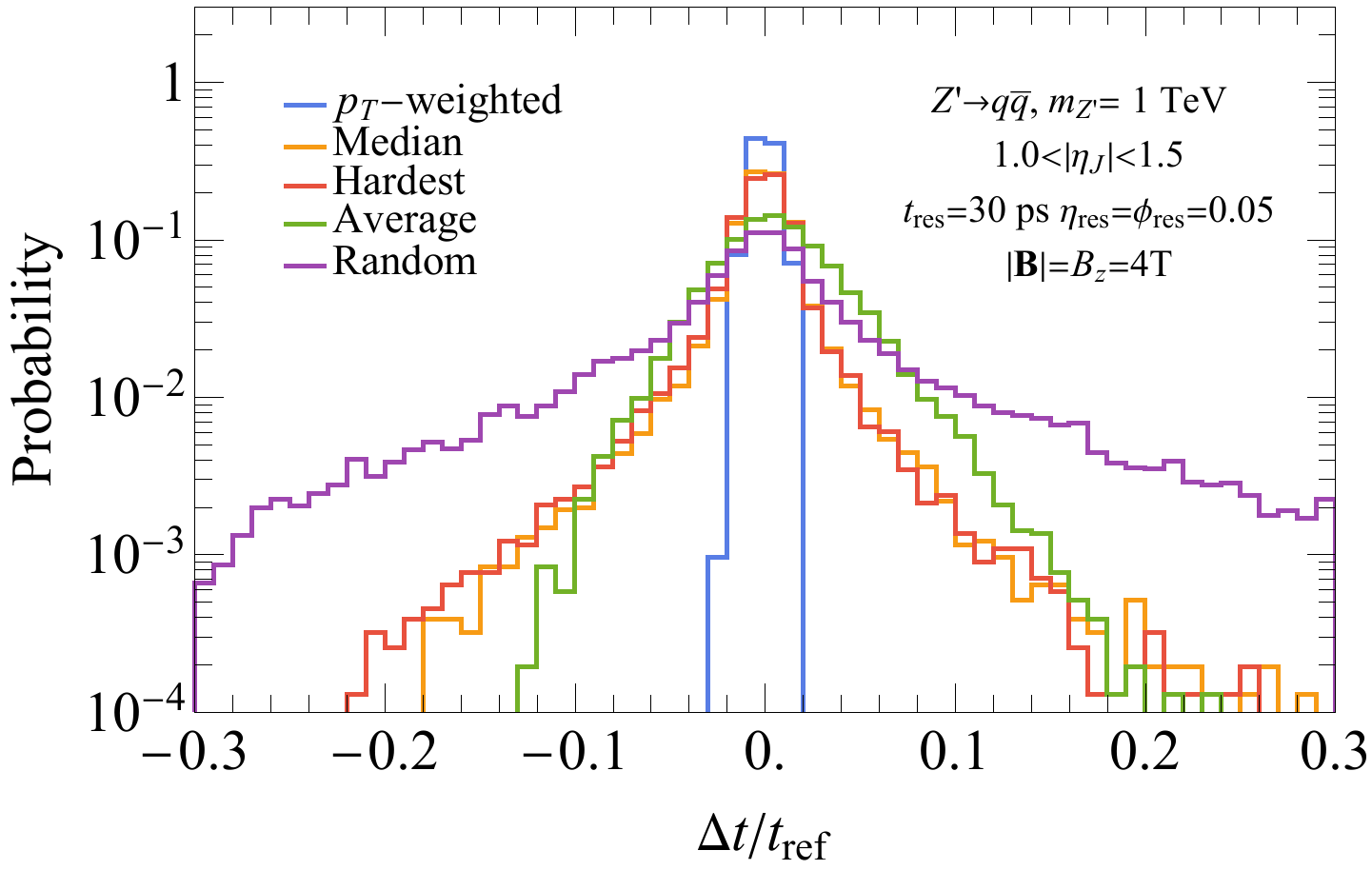}
  \caption{The relative time difference distribution for $|\eta_J|<0.5$ (left) and $1.0<|\eta_J|<1.5$ (right) with a magnetic field. The $|\eta_J|<0.5$ plot can be compared with Fig.~\ref{fig:relTD_prompt_05_with_resolution} (right) and the $1.0<|\eta_J|<1.5$ plot can be compared with Fig.~\ref{fig:relTD_prompt_20_with_resolution} (right).}
  \label{fig:relTD_prompt_with_bfield}
\end{center}
\end{figure}

Lastly, we considered the impact of including a $4$T magnetic field. In this case, the particles were hadronized (in order to get the correct electric charge of the hadrons), and at the same time and spatial resolution was applied. The effect on the relative time distribution for both bins are shown in Fig.~\ref{fig:relTD_prompt_with_bfield}. There is a very slight positive pull in the $|\eta_J|<0.5$ bin (left) while the $1.0<|\eta_J|<1.5$ bin (right) has no noticeable change. This difference is due to the $p_T>0.5~{\rm GeV}$ cut imposed on the constituents. As $\eta$ increases, the energy required to pass the $p_T$ cut also increases. The shift in arrival time due to the change in path length is inversely proportional to the energy.

%%%%%%%%%%%%%%%%%%
%%%%%%%%%%%%%%%%%%

%%%%%%%%%%%%%%%%%%%%%%%%%%%%%%%%%%%%%%%%%%%%%%%%%%%%%%%%%%%%%%%%%%
%%%%%%%%%%%%%%%%%%%%%%%%%%%%%%%%%%%%%%%%%%%%%%%%%%%%%%%%%%%%%%%%%%
\section{Hadronization}
\label{app:hadronization}

\begin{figure} [H]
\begin{center}
  \includegraphics[width=0.3\textwidth]{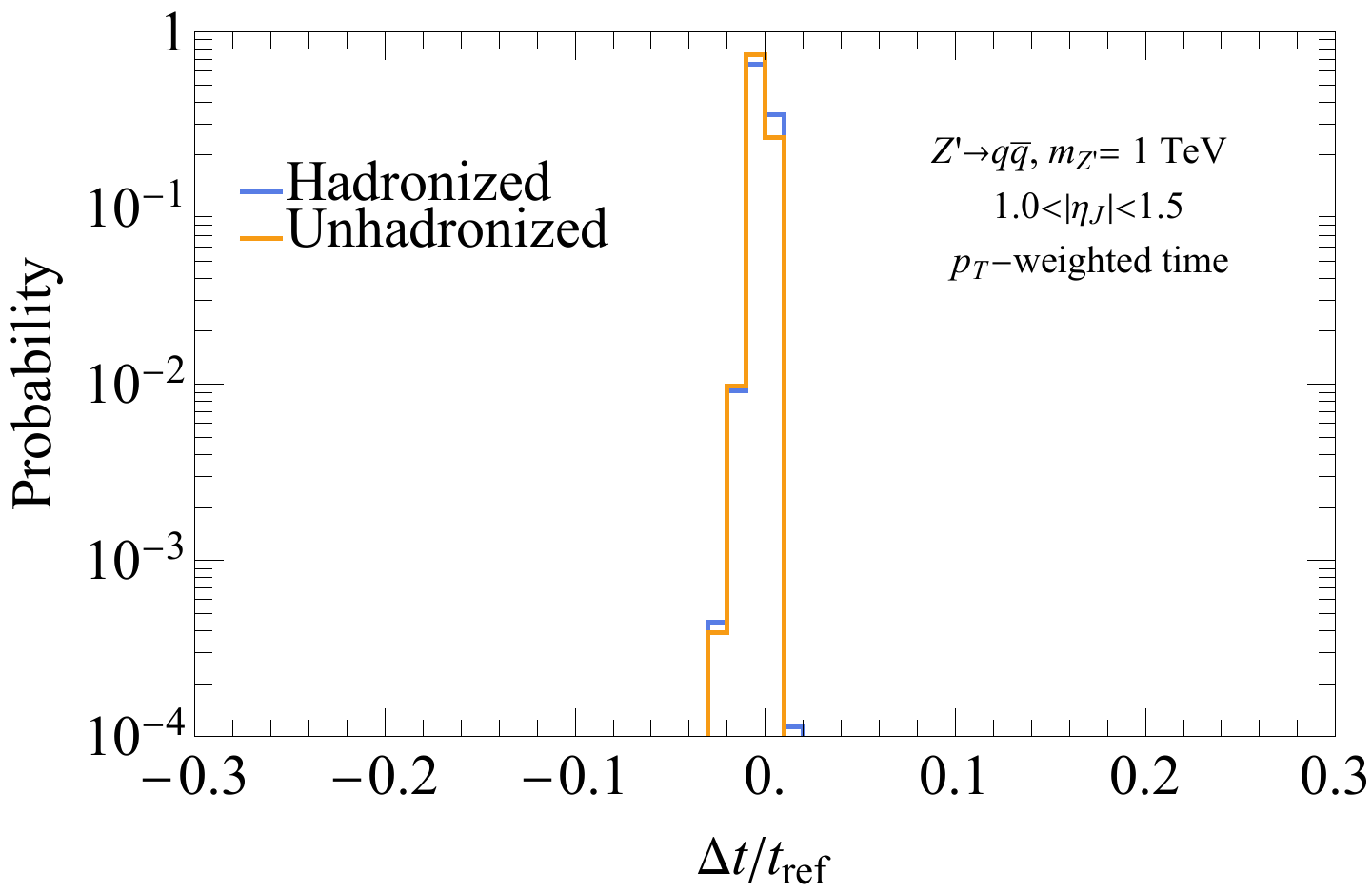} \quad
  \includegraphics[width=0.3\textwidth]{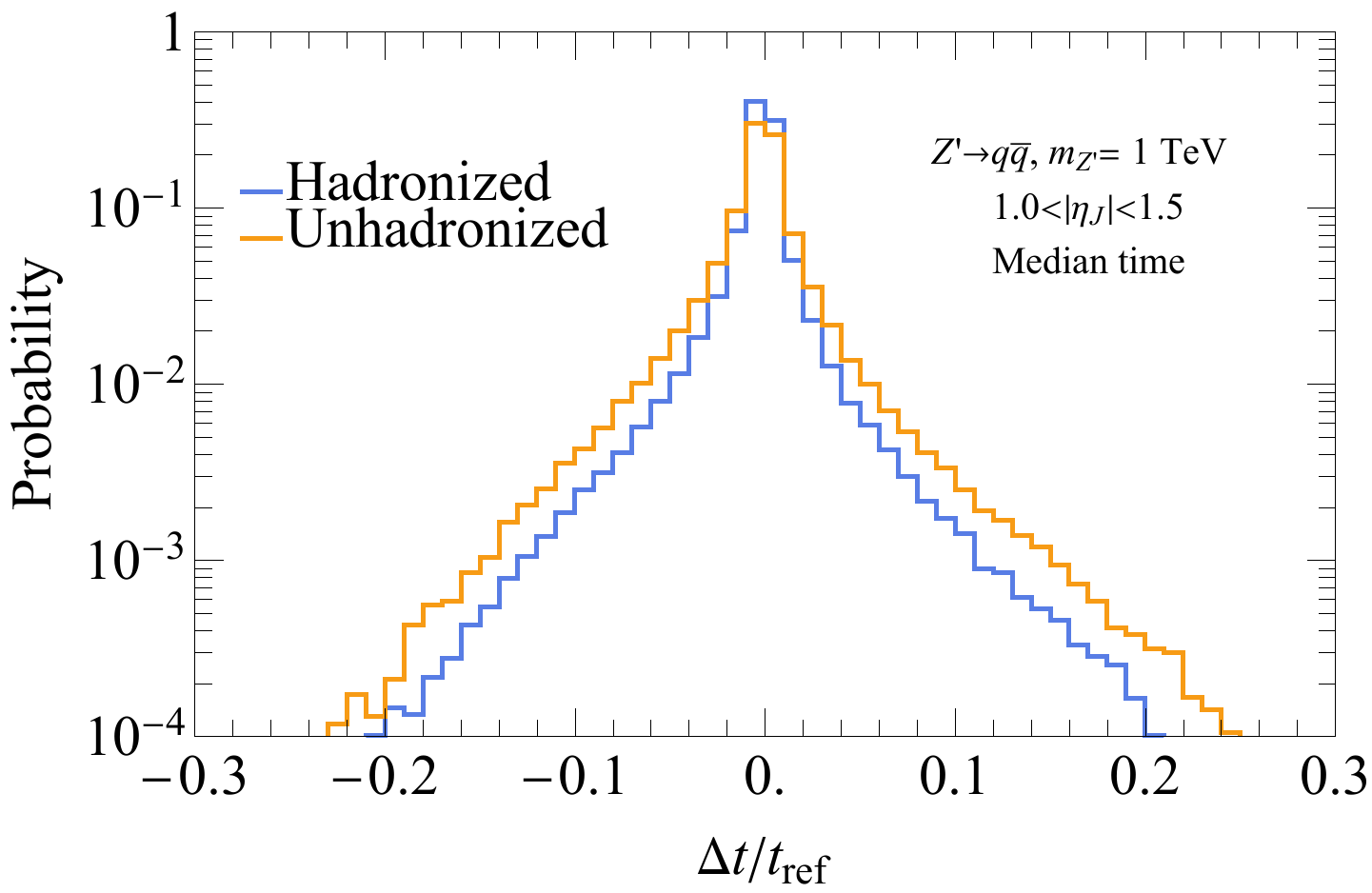} \quad
  \includegraphics[width=0.3\textwidth]{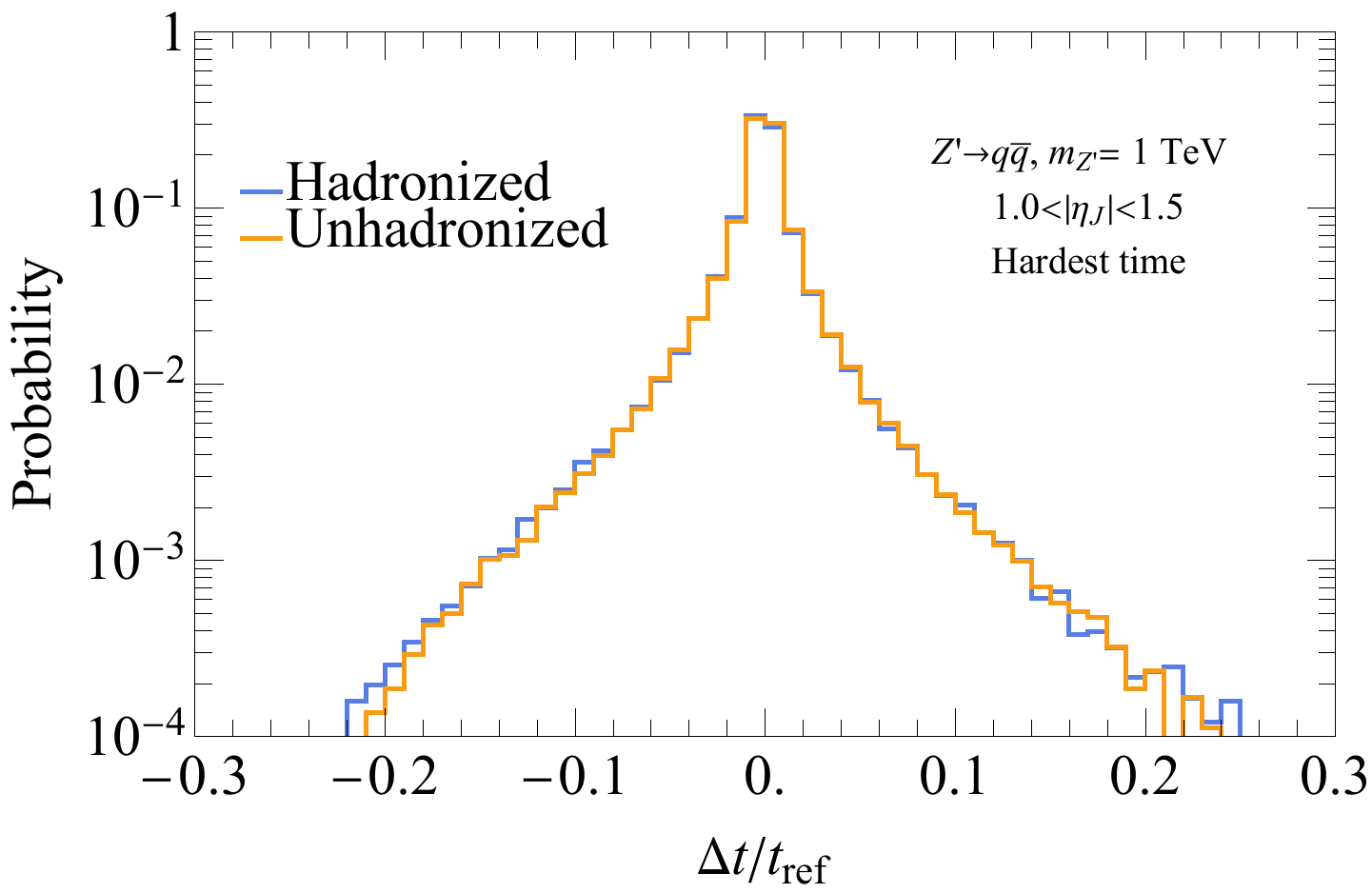}
  \caption{The relative time difference distributions with and without hadronization for the $p_T$-weighted time (left), median time (center), and hardest time (right) with $1.0 < |\eta_J| < 1.5$. }
  \label{fig:relTD_prompt_no_had}
\end{center}
\end{figure}

In this study the events are not hadronized to ensure that Pythia assigns the correct vertex to each delayed particle.  Fig.~\ref{fig:relTD_prompt_no_had} compares the relative time difference for prompt jets with and without hadronization for the $p_T$-weighted time (left), median time (center), and hardest time (right) in the range $1.0 < |\eta_J| < 1.5$.  Of the three, only the median time shows a slight observable change.

%%%%%%%%%%%%%%%%%%
%%%%%%%%%%%%%%%%%%

%%%%%%%%%%%%%%%%%%%%%%%%%%%%%%%%%%%%%%%%%%%%%%%%%%%%%%%%%%%%%%%%%%%%%%%%%%%%%%%%%%%
%%%%%%%%%%%%%%%%%%%%%%%%%%%%%%%%%%%%%%%%%%%%%%%%%%%%%%%%%%%%%%%%%%%%%%%%%%%%%%%%%%%
\section{Parameter Scans}
\label{app:para_scan}

%%%%%%%%%%%%%%%%%%
\begin{figure}%[H]
  \begin{center}
    \includegraphics[width=\textwidth]{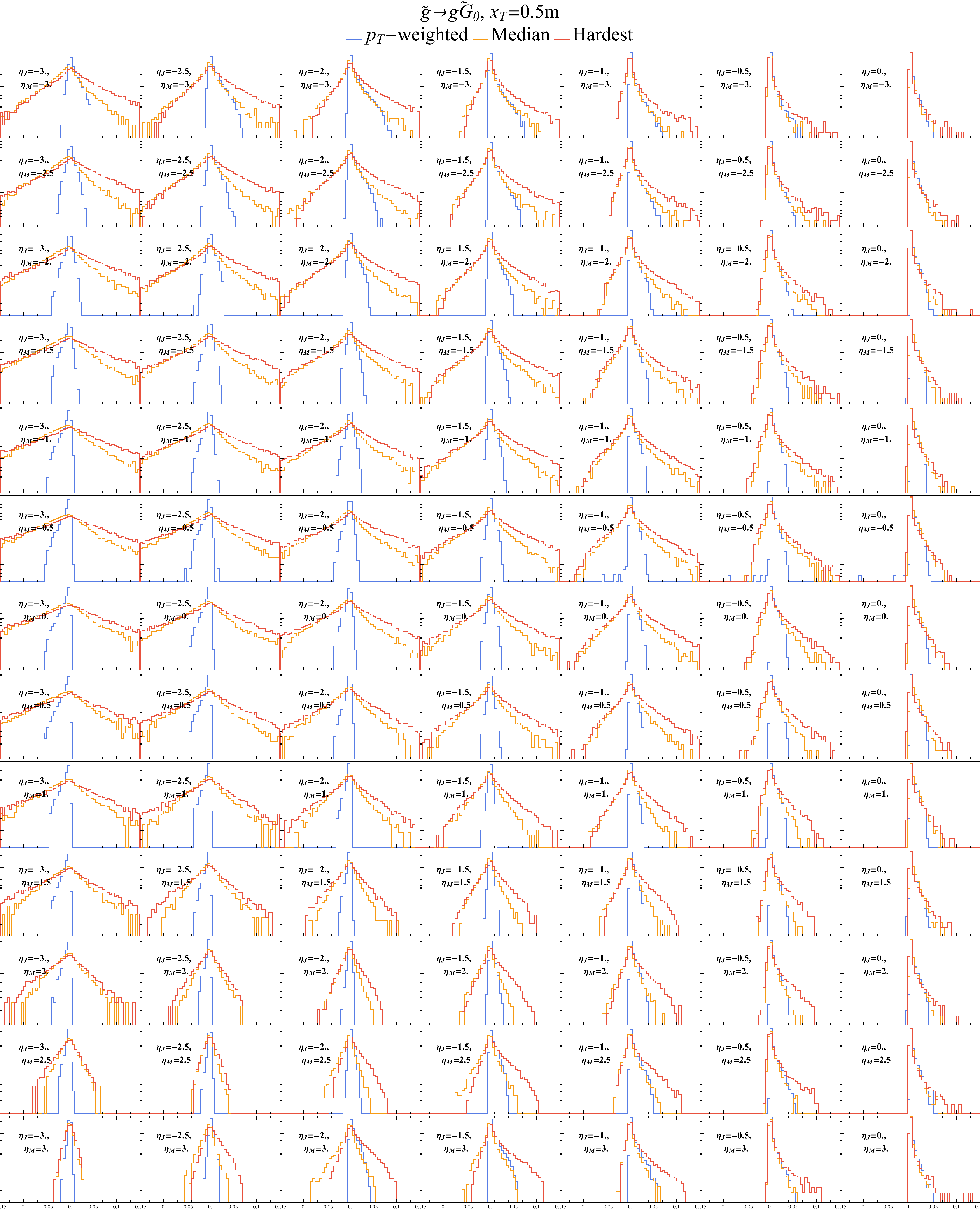}
    \caption{The relative time difference distribution for the $p_T$-weighted (blue), median (yellow), and hardest (red) times as a function of $\eta_J$ ($x$-axis) and of $\eta_M$ ($y$-axis) with a transverse decay location of $x_{T,M}=0.5~{\rm m}$. The vertical axis in each plot is in log-scale and ranges from $10^{-5}$ to $1$.}
    \label{fig:relTD_delayed_grid_ptw_median_hardest}
  \end{center}
  \end{figure}
%%%%%%%%%%%%%%%%%%
  
%%%%%%%%%%%%%%%%%%
\begin{figure}%[H]
  \begin{center}
    \includegraphics[width=\textwidth]{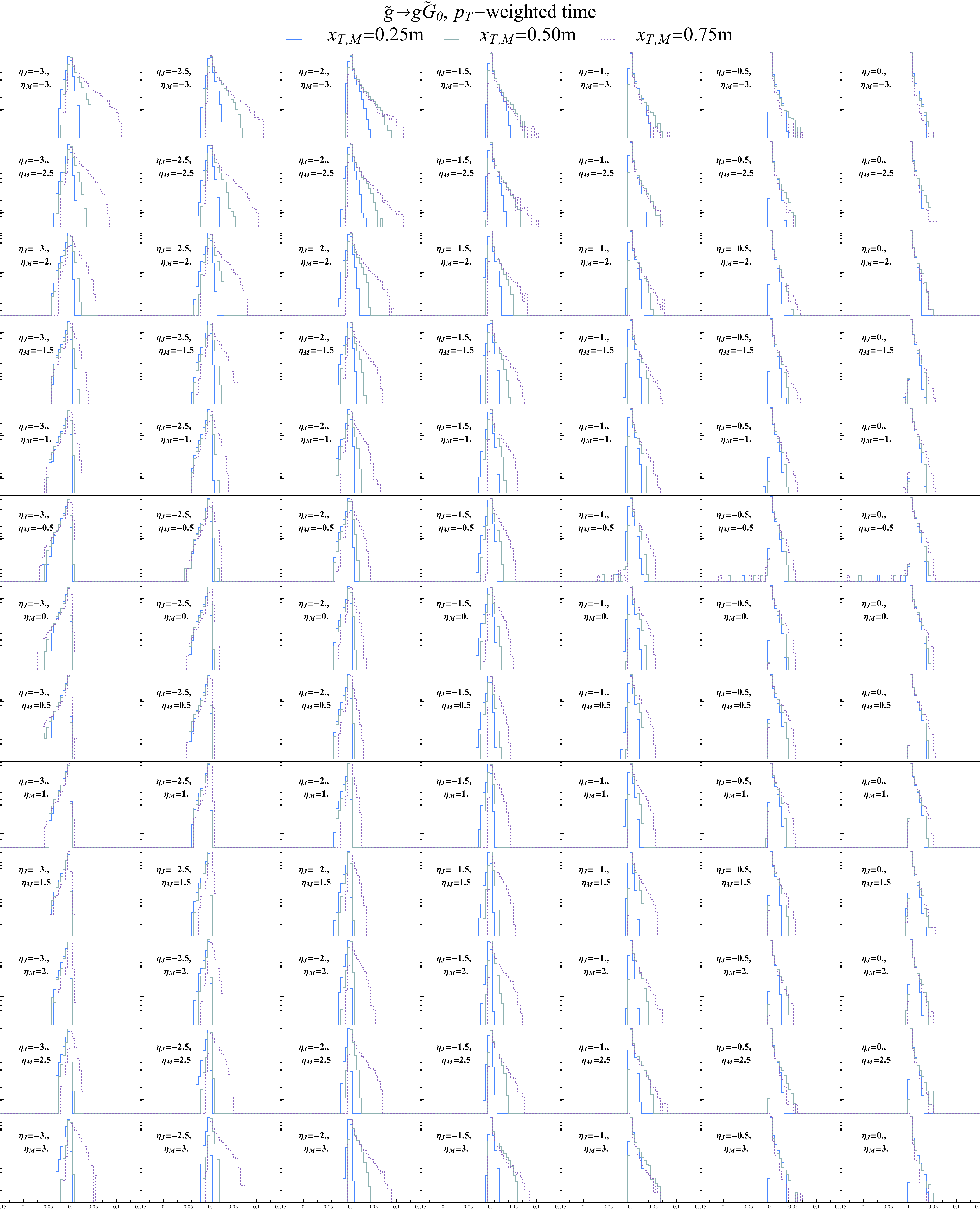}
    \caption{The relative time difference distribution for $x_{T,M} = 0.25~{\rm m}$ (blue), $x_{T,M} = 0.50~{\rm m}$ (gray), and $x_{T,M} = 0.75~{\rm m}$ (purple dashed) times as a function of $\eta_J$ ($x$-axis) and of $\eta_M$ ($y$-axis) for the $p_T$-weighted time. The vertical axis in each plot is in log-scale and ranges from $10^{-5}$ to 1.}
    \label{fig:relTD_delayed_grid_ptw}
  \end{center}
\end{figure}
%%%%%%%%%%%%%%%%%%
  
In Fig.~\ref{fig:relTD_delayed_grid_ptw_median_hardest} we show a scan over $\eta_J$ in the $x$-direction and $\eta_{M}$ in the $y$-direction for the $p_T$-weighted, median, and hardest times.  Here, we observe the general trend primarily follows Eq.~\eqref{eq:tdiff-delayed}.   Changing $\eta_J$ we see that distribution width tends to track with the corresponding pseudorapidity for the prompt distribution.  The slight narrowing at large $|\eta_{M}|$ is due to the changing daughter time fraction, as in Eq.~\eqref{eq:dtf-prefactor}.  The few deviations from this pattern are caused by larger changes in the effective radius and additionally from the observed kinematics for the $p_T$-weighted time.  
  
Fig.~\ref{fig:relTD_delayed_grid_ptw} shows a scan over $\eta_J$ in the $x$-direction and $\eta_{M}$ in the $y$-direction for different values of $x_{T,M}$. Here, we see the same behavior as Fig.~\ref{fig:diff_xt}. The distribution either narrows or broadens with $x_{T,M}$. The few deviations from this occurs whenever $R_\text{eff}$ is very small ({\it e.g.} at ($\eta_J,\eta_M)=(3,-3)$ for $x_{T,M}=0.5~\rm{m}$). The distributions are narrow over the full parameter space.

\bibliographystyle{JHEP}
\bibliography{refs}

\providecommand{\href}[2]{#2}\begingroup\raggedright\begin{thebibliography}{10}

\bibitem{CMS:2016kce}
{\scshape CMS} collaboration, \emph{{Search for long-lived charged particles in
  proton-proton collisions at $\sqrt s=$ 13 TeV}},
  \href{https://doi.org/10.1103/PhysRevD.94.112004}{\emph{Phys. Rev. D}
  {\bfseries 94} (2016) 112004}
  [\href{https://arxiv.org/abs/1609.08382}{{\ttfamily 1609.08382}}].

\bibitem{ATLAS:2019gqq}
{\scshape ATLAS} collaboration, \emph{{Search for heavy charged long-lived
  particles in the ATLAS detector in 36.1 fb$^{-1}$ of proton-proton collision
  data at $\sqrt{s} = 13$ TeV}},
  \href{https://doi.org/10.1103/PhysRevD.99.092007}{\emph{Phys. Rev. D}
  {\bfseries 99} (2019) 092007}
  [\href{https://arxiv.org/abs/1902.01636}{{\ttfamily 1902.01636}}].

\bibitem{CMS:2017kku}
{\scshape CMS} collaboration, \emph{{Search for decays of stopped exotic
  long-lived particles produced in proton-proton collisions at $\sqrt{s}=$ 13
  TeV}}, \href{https://doi.org/10.1007/JHEP05(2018)127}{\emph{JHEP} {\bfseries
  05} (2018) 127} [\href{https://arxiv.org/abs/1801.00359}{{\ttfamily
  1801.00359}}].

\bibitem{ATLAS:2021mdj}
{\scshape ATLAS} collaboration, \emph{{A search for the decays of stopped
  long-lived particles at $\sqrt{s}=13$ TeV with the ATLAS detector}},
  \href{https://arxiv.org/abs/2104.03050}{{\ttfamily 2104.03050}}.

\bibitem{ATLAS:2014kbb}
{\scshape ATLAS} collaboration, \emph{{Search for nonpointing and delayed
  photons in the diphoton and missing transverse momentum final state in 8 TeV
  $pp$ collisions at the LHC using the ATLAS detector}},
  \href{https://doi.org/10.1103/PhysRevD.90.112005}{\emph{Phys. Rev. D}
  {\bfseries 90} (2014) 112005}
  [\href{https://arxiv.org/abs/1409.5542}{{\ttfamily 1409.5542}}].

\bibitem{CMS:2019zxa}
{\scshape CMS} collaboration, \emph{{Search for long-lived particles using
  delayed photons in proton-proton collisions at $\sqrt{s}=$ 13 TeV}},
  \href{https://doi.org/10.1103/PhysRevD.100.112003}{\emph{Phys. Rev. D}
  {\bfseries 100} (2019) 112003}
  [\href{https://arxiv.org/abs/1909.06166}{{\ttfamily 1909.06166}}].

\bibitem{Liu:2018wte}
J.~Liu, Z.~Liu and L.-T.~Wang, \emph{{Enhancing Long-Lived Particles Searches
  at the LHC with Precision Timing Information}},
  \href{https://doi.org/10.1103/PhysRevLett.122.131801}{\emph{Phys. Rev. Lett.}
  {\bfseries 122} (2019) 131801}
  [\href{https://arxiv.org/abs/1805.05957}{{\ttfamily 1805.05957}}].

\bibitem{Klimek:2019cny}
M.D.~Klimek, \emph{{The Time Substructure of Jets and Boosted Object Tagging}},
   \href{https://arxiv.org/abs/1911.11235}{{\ttfamily 1911.11235}}.

\bibitem{Flowers:2019gvj}
Z.~Flowers, Q.~Meier, C.~Rogan, D.W.~Kang and S.C.~Park, \emph{{Timing
  information at HL-LHC: Complete determination of masses of Dark Matter and
  Long lived particle}},
  \href{https://doi.org/10.1007/JHEP03(2020)132}{\emph{JHEP} {\bfseries 03}
  (2020) 132} [\href{https://arxiv.org/abs/1903.05825}{{\ttfamily
  1903.05825}}].

\bibitem{Kang:2019ukr}
Z.~Flowers, Q.~Meier, C.~Rogan, D.W.~Kang and S.C.~Park, \emph{{Timing
  information at HL-LHC: Complete determination of masses of Dark Matter and
  Long lived particle}},
  \href{https://doi.org/10.1007/JHEP03(2020)132}{\emph{JHEP} {\bfseries 03}
  (2020) 132} [\href{https://arxiv.org/abs/1903.05825}{{\ttfamily
  1903.05825}}].

\bibitem{Banerjee:2019ktv}
S.~Banerjee, B.~Bhattacherjee, A.~Goudelis, B.~Herrmann, D.~Sengupta and
  R.~Sengupta, \emph{{Determining the lifetime of long-lived particles at the
  HL-LHC}}, \href{https://doi.org/10.1140/epjc/s10052-021-08945-9}{\emph{Eur.
  Phys. J. C} {\bfseries 81} (2021) 172}
  [\href{https://arxiv.org/abs/1912.06669}{{\ttfamily 1912.06669}}].

\bibitem{Bae:2020dwf}
K.J.~Bae, M.~Park and M.~Zhang, \emph{{Demystifying freeze-in dark matter at
  the LHC}}, \href{https://doi.org/10.1103/PhysRevD.101.115036}{\emph{Phys.
  Rev. D} {\bfseries 101} (2020) 115036}
  [\href{https://arxiv.org/abs/2001.02142}{{\ttfamily 2001.02142}}].

\bibitem{ElHedri:2018atj}
S.~El~Hedri and M.~de~Vries, \emph{{Cornering Colored Coannihilation}},
  \href{https://doi.org/10.1007/JHEP10(2018)102}{\emph{JHEP} {\bfseries 10}
  (2018) 102} [\href{https://arxiv.org/abs/1806.03325}{{\ttfamily
  1806.03325}}].

\bibitem{Cerri:2018rkm}
O.~Cerri, S.~Xie, C.~Pena and M.~Spiropulu, \emph{{Identification of Long-lived
  Charged Particles using Time-Of-Flight Systems at the Upgraded LHC
  detectors}}, \href{https://doi.org/10.1007/JHEP04(2019)037}{\emph{JHEP}
  {\bfseries 04} (2019) 037}
  [\href{https://arxiv.org/abs/1807.05453}{{\ttfamily 1807.05453}}].

\bibitem{Abada:2018sfh}
A.~Abada, N.~Bernal, M.~Losada and X.~Marcano, \emph{{Inclusive Displaced
  Vertex Searches for Heavy Neutral Leptons at the LHC}},
  \href{https://doi.org/10.1007/JHEP01(2019)093}{\emph{JHEP} {\bfseries 01}
  (2019) 093} [\href{https://arxiv.org/abs/1807.10024}{{\ttfamily
  1807.10024}}].

\bibitem{Frugiuele:2018coc}
C.~Frugiuele, E.~Fuchs, G.~Perez and M.~Schlaffer, \emph{{Relaxion and light
  (pseudo)scalars at the HL-LHC and lepton colliders}},
  \href{https://doi.org/10.1007/JHEP10(2018)151}{\emph{JHEP} {\bfseries 10}
  (2018) 151} [\href{https://arxiv.org/abs/1807.10842}{{\ttfamily
  1807.10842}}].

\bibitem{Kribs:2018ilo}
G.D.~Kribs, A.~Martin, B.~Ostdiek and T.~Tong, \emph{{Dark Mesons at the LHC}},
  \href{https://doi.org/10.1007/JHEP07(2019)133}{\emph{JHEP} {\bfseries 07}
  (2019) 133} [\href{https://arxiv.org/abs/1809.10184}{{\ttfamily
  1809.10184}}].

\bibitem{Berlin:2018jbm}
A.~Berlin and F.~Kling, \emph{{Inelastic Dark Matter at the LHC Lifetime
  Frontier: ATLAS, CMS, LHCb, CODEX-b, FASER, and MATHUSLA}},
  \href{https://doi.org/10.1103/PhysRevD.99.015021}{\emph{Phys. Rev. D}
  {\bfseries 99} (2019) 015021}
  [\href{https://arxiv.org/abs/1810.01879}{{\ttfamily 1810.01879}}].

\bibitem{Xu:2018ofw}
L.-X.~Xu, J.-H.~Yu and S.-H.~Zhu, \emph{{Minimal neutral naturalness model}},
  \href{https://doi.org/10.1103/PhysRevD.101.095014}{\emph{Phys. Rev. D}
  {\bfseries 101} (2020) 095014}
  [\href{https://arxiv.org/abs/1810.01882}{{\ttfamily 1810.01882}}].

\bibitem{Belanger:2018sti}
G.~B\'elanger et~al., \emph{{LHC-friendly minimal freeze-in models}},
  \href{https://doi.org/10.1007/JHEP02(2019)186}{\emph{JHEP} {\bfseries 02}
  (2019) 186} [\href{https://arxiv.org/abs/1811.05478}{{\ttfamily
  1811.05478}}].

\bibitem{Evans:2018jmd}
J.A.~Evans and M.A.~Luty, \emph{{Stopping Quirks at the LHC}},
  \href{https://doi.org/10.1007/JHEP06(2019)090}{\emph{JHEP} {\bfseries 06}
  (2019) 090} [\href{https://arxiv.org/abs/1811.08903}{{\ttfamily
  1811.08903}}].

\bibitem{Kilic:2018sew}
C.~Kilic, S.~Najjari and C.B.~Verhaaren, \emph{{Discovering the Twin Higgs
  Boson with Displaced Decays}},
  \href{https://doi.org/10.1103/PhysRevD.99.075029}{\emph{Phys. Rev. D}
  {\bfseries 99} (2019) 075029}
  [\href{https://arxiv.org/abs/1812.08173}{{\ttfamily 1812.08173}}].

\bibitem{Delgado:2018qxq}
A.~Delgado, A.~Martin and M.~Quir\'os, \emph{{Higgsino Dark Matter in an
  economical Scherk-Schwarz setup}},
  \href{https://doi.org/10.1103/PhysRevD.99.075015}{\emph{Phys. Rev. D}
  {\bfseries 99} (2019) 075015}
  [\href{https://arxiv.org/abs/1812.08019}{{\ttfamily 1812.08019}}].

\bibitem{Liu:2019ayx}
J.~Liu, Z.~Liu, L.-T.~Wang and X.-P.~Wang, \emph{{Seeking for sterile neutrinos
  with displaced leptons at the LHC}},
  \href{https://doi.org/10.1007/JHEP07(2019)159}{\emph{JHEP} {\bfseries 07}
  (2019) 159} [\href{https://arxiv.org/abs/1904.01020}{{\ttfamily
  1904.01020}}].

\bibitem{Chakraborti:2019ohe}
S.~Chakraborti, V.~Martin and P.~Poulose, \emph{{Freeze-in and freeze-out of
  dark matter with charged long-lived partners}},
  \href{https://doi.org/10.1088/1475-7516/2020/03/057}{\emph{JCAP} {\bfseries
  03} (2020) 057} [\href{https://arxiv.org/abs/1904.09945}{{\ttfamily
  1904.09945}}].

\bibitem{Serra:2019omd}
J.~Serra, S.~Stelzl, R.~Torre and A.~Weiler, \emph{{Hypercharged Naturalness}},
  \href{https://doi.org/10.1007/JHEP10(2019)060}{\emph{JHEP} {\bfseries 10}
  (2019) 060} [\href{https://arxiv.org/abs/1905.02203}{{\ttfamily
  1905.02203}}].

\bibitem{Mason:2019okp}
J.D.~Mason, \emph{{Time-Delayed Electrons from Higgs Decays to Right-Handed
  Neutrinos}}, \href{https://doi.org/10.1007/JHEP07(2019)089}{\emph{JHEP}
  {\bfseries 07} (2019) 089}
  [\href{https://arxiv.org/abs/1905.07772}{{\ttfamily 1905.07772}}].

\bibitem{Du:2019mlc}
M.~Du, Z.~Liu and V.Q.~Tran, \emph{{Enhanced Long-Lived Dark Photon Signals at
  the LHC}}, \href{https://doi.org/10.1007/JHEP05(2020)055}{\emph{JHEP}
  {\bfseries 05} (2020) 055}
  [\href{https://arxiv.org/abs/1912.00422}{{\ttfamily 1912.00422}}].

\bibitem{Zabi:2020gjd}
{\scshape CMS} collaboration, \emph{{The Phase-2 Upgrade of the CMS Level-1
  Trigger}}, .

\bibitem{Shuve:2020evk}
B.~Shuve and D.~Tucker-Smith, \emph{{Baryogenesis and Dark Matter from
  Freeze-In}}, \href{https://doi.org/10.1103/PhysRevD.101.115023}{\emph{Phys.
  Rev. D} {\bfseries 101} (2020) 115023}
  [\href{https://arxiv.org/abs/2004.00636}{{\ttfamily 2004.00636}}].

\bibitem{Yuan:2020eeu}
C.~Yuan, H.~Zhang, Y.~Zhao and G.~Chen, \emph{{Producing and detecting
  long-lived particles at different experiments at the LHC}},
  \href{https://arxiv.org/abs/2004.08820}{{\ttfamily 2004.08820}}.

\bibitem{Liu:2020vur}
J.~Liu, Z.~Liu, L.-T.~Wang and X.-P.~Wang, \emph{{Enhancing Sensitivities to
  Long-lived Particles with High Granularity Calorimeters at the LHC}},
  \href{https://doi.org/10.1007/JHEP11(2020)066}{\emph{JHEP} {\bfseries 11}
  (2020) 066} [\href{https://arxiv.org/abs/2005.10836}{{\ttfamily
  2005.10836}}].

\bibitem{Fuchs:2020cmm}
E.~Fuchs, O.~Matsedonskyi, I.~Savoray and M.~Schlaffer, \emph{{Collider
  searches for scalar singlets across lifetimes}},
  \href{https://doi.org/10.1007/JHEP04(2021)019}{\emph{JHEP} {\bfseries 04}
  (2021) 019} [\href{https://arxiv.org/abs/2008.12773}{{\ttfamily
  2008.12773}}].

\bibitem{Gershtein:2020mwi}
Y.~Gershtein, S.~Knapen and D.~Redigolo, \emph{{Probing naturally light
  singlets with a displaced vertex trigger}},
  \href{https://arxiv.org/abs/2012.07864}{{\ttfamily 2012.07864}}.

\bibitem{Borsato:2021aum}
M.~Borsato et~al., \emph{{Unleashing the full power of LHCb to probe Stealth
  New Physics}},  \href{https://arxiv.org/abs/2105.12668}{{\ttfamily
  2105.12668}}.

\bibitem{Cheung:2021utb}
K.~Cheung, K.~Wang and Z.S.~Wang, \emph{{Time-delayed electrons from neutral
  currents at the LHC}},  \href{https://arxiv.org/abs/2107.03203}{{\ttfamily
  2107.03203}}.

\bibitem{Dienes:2021cxr}
K.R.~Dienes, D.~Kim, T.~Leininger and B.~Thomas, \emph{{Tumblers: A Novel
  Collider Signature for Long-Lived Particles}},
  \href{https://arxiv.org/abs/2108.02204}{{\ttfamily 2108.02204}}.

\bibitem{Bhattacherjee:2020nno}
B.~Bhattacherjee, S.~Mukherjee, R.~Sengupta and P.~Solanki, \emph{{Triggering
  long-lived particles in HL-LHC and the challenges in the rst stage of the
  trigger system}}, \href{https://doi.org/10.1007/JHEP08(2020)141}{\emph{JHEP}
  {\bfseries 08} (2020) 141}
  [\href{https://arxiv.org/abs/2003.03943}{{\ttfamily 2003.03943}}].

\bibitem{Allaire:2018bof}
C.~Allaire et~al., \emph{{Beam test measurements of Low Gain Avalanche Detector
  single pads and arrays for the ATLAS High Granularity Timing Detector}},
  \href{https://arxiv.org/abs/1804.00622}{{\ttfamily 1804.00622}}.

\bibitem{Collaboration:2296612}
\emph{{TECHNICAL PROPOSAL FOR A MIP TIMING DETECTOR IN THE CMS EXPERIMENT PHASE
  2 UPGRADE}},  Tech. Rep.
  \href{https://cds.cern.ch/record/2296612}{CERN-LHCC-2017-027. LHCC-P-009},
  CERN, Geneva (Dec, 2017).

\bibitem{Aaij:2244311}
{\scshape LHCb Collaboration} collaboration, \emph{{Expression of Interest for
  a Phase-II LHCb Upgrade: Opportunities in flavour physics, and beyond, in the
  HL-LHC era}},  Tech. Rep.
  \href{http://cds.cern.ch/record/2244311}{CERN-LHCC-2017-003}, CERN, Geneva
  (Feb, 2017).

\bibitem{Sirunyan:2019gut}
{\scshape CMS} collaboration, \emph{{Search for long-lived particles using
  nonprompt jets and missing transverse momentum with proton-proton collisions
  at $\sqrt{s} =$ 13 TeV}},
  \href{https://doi.org/10.1016/j.physletb.2019.134876}{\emph{Phys. Lett. B}
  {\bfseries 797} (2019) 134876}
  [\href{https://arxiv.org/abs/1906.06441}{{\ttfamily 1906.06441}}].

\bibitem{Salam:2009jx}
G.P.~Salam, \emph{{Towards Jetography}},
  \href{https://doi.org/10.1140/epjc/s10052-010-1314-6}{\emph{Eur. Phys. J. C}
  {\bfseries 67} (2010) 637} [\href{https://arxiv.org/abs/0906.1833}{{\ttfamily
  0906.1833}}].

\bibitem{CMS:TrackEff}
{\scshape CMS} collaboration, C.~Collaboartion, \emph{Cms tracking pog
  performance plots for 2017 with phase i pixel detector},  May, 2017.

\bibitem{Cacciari:2008gn}
M.~Cacciari, G.P.~Salam and G.~Soyez, \emph{{The Catchment Area of Jets}},
  \href{https://doi.org/10.1088/1126-6708/2008/04/005}{\emph{JHEP} {\bfseries
  04} (2008) 005} [\href{https://arxiv.org/abs/0802.1188}{{\ttfamily
  0802.1188}}].

\bibitem{Krohn:2009zg}
D.~Krohn, J.~Thaler and L.-T.~Wang, \emph{{Jets with Variable R}},
  \href{https://doi.org/10.1088/1126-6708/2009/06/059}{\emph{JHEP} {\bfseries
  06} (2009) 059} [\href{https://arxiv.org/abs/0903.0392}{{\ttfamily
  0903.0392}}].

\bibitem{Sjostrand:2014zea}
T.~Sj\"ostrand, S.~Ask, J.R.~Christiansen, R.~Corke, N.~Desai, P.~Ilten et~al.,
  \emph{{An introduction to PYTHIA 8.2}},
  \href{https://doi.org/10.1016/j.cpc.2015.01.024}{\emph{Comput. Phys. Commun.}
  {\bfseries 191} (2015) 159}
  [\href{https://arxiv.org/abs/1410.3012}{{\ttfamily 1410.3012}}].

\bibitem{Cacciari:2008gp}
M.~Cacciari, G.P.~Salam and G.~Soyez, \emph{{The anti-$k_t$ jet clustering
  algorithm}}, \href{https://doi.org/10.1088/1126-6708/2008/04/063}{\emph{JHEP}
  {\bfseries 04} (2008) 063} [\href{https://arxiv.org/abs/0802.1189}{{\ttfamily
  0802.1189}}].

\bibitem{Cacciari:2011ma}
M.~Cacciari, G.P.~Salam and G.~Soyez, \emph{{FastJet User Manual}},
  \href{https://doi.org/10.1140/epjc/s10052-012-1896-2}{\emph{Eur. Phys. J. C}
  {\bfseries 72} (2012) 1896}
  [\href{https://arxiv.org/abs/1111.6097}{{\ttfamily 1111.6097}}].

\bibitem{Alwall:2014hca}
J.~Alwall, R.~Frederix, S.~Frixione, V.~Hirschi, F.~Maltoni, O.~Mattelaer
  et~al., \emph{{The automated computation of tree-level and next-to-leading
  order differential cross sections, and their matching to parton shower
  simulations}}, \href{https://doi.org/10.1007/JHEP07(2014)079}{\emph{JHEP}
  {\bfseries 07} (2014) 079} [\href{https://arxiv.org/abs/1405.0301}{{\ttfamily
  1405.0301}}].

\bibitem{CMS:1994prop}
{\scshape CMS} collaboration, \emph{{CMS, the Compact Muon Solenoid: Technical
  proposal}}, .

\bibitem{ATLAS:1994prop}
{\scshape ATLAS Collaboration} collaboration, \emph{{ATLAS: technical proposal
  for a general-purpose pp experiment at the Large Hadron Collider at CERN}},
  LHC technical proposal, CERN, Geneva (1994).

\bibitem{Butler:2019rpu}
{\scshape CMS} collaboration, \emph{{A MIP Timing Detector for the CMS Phase-2
  Upgrade}}, .

\bibitem{Sirunyan:2017ulk}
{\scshape CMS} collaboration, \emph{{Particle-flow reconstruction and global
  event description with the CMS detector}},
  \href{https://doi.org/10.1088/1748-0221/12/10/P10003}{\emph{JINST} {\bfseries
  12} (2017) P10003} [\href{https://arxiv.org/abs/1706.04965}{{\ttfamily
  1706.04965}}].

\bibitem{Krohn:2009th}
D.~Krohn, J.~Thaler and L.-T.~Wang, \emph{{Jet Trimming}},
  \href{https://doi.org/10.1007/JHEP02(2010)084}{\emph{JHEP} {\bfseries 02}
  (2010) 084} [\href{https://arxiv.org/abs/0912.1342}{{\ttfamily 0912.1342}}].

\bibitem{Krohn:2013lba}
D.~Krohn, M.D.~Schwartz, M.~Low and L.-T.~Wang, \emph{{Jet Cleansing: Pileup
  Removal at High Luminosity}},
  \href{https://doi.org/10.1103/PhysRevD.90.065020}{\emph{Phys. Rev. D}
  {\bfseries 90} (2014) 065020}
  [\href{https://arxiv.org/abs/1309.4777}{{\ttfamily 1309.4777}}].

\bibitem{Berta:2014eza}
P.~Berta, M.~Spousta, D.W.~Miller and R.~Leitner, \emph{{Particle-level pileup
  subtraction for jets and jet shapes}},
  \href{https://doi.org/10.1007/JHEP06(2014)092}{\emph{JHEP} {\bfseries 06}
  (2014) 092} [\href{https://arxiv.org/abs/1403.3108}{{\ttfamily 1403.3108}}].

\bibitem{Bertolini:2014bba}
D.~Bertolini, P.~Harris, M.~Low and N.~Tran, \emph{{Pileup Per Particle
  Identification}}, \href{https://doi.org/10.1007/JHEP10(2014)059}{\emph{JHEP}
  {\bfseries 10} (2014) 059} [\href{https://arxiv.org/abs/1407.6013}{{\ttfamily
  1407.6013}}].

\bibitem{Cacciari:2014gra}
M.~Cacciari, G.P.~Salam and G.~Soyez, \emph{{SoftKiller, a particle-level
  pileup removal method}},
  \href{https://doi.org/10.1140/epjc/s10052-015-3267-2}{\emph{Eur. Phys. J. C}
  {\bfseries 75} (2015) 59} [\href{https://arxiv.org/abs/1407.0408}{{\ttfamily
  1407.0408}}].

\bibitem{Komiske:2017ubm}
P.T.~Komiske, E.M.~Metodiev, B.~Nachman and M.D.~Schwartz, \emph{{Pileup
  Mitigation with Machine Learning (PUMML)}},
  \href{https://doi.org/10.1007/JHEP12(2017)051}{\emph{JHEP} {\bfseries 12}
  (2017) 051} [\href{https://arxiv.org/abs/1707.08600}{{\ttfamily
  1707.08600}}].

\end{thebibliography}\endgroup
\end{document}